\documentclass[astrosymb,twocolumn]{aastex631}

\usepackage{gensymb}
\usepackage{lmodern}
\usepackage{amsmath}
\usepackage{natbib}
\usepackage[shortlabels]{enumitem}
\usepackage{amssymb}
\usepackage{xfrac}
\usepackage{calc}
\usepackage{booktabs}
\usepackage{hyperref}
\usepackage[nameinlink]{cleveref}
\setcitestyle{citesep={;}}
\setcitestyle{aysep={}}

\creflabelformat{equation}{#2\textup{#1}#3}

\newcommand{\RNum}[1]{\uppercase\expandafter{{\scshape\romannumeral #1\relax}}}
\defcitealias{Rajpaul2015}{R15}
\defcitealias{Aigrain2012}{A12}

\shorttitle{Gaussian Process Models Impact Planetary Properties}
\shortauthors{Tran \& Bowler}
\graphicspath{{./}{Figures/}}

\begin{document}

\title{Gaussian Process Models Impact the Inferred Properties of Giant Planets around Active Stars}

\correspondingauthor{Quang H. Tran}
\author[0000-0001-6532-6755]{Quang H. Tran}
\email{quangtran@utexas.edu}
\affiliation{Department of Astronomy, The University of Texas at Austin, 2515 Speedway, Stop C1400, Austin, TX 78712, USA}

\author[0000-0003-2649-2288]{Brendan P. Bowler}
\affiliation{Department of Astronomy, The University of Texas at Austin, 2515 Speedway, Stop C1400, Austin, TX 78712, USA}

\accepted{August 2024}
\submitjournal{AJ}

\begin{abstract}
    The recent development of statistical methods that can distinguish between stellar activity and dynamical signals in radial velocity (RV) observations has facilitated the discovery and characterization of planets orbiting young stars. One such technique, Gaussian process (GP) regression, has been regularly employed to improve the detection of a growing number of planets, but the impact of this model for mitigating stellar activity has not been uniformly analyzed for a large sample with real observations. The goal of this study is to investigate how GPs can affect the inferred parameters of RV-detected planets. We homogeneously analyze how two commonly adopted GP frameworks, a GP trained on RVs alone and a GP pretrained on photometry and then applied to RVs, can influence the inferred physical and orbital parameters compared to a traditional Keplerian orbit fit. Our sample comprises 17 short-period giant planets orbiting stars that exhibit a broad range of activity levels. We find that the decision to adopt GPs, as well as the choice of GP framework, can result in variations of inferred parameters such as minimum planet mass and eccentricity by up to 67\% and 95\%, respectively. This implies that the method for modeling stellar activity in RVs of young planet-hosting stars can have widespread ramifications on the interpretation of planet properties including their masses, densities, circularization timescales, and tidal quality factors. When mitigating stellar activity with GPs, we recommend carrying out comparative tests between different models to assess the sensitivity of planet physical and orbital parameters to these choices.
\end{abstract}

\section{Introduction}

The radial velocity (RV) method has been instrumental in the discovery and characterization of hundreds of exoplanets. RVs have been especially important to determine the masses of planets identified with ground- and space-based transit surveys, in particular NASA's Kepler, K2, and TESS missions \citep{Koch2010, Howell2014, Ricker2015}. For instance, masses directly impact inferences about average density, interior structure, and bulk composition in a way that cannot be probed with radii alone. Masses are also important for understanding planet demographics such as the cutoff between rocky and gaseous planets \citep[e.g.,][]{Weiss2014, Rogers2015}, atmospheric mass loss and the radius valley \citep[e.g.,][]{Fulton2017, Neil2020, Luque2022}, and planetary mass-metallicity trends \citep[e.g.,][]{Thorngren2016, Welbanks2019}.

Masses across a range of ages provide further constraints on how planets contract and cool after formation, and how their orbits evolve and migrate. As a result, there has been a growing effort to study young planets with radial velocities. In particular, young planets provide valuable information about evolutionary processes such as atmospheric mass loss, migration, tidal circularization, tidal realignment, and tidal decay. Several RV planet search programs have targeted young nearby stars \citep[e.g.,][]{Paulson2006, Bailey2012, Crockett2012, Lagrange2013, Grandjean2020, Tran2021, Grandjean2021, Zakhozhay2022a}, which has led to a growing number of RV-detected candidate young giant planets \citep[e.g.,][]{Donati2016, Johns-Krull2016, Yu2017a, Tran2024}. Similarly, several young, large transiting planets have also been found in transit surveys \citep[e.g.,][]{Mann2016, Newton2019, David2019, Rizzuto2020, Bouma2020}, which could be giant planets or sub-Saturns in the process of cooling and shrinking. Masses are needed to determine their true nature, but activity-based RV variations at the level of 0.1--1~km s$^{-1}$ has made it challenging to establish reliable mass measurements.

At early ages, starspots with large surface covering fractions induce asymmetries in spectral line profile shapes and produce strong rotationally modulated RV variations. These fluctuations are often quasiperiodic and can both mask and mimic dynamical signals induced by planets \citep[e.g.,][]{Queloz2001b, Boisse2011, Aigrain2012, Rajpaul2015, Haywood2016, Dumusque2018}. The contribution from this astrophysical noise, or ``jitter'', coupled with measurement uncertainties---which depend on properties of the instrument and characteristics of the host star---can reach amplitudes as high as hundreds to thousands of m s$^{-1}$ for stars between $\sim$1--100 Myr \citep{Crockett2012, Hillenbrand2015, Johns-Krull2016, Yu2017b, Tran2021}.

A robust set of computational techniques and observing strategies have been developed to distinguish stellar activity from dynamically induced variations. These techniques include modeling rotational modulations as sinusoids and simultaneously fitting for both activity and Keplerian signals \citep[e.g.,][]{Boisse2011, Pepe2013}, the ``floating chunk'' method, which corrects for offsets introduced by activity in different segments of the RV curve \citep{Hatzes2011}, and correcting for RV trends caused by activity using spectroscopic indicators \citep[e.g.,][]{Lovis2011, Dumusque2011b, Meunier2013, Robertson2014}.

Other observationally oriented strategies include selection of old, slowly rotating Sun-like stars to lower the contributions of activity, data driven observing cadence strategies \citep[e.g.,][]{Gupta2021, Petigura2022}, simultaneous reconstruction of large-scale surface brightness maps with the Doppler imaging technique \citep[e.g.,][]{Artigau2014, Donati2016, Yu2017a, Donati2020}, distinguishing between stellar activity and planetary signals with multiwavelength observations \citep[e.g.,][]{Crockett2012, Marchwinski2015, Carleo2020, Robertson2020}, and even dedicated telescope architectures for optimal RV follow-up campaigns \citep[e.g.,][]{Newman2022, Luhn2022}.

One of the most promising pathways forward that has gained attention over the past decade has been the development of modeling techniques---in particular Gaussian process (GP) regression \citep{Rasmussen2006, Aigrain2022}. GPs are non-parametric models that describe the covariance between pairs of samples drawn from a stochastic process. As GPs do not require a deterministic expression to define a model, they are an especially fruitful approach for systems affected by physical mechanisms that generate quasiperiodic signals, such as the emergence and evolution of starspots.

GPs have been adopted in a number of ways to mitigate stellar activity signals in RV time series. These include treating the GP model as a red-noise filter with \citep[e.g.,][]{Grunblatt2015, Dai2017} and without \citep[e.g.,][]{Baluev2013, Affer2016, Faria2020} supplementary data such as photometry, and joint models of RVs in a multidimensional GP framework \citep[e.g.,][]{Rajpaul2015, Jones2017, Gilbertson2020, Tran2023}.

However, a systematic study of the impact of GPs on the inferred properties of RV-detected planets spanning a wide range of host-star activity levels and planet masses has not yet been conducted. Thus, while there has been demonstrated success at the individual system level, a principled approach into how best to utilize GPs and what potential effects might be imparted on inferred planet properties, remains untested.

In this study, we analyze a sample of active systems hosting giant planets to test the limits and performance of two different GP methods. The goal of this work is to evaluate how inferred orbital and physical properties of planets are impacted as a result of different model choices. In particular, this study aims to investigate how jointly employing GP stellar activity models compares to inferred planet properties using a traditional orbit fit that does not account for correlated activity signals. In \autoref{sec:targets}, we describe the construction of our sample of active stars hosting giant planets. In \autoref{sec:application}, we briefly summarize the adopted GP techniques and kernel choice. We discuss the results of each modeling approach in \autoref{sec:results} and their implications on the interpretation of planetary parameters in \autoref{sec:discuss}. Finally, an overview of results and future directions is presented in \autoref{sec:summary}.

\section{Radial Velocities and Light Curves}
\label{sec:targets}

The goal of this study is to examine how the use of GPs trained on RV time series and light curves can impact inferred planet properties, particularly of planets around active stars. We therefore require a sample comprising systems with a known RV-detected planet, high-cadence RVs spanning multiple orbital period cycles, and a light curve exhibiting clear and persistent stellar rotational modulations.

To simultaneously model stellar activity and a planet with RVs, the amplitude of the planet should ideally be comparable to or larger than the amplitude of the stellar jitter. If activity modulations are too strong, they may completely obscure the planet signal. Short-period giant planets have RV semi-amplitudes reaching hundreds of m s$^{-1}$ and produce the most readily detectable signals. Systems with clear and persistent photometric variability, host a known close-in giant planet, and have high-precision RV measurements are therefore the most promising targets for this study.

\subsection{Sample Selection}\label{sec:sample_and_data}

A list of systems for this analysis was compiled by querying the NASA Exoplanet Archive\footnote{\label{note1}\href{https://exoplanetarchive.ipac.caltech.edu/index.html}{https://exoplanetarchive.ipac.caltech.edu/index.html}.} for confirmed planets as of June 2022. We first select systems with minimum mass measurements constrained by RV observations, resulting in an initial list of 3792 entries. A brightness cut ($V < 16.0$ mag) is then applied to limit the contribution of photon noise on photometric variability, removing 132 systems.

\begin{figure}[!t]
    \centering
    \includegraphics[width=1\linewidth]{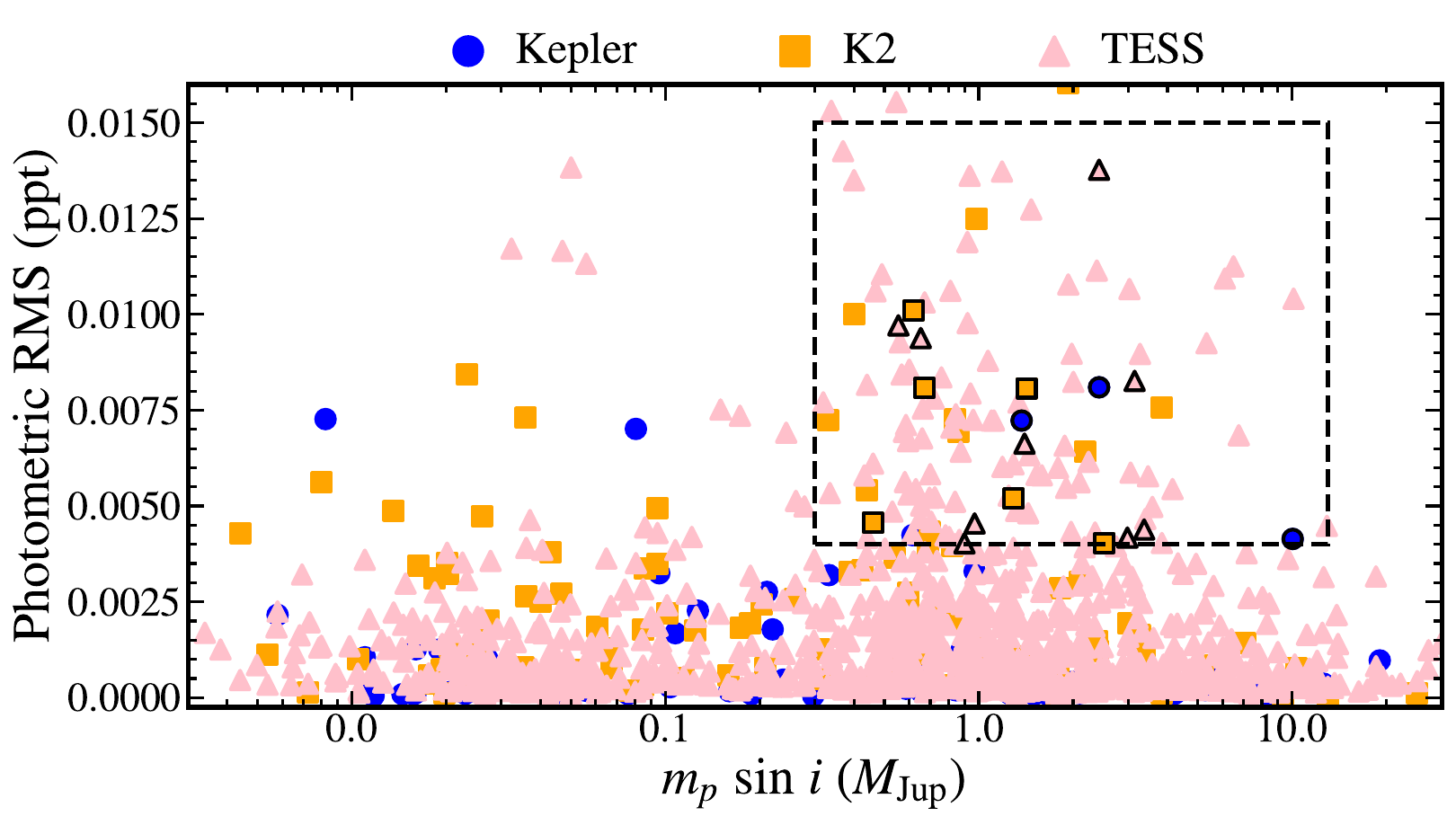}
    \caption{Photometric RMS as a function of planet minimum mass for systems with light curves available from Kepler (blue circles), K2 (orange squares), or TESS (pink triangles) and masses determined from RVs. Stars can be plotted more than once if they have light curves available from multiple surveys. The dashed box represents our selection cut and symbols with a black outline highlight systems in our final sample, which include single giant planets between 0.3 and 13 $M_\mathrm{Jup}$, photometrically active stars with light curve RMS levels between 0.004 and 0.015 ppt, and stars with light curves that exhibit clear and persistent modulations.}
    \label{fig:LC_RMS}
\end{figure}

Next, we use the \texttt{lightkurve} \citep{Lightkurve2018} software package to search for available Kepler, K2, and TESS light curves for each target in the MAST data archive.\footnote{All Kepler, K2, and TESS data used is available at MAST:\dataset[10.17909/T98304]{http://dx.doi.org/10.17909/T98304}, \dataset[10.17909/T9WS3R]{http://dx.doi.org/10.17909/T9WS3R}, and \dataset[10.17909/t9-nmc8-f686]{http://dx.doi.org/10.17909/t9-nmc8-f686} \citep{doiKepler2016, doiK22016, doiTESS2021}.} If the system was observed in any of those surveys, we apply a Savitzky-Golay filter \citep{Savitzky1964} to the available light curve with a large window length to remove transiting events, outliers from flares and instrumental effects, and high-frequency variations but retain long-term stellar activity trends. Finally, we calculate the photometric RMS from this processed light curve as a metric of the overall variability amplitude for that star. Some systems have multiple light curves and therefore multiple RMS measurements. We visually inspect each light curve to check for clear and persistent modulations and adopt an empirically motivated ``variability'' cut of photometric RMS between 0.004 and 0.015 parts-per-thousand (ppt). Systems with an RMS lower than 0.004 ppt are typically dominated by photon noise and systems with an RMS higher than 0.015 ppt appear to be regularly contaminated by artifacts from the instrument or spacecraft (e.g., photometric trends from spacecraft motion, pointing offsets, and differential velocity aberration). Stars with light curves from multiple instruments are retained if at least one light curve has an RMS within these boundaries. The photometric RMS of each light curve is shown as a function of the reported minimum planet mass in \autoref{fig:LC_RMS}.

For this study, we select single planets with minimum masses of $0.3$~$M_\mathrm{Jup} < m_p \sin i < 13$~$M_\mathrm{Jup}$.\footnote{Kepler-539 has a wide outer companion detected through transit timing variations ($m_c = 2.4$~$M_\mathrm{Jup}$, orbital period $P_c > 1000$~d). We choose to include this system but not model the outer planet, similar to the analysis of \citet{Mancini2016}. The wide orbit and mass translates to an RV signal comparable to the RV measurement uncertainties ($K_p \lesssim 50$ m s$^{-1}$). In comparison, Kepler-539 b has an RV semi-amplitude of $\sim$132~m s$^{-1}$. As a result, we consider Kepler-539 a single-planet system for these purposes. Qatar 2 also has an wide outer companion ($P_c > 1$~yr) that is modeled as a quadratic acceleration in \citet{Bryan2012}. We include this system in our analysis with curvature in our model as the signal of the short-period planet Qatar 2 b is distinct from the outer planet in the RVs.} This limits our sample to giant planets with RV semi-amplitudes generally greater than the expected RV stellar activity signal, giving us the greatest leverage to disentangle those effects from the planetary signals using GPs. Altogether our final sample comprises 17 short-period giant planet hosts spanning F, G, and K spectral types. \autoref{tab:systems} reports the origin of the light curve used for this target selection and subsequent analysis, the instruments and facilities from which the RV observations were obtained, the original reference from which the RVs are adopted, and the total number of published RVs used. Note that for the target selection, the entire light curve is analyzed, while the ``LC Window'' column refers to the section of the photometry used in the modeling described in \autoref{sec:gp+lc}. Using our estimated rotation periods from the light curves (see \autoref{sec:gp+lc}) and gyrochronology relations from \citet{Angus2019}, we find that the majority of systems in our sample are relatively young ($<$1~Gyr), with a median age of approximately 700~Myr.

\section{Approaches to Modeling Stellar Activity}
\label{sec:application}

To motivate the approaches we adopt in this study, we first summarize how GPs have typically been employed to model stellar activity in astronomical time series data. A detailed review of GPs can be found in \citet{Rasmussen2006} and an in-depth overview of the specific application of GPs to astronomical data is presented in \citet{Aigrain2022}.

GPs are a class of stochastic processes that describe the probability distribution of a set of random functions as a multivariate Gaussian distribution. As a Bayesian nonparametric regression tool, GPs model the covariance between pairs of observations with a ``kernel" function, which governs how malleable the GP is to variations in the observations. The regression model consists of the kernel hyperparameters\footnote{Hyperparameters here refer to the adopted GP kernel terms (see \autoref{sec:qp-kernel}). The term ``hyperparameter" is used because the real GP parameters are the family of functions describing the modeled process. These functions are marginalized over and almost never directly inferred. GP kernel variables are therefore latent, or unobserved, parameters.} and the parameters of a deterministic mean function; for instance, in the case of modeling a planetary companion with RVs, these would be the five parameters of the Keplerian model (e.g., \autoref{sec:rv-only}).

GPs have been routinely employed to model high-cadence stellar photometry. Light curves from space-based facilities like Kepler and TESS are densely sampled, have long observational baselines, and therefore are excellent use cases for this technique. Examples of how GPs have been applied to photometry include the modeling of instrumental systematics \citep[e.g.,][]{Gibson2012, Dawson2014, Evans2015, Aigrain2016}, stellar granulation and oscillations \citep[e.g.,][]{Brewer2009, Barclay2015, Pereira2019, Barros2020}, and rotationally modulated stellar activity \citep[e.g.,][]{Haywood2014, Vanderburg2015, Angus2018, Kosiarek2020}.

Similarly, GP regression has been used to disentangle activity-driven variability from planetary signals in RV time series, for example as a correlated red-noise filter to model RV residuals \citep[e.g.,][]{Baluev2013, Lopez-Morales2016, Damasso2017}. However, a single GP applied to RVs alone can be susceptible to over-fitting---applying an overly complex and flexible generative model to observations in a way that may not be justified by the data or the physical processes at play. As a result, GPs can impact the inferred properties of a real planet, leading to the introduction of systematic biases \citep[e.g.,][]{Blunt2023}.

Common approaches that attempt to correct for this drawback make use of ancillary time series, such as spectroscopic indicators measured concurrently with RVs, like the chromospheric activity index log$\:R^\prime_\mathrm{HK}$ or the cross-correlation function bisector inverse span, or high-precision photometry taken either synchronously (temporally coincident with the RV observations) or asynchronously (sampling a non-overlapping observational baseline). One method is a serial training approach, typically performed with photometry. In this scenario, model constraints are improved by pretraining the GP on a star's light curve. The GP hyperparameter posteriors are adopted as Bayesian priors for the GP kernel when fitting the RVs. This approach rests on the assumption that activity signals imprint common features between different time series datasets, with the correlation strength increasing for observations taken closer together in time, and simultaneous light curves and RVs sharing the most physically related activity signals. Here, a single GP can be used to model the stellar activity in the RVs, but it is first informed by the photometric data. This serial training approach has been used to recover and improve the significance of planetary signals for a handful of systems \citep[e.g.,][]{Grunblatt2015, Dai2017}.

Despite these encouraging results, this framework also has limitations that have prevented its more universal adoption. In particular, active regions can evolve over a range of timescales spanning less than a rotation period to the longer-term magnetic activity cycle of the star. \citet{Kosiarek2020} found that spot parameters from a GP model trained on solar photometric data can fluctuate substantially. This suggests that the extent to which training a GP on photometry can enhance the modeling of stellar activity in RVs depends to some degree on the simultaneity between the two datasets.

\begin{longrotatetable}
%\begin{rotatetable*}
\begin{deluxetable*}{cccccccccccccc}
    \setlength{\tabcolsep}{4pt}
    \centerwidetable
    \tablecaption{Host Star Light Curves (LCs) and Radial Velocities. \label{tab:systems}}
    \tablehead{\colhead{Host Star} &
    \colhead{LC} &
    \colhead{Overlap?\tablenotemark{a}} & \colhead{LC Window} & \colhead{LC Binning} & \colhead{Transit?} &
    \colhead{LC RMS} & \colhead{$N_\mathrm{RV}$} & \colhead{$N_\mathrm{instr}$} & \colhead{Instr.\tablenotemark{b}} & \colhead{RV} & \colhead{$P_\mathrm{rot}$} & \colhead{$P_\mathrm{rot}$} \\
    \colhead{} & \colhead{} & \colhead{} & \colhead{(BJD -- 2450000 d)} & \colhead{(hr)} & \colhead{} & \colhead{(ppt)} & \colhead{} & \colhead{} & \colhead{} & \colhead{Ref.} & \colhead{(d)} & \colhead{Ref.}}
    \startdata
    GJ 3021 & TESS S1 & N & 8325.3--8353.2 & 0.5 & N & 0.004 & 61 & 1 & A & 1 & 4.1 & 1 \\
    HATS-29 & TESS S13 & N & 8653.9--8682.4 & 0.5 & Y & 0.009 & 16 & 3 & A, B, C & 2 & 0.63 & 3 \\
    HAT-P-54 & K2 C0 & N & 6772.1--6805.2 & 1.0 & Y & 0.005 & 17 & 2 & D, E & 4 & 15.6 & 4 \\
    HD 12484 & TESS S4 & N & 8410.9--8436.5 & 0.5 & N & 0.004 & 65 & 1 & F & 5 & 7.0 & 5 \\
    HD 102195 & K2 C1 & Y & 6810.5--6890.3 & 2.0 & N & 0.005 & 98 & 5 & B, G, G, H, I & 6, 7, 8 & 12.3 & 6 \\
    HD 103720 & K2 C1 & N & 6810.3--6890.3 & 2.0 & N & 0.010 & 71 & 1 & B & 9 & 17.0 & 9 \\
    Kepler-17 & Kepler Q6 & Y & 5420.8--5462.3 & 1.0 & Y & 0.008 & 13 & 1 & H & 10 & 11.9 & 10 \\
    Kepler-43 & Kepler Q10 & Y & 5742.5--5817.3 & 2.0 & Y & 0.008\tablenotemark{c} & 22 & 2 & F, J & 11, 12 & 12.9 & 11 \\
    Kepler-75 & Kepler Q15 & Y & 6105.6--6242.4 & 3.0 & Y & 0.004 & 11 & 2 & F, K & 13 & 19.2 & 13 \\
    Kepler-77 & Kepler Q9-15 & Y & 5692.0--6224.6 & 12.0 & Y & 0.010\tablenotemark{c} & 15 & 2 & J, L & 14 & 36.0 & 14 \\
    Kepler-447 & Kepler Q14-15 & Y & 6129.4--6304.1 & 3.0 & Y & 0.007 & 21 & 3 & M, M, M & 15 & 6.5 & 15 \\
    Kepler-539 & Kepler Q14-17 & Y & 6107.2--6424.0 & 6.0 & Y & 0.005\tablenotemark{c} & 20 & 1 & M & 16 & 11.8 & 16 \\
    K2-29 & K2 C4 & N & 7064.1--7132.7 & 2.0 & Y & 0.008 & 34 & 5 & F, J, K, K, N & 17, 18 & 10.78 & 17 \\
    K2-237 & K2 C11 & N & 7682.8--7730.5 & 1.0 & Y & 0.007\tablenotemark{c} & 29 & 4 & A, B, B, J & 19, 20 & 5.1 & 19 \\
    K2-260 & K2 C13 & N & 7821.5--7901.2 & 1.0 & Y & 0.008 & 20 & 1 & J & 21 & 2.2 & 21 \\
    Qatar-2 & K2 C6 & N & 7217.8--7296.4 & 2.0 & N & 0.004 & 44 & 1 & D & 22 & 18.3 & 23 \\
    WASP-180 A\tablenotemark{d} & TESS S34 & N & 9229.1--9254.0 & 0.5 & Y & 0.004 & 15 & 2 & A, B & 24 & 4.6 & 24
    \enddata
    \tablenotetext{a}{This ``Yes/No'' flag refers to whether the photometry and radial velocities used in the modeling were obtained in partially overlapping observational baselines.}
    \tablenotetext{b}{This column refers the instruments corresponding to observations for each system. Repeated instances refer to instruments with velocity breaks or used in different studies.}
    \tablenotetext{c}{For these systems, reported photometric RMS values are calculated with the TESS light curves, while the analysis is conducted using the light curve reported in the ``LC'' column.}
    \tablenotetext{d}{In model fits of WASP-180 A, we exclude RVs taken over the course of a single night as part of a Rossiter-McLaughlin measurement.}
    \tablecomments{Instruments and facilities of datasets for each system: (A) CORALIE on the 1.2-m Euler Swiss telescope at La Silla Observatory (LSO) \citep{Udry2000, Queloz2001a}, (B) HARPS (High Accuracy Radial velocity Planet Searcher) on the European Southern Observatory 3.6-m telescope at LSO \citep{Mayor2003}, (C) CYCLOPS2 and UCLES (University College London Echelle Spectrograph) on the 3.9-m Anglo-Australian Telescope at the Australian Astronomical Observatory \citep{Diego1990, Horton2012}, (D) TRES (Tillinghast Reflector Echelle Spectrograph) on the Tillinghast Reflector 1.5-m telescope at the Fred L. Whipple Observatory, (E) HIRES (High Resolution Echelle Spectrometer) on the Keck I 10-m telescope \citep{Vogt1994}, (F) SOPHIE (Spectrographe pour l’Observation des Phénomènes des Intérieurs stellaires et des Exoplanètes) on the 1.93-m telescope at CNRS (Observatoire de Haute-Provence) \citep{Bouchy2006, Perruchot2008}, (G) ET (Exoplanet Tracker) on the 0.9-m coud\'{e} feed and 2.1-m telescopes at Kitt Peak National Observatory (KPNO) \citep{vanEyken2004}, (H) HRS (High Resolution Spectrograph) on the 9-m Hobby-Eberly Telescope at McDonald Observatory (MO) \citep{Tull1998}, (I) CHIRON on the SMARTS (Small and Moderate Aperture Research Telescope System) 1.5-m telescope at Cerro Tololo Interamerican Observatory \citep{Tokovinin2013}, (J) FIES (Fibre–fed Echelle Spectrograph) on the 2.5-m Nordic Optical Telescope at Roque de los Muchachos Observatory (ORM) \citep{Diego2010}, (K) HARPS-N (High Accuracy Radial velocity Planet Searcher - North) on the 3.6-m Telescopio Nazionale Galileo at ORM) \citep{Cosentino2012}, (L) Sandiford \'{E}chelle Spectrograph on the 2.1-m Struve Telescope at MO \citep{McCarthy1993}, (M) CAFE (Calar Alto Fiber-fed \'{E}chelle spectrograph) on the 2.2-m telescope at Calar Alto Observatory \citep{Aceituno2013}, (N) TS23 (Robert G. Tull coud\'{e} spectrograph) on the 2.7-m Harlan J. Smith Telescope at MO \citep{Tull1995}.}
    \tablerefs{(1) \citet{Naef2001}, (2) \citet{Espinoza2016}, (3) \citet{CantoMartins2020}, (4) \citet{Bakos2015}, (5) \citet{Hebrard2016}, (6) \citet{Ge2006}, (7) \citet{Melo2007}, (8) \citet{Paredes2021}, (9) \citet{Moutou2015}, (10) \citet{Desert2011}, (11) \citet{Bonomo2012}, (12) \citet{Endl2014}, (13) \citet{Hebrard2013}, (14) \citet{Gandolfi2013}, (15) \citet{Lillo-Box2015}, (16) \citet{Mancini2016}, (17) \citet{Johnson2016}, (18) \citet{Santerne2016}, (19) \citet{Soto2018}, (20) \citet{Smith2019}, (21) \citet{Johnson2018}, (22) \citet{Bryan2012}, (23) \citet{Reinhold2020}, (24) \citet{Temple2019}.}
\end{deluxetable*}
%\end{rotatetable*}
\end{longrotatetable}

\noindent{}Furthermore, there is evidence that even for contemporaneous observations, at least one GP kernel hyperparameter can differ significantly when trained on photometry and on RVs \citep{Kosiarek2020, Barragan2022, Nicholson2022}.

Here, we carry out a systematic investigation on the performance of a single GP model for a sample of 17 active stars hosting giant planets (\autoref{tab:systems}). In particular, we aim to assess the degree to which traditional GP models can affect both the parameter inference and uncertainties of planetary properties. For each system, we carry out Keplerian orbit fits of the known planet while varying the GP model in the following way:
\begin{enumerate}[label={\textit{Model \theenumi.}},itemsep=0ex,partopsep=1ex,parsep=0ex,labelindent=12pt,labelwidth=\widthof{Model\theenumi.}+\labelsep,leftmargin=!]
    \item Keplerian-only (RVs)
    \item Keplerian + GP (RVs)
    \item Keplerian + Pretrained GP (Light curve then RVs)
\end{enumerate}
Here, Model 1 refers to fitting only a Keplerian model to the RV data without the use of GPs to remove stellar activity signals. Model 2 is a simultaneous fit to the RVs with both a GP stellar activity model and a Keplerian model. Model 3 is the same as Model 2, but the GP hyperparameter priors are conditioned on the photometric data. In this way these three approaches encompass the range of models typically adopted when fitting planets orbiting active stars. By carrying out these fits in a uniform fashion on a substantial sample, we aim to better understand how GP regression can impact inferred planetary parameters---and in particular planet mass (through the RV semi-amplitude) and eccentricity. Additional details about each model are described below.

\begin{deluxetable}{llll}
    \centering
    \setlength{\tabcolsep}{4pt}
    \tablecaption{Adopted priors of the Keplerian model fit to RV data. \label{tab:rv_priors}}
    \tablehead{\colhead{System Name} & \colhead{$T_{0}$} & \colhead{$P$} & \colhead{$K$} \\
    \colhead{} & \colhead{(BJD $-$ 2450000 d)} & \colhead{(d)} & \colhead{(m s$^{-1}$)}}
    \startdata
    GJ 3021 b & $\mathcal{U} \left[1535.86, 1605.86 \right]$ & $\mathcal{U} \left[1, 230 \right]$ & $\mathcal{U} \left[1.1, 500 \right]$ \\
    HATS-29 b & $\mathcal{U} \left[7031.46, 7032.46 \right]$ & $\mathcal{U} \left[1, 10 \right]$ & $\mathcal{U} \left[1.3, 250 \right]$ \\
    HAT-P-54 b & $\mathcal{U}\left[6298.31, 6230.31\right]$ & $\mathcal{U}\left[1, 10 \right]$ & $\mathcal{U} \left[2.2, 500 \right]$ \\
    HD 12484 b & $\mathcal{U} \left[6674.0, 6724.0 \right]$ & $\mathcal{U} \left[1, 125 \right]$ & $\mathcal{U} \left[1.0, 500 \right]$ \\
    HD 102195 b & $\mathcal{U} \left[3731.2, 3734.2 \right]$ & $\mathcal{U} \left[1, 8.5 \right]$ & $\mathcal{U} \left[1.0, 250 \right]$ \\
    HD 103720 b & $\mathcal{U} \left[5382.46, 5392.46 \right]$ & $\mathcal{U} \left[1, 25 \right]$ & $\mathcal{U} \left[2.0, 500 \right]$ \\
    K2-29 b & $\mathcal{U} \left[7381.2, 7386.2 \right]$ & $\mathcal{U} \left[1, 12 \right]$ & $\mathcal{U} \left[1.9, 250 \right]$ \\
    K2-237 b & $\mathcal{U}\left[7683.81, 7685.81 \right]$ & $\mathcal{U}\left[1, 8 \right]$ & $\mathcal{U}\left[3.5, 500 \right]$ \\
    K2-260 b & $\mathcal{U} \left[7818.74, 7822.74 \right]$ & $\mathcal{U} \left[1, 7 \right]$ & $\mathcal{U} \left[9.9, 500 \right]$ \\
    Kepler-17 b & $\mathcal{U} \left[5185.18, 5186.18 \right]$ & $\mathcal{U} \left[1, 3 \right]$ & $\mathcal{U} \left[6.4, 2000\right]$ \\
    Kepler-43 b & $\mathcal{U} \left[4962.42, 4968.42 \right]$ & $\mathcal{U} \left[1, 15 \right]$ & $\mathcal{U} \left[2.1, 500 \right]$ \\
    Kepler-75 b & $\mathcal{U} \left[5725.44, 5755.44 \right]$ & $\mathcal{U} \left[1, 30 \right]$ & $\mathcal{U} \left[4.2, 2000 \right]$ \\
    Kepler-77 b & $\mathcal{U} \left[5093.87, 5097.87 \right]$ & $\mathcal{U} \left[1, 12 \right]$ & $\mathcal{U} \left[2.3, 250 \right]$ \\
    Kepler-447 b & $\mathcal{U} \left[4967.76, 4972.76 \right]$ & $\mathcal{U} \left[1, 16 \right]$ & $\mathcal{U} \left[3.6, 500 \right]$ \\
    Kepler-539 b & $\mathcal{U} \left[5568.87, 5668.87 \right]$ & $\mathcal{U} \left[115, 130 \right]$ & $\mathcal{U} \left[3.3, 250 \right]$ \\
    Qatar-2 b & $\mathcal{U}\left[5623.27, 5625.27 \right]$ & $\mathcal{U}\left[1, 7 \right]$ & $\mathcal{U}\left[2.8, 1000 \right]$ \\
    WASP-180 A b & $\mathcal{U} \left[7761.82, 7764.82 \right]$ & $\mathcal{U} \left[1, 8 \right]$ & $\mathcal{U} \left[3.1, 500 \right]$
    \enddata
\end{deluxetable}

\subsection{Model 1: Keplerian-only Fit}
\label{sec:rv-only}

The first approach is to fit the RVs using only a Keplerian model. We assemble all published RV measurements as of June 2022 to use in the orbit fit. \autoref{tab:systems} reports details of the adopted RV datasets for each system. The RVs are fit with a Keplerian model using the open software package \texttt{pyaneti} \citep{Barragan2019a, Barragan2022}. \texttt{pyaneti} models five parameters: the time of inferior conjunction $T_{0}$, orbital period $P$, RV semi-amplitude $K$, and parameterized forms of the eccentricity and argument of periastron, $\sqrt{e} \sin \omega$ and $\sqrt{e} \cos \omega$.\footnote{$\omega$ refers to the  argument of periastron of the star, $\omega_*$.}

Uniform priors are adopted for all orbital parameters. $T_{0}$ is informed by $T_{0,\mathrm{ref}}$---a previously inferred time of inferior conjunction from the literature---when this value is reported. For non-transiting systems, the times of periastron passage, $T_\mathrm{peri}$, are typically reported. In these instances, we adopt broader limits on $T_{0}$ to account for differences between $T_\mathrm{peri}$ and $T_{0}$. In general, the adopted priors on $T_{0}$ span the orbital period reported in the original reference, $P_{\mathrm{ref}}$. 12 of the 17 giant planets transit their host stars (see \autoref{tab:systems}); for these systems, $T_{0}$ is constrained by mid-transit times. For $P$, we adopt an uninformative prior bounded by a lower limit of one day and an upper limit of approximately three times $P_{\mathrm{ref}}$.\footnote{For Kepler-539, we impose a tighter prior on $P$ based on constraints from the transit (see \autoref{tab:rv_priors}).} The RV semi-amplitude $K$ prior has a lower limit of one-tenth the average RV measurement uncertainty and an upper limit between 0.25 and 2.0 km s$^{-1}$ (typically 0.5 km s$^{-1}$); the specific choice is selected to improve convergence. $\sqrt{e} \sin \omega$ and $\sqrt{e} \cos \omega$ are both allowed to vary between $-1$ and 1. For every spectrograph, \texttt{pyaneti} also fits for velocity offset, $\gamma$, and RV jitter, $\sigma_\mathrm{RV}$, terms. These parameters use the default \texttt{pyaneti} priors, $\gamma_\mathrm{spec} = \mathcal{U}[\mathrm{RV}_\mathrm{spec, min}, \;\mathrm{RV}_\mathrm{spec, max}]$ km s$^{-1}$ and $\sigma_\mathrm{RV} = \mathcal{J}(1.0, 1000.0)$ m s$^{-1}$.\footnote{$\mathcal{U}[a, b]$ is the uniform distribution inclusive of lower limit $a$ and upper limit $b$. $\mathcal{J}(a, b)$ is the modified Jeffreys prior as defined in Equation 6 of  \citet{Gregory2005}, $\mathcal{P}(x) = \{(a + x) \cdot \ln \left[\left(a + b\right) / a\right]\}^{-1}$. This prior behaves as a uniform distribution for $x \gg a$ and a Jeffreys prior for $x \ll a$, and is designed to assign equal probability in each decade.} The specific priors for $T_0$, $P$, and $K$ adopted for each system are reported in \autoref{tab:rv_priors}.

Parameter posterior distributions are sampled with a Markov chain Monte Carlo (MCMC) Metropolis-Hasting algorithm following \citet{Sharma2017}. We use 30 walkers and $10^5$ steps with a thinning factor of 10 to sample the posteriors of the Keplerian model parameters. Furthermore, we use the default \texttt{pyaneti} burn-in period, which is equivalent to the number of steps times the thinning factor (in this case, $10^6$ iterations). Posterior convergence is confirmed using the Gelman-Rubin (GR) diagnostic test where all chains have values under 1.02 \citep{Gelman1992}.

\vspace{-2em}

\begin{deluxetable*}{llll}
    \setlength{\tabcolsep}{14pt}
    \centerwidetable
    \tablecaption{Adopted priors of the QP GP hyperparameters for the MCMC fit to the photometric data and Models 2 and 3 fits to RV data. \label{tab:gp_priors}}
    \tablehead{\colhead{Parameter} & \colhead{Prior$_\mathrm{Model \; 2}$\tablenotemark{a}} & \colhead{Prior$_\mathrm{LC}$} & \colhead{Prior$_\mathrm{Model \; 3}$\tablenotemark{b, c}}}
    \startdata
    $A$ & $\mathcal{U}\left[10^{-3}, \; 100\right]$ & $\mathcal{U}\left[10^{-3}, \; 100\right]$ & $\mathcal{U}\left[10^{-3}, \; 100\right]$ \\
    $l_e$ & $\mathcal{U}\left[10^{-3}, \; 50\right]$ & $\mathcal{U}\left[10^{-3}, \; 50\right]$ & $\mathcal{N}\left(l_{e, \mathrm{Med \; LC}}, \; \sigma_{l_{e, \mathrm{LC}}} \right)$ \\
    $l_p$ & $\mathcal{U}\left[10^{-3}, \; 50\right]$ & $\mathcal{U}\left[10^{-3}, \; 50\right]$ & $\mathcal{N}\left(l_{p, \mathrm{Med \; LC}}, \; \sigma_{l_{p, \mathrm{LC}}} \right)$ \\
    $P_\mathrm{GP}$ & $\mathcal{U}\left[10^{-3}, \; 50\right]$ & $\mathcal{U}\left[P_\mathrm{rot}/3, \; 3P_\mathrm{rot}\right]$\tablenotemark{d} & $\mathcal{N}\left(P_\mathrm{GP, Med \; LC}, \; \sigma_{P_\mathrm{GP, LC}} \right)$ \\
    $\sigma$ & $\mathcal{U}\left[0,\; 100\right]$ m s$^{-1}$ & $\mathcal{U}\left[\mathrm{ln}(10^{-12}), \; \mathrm{ln}(2)\right]$ ppt & $\mathcal{U}\left[0,\; 100\right]$ m s$^{-1}$
    \enddata
    \tablenotetext{a}{$\mathcal{U}\left[a, b\right]$ refers to a uniform distribution inclusive of lower bound $a$ and upper bound $b$.}
    \tablenotetext{b}{$\mathcal{N}\left(\mu, \sigma\right)$ refers to a Gaussian distribution with mean $\mu$ and standard deviation $\sigma$.}
    \tablenotetext{c}{The values $l_{e, \mathrm{Med \; LC}}$, $l_{p, \mathrm{Med \; LC}}$, and $P_\mathrm{GP, Med \; LC}$ and $\sigma_{l_{e, \mathrm{LC}}}$, $\sigma_{l_{p, \mathrm{LC}}}$, and $\sigma_{P_\mathrm{GP, LC}}$ refer to the median and standard deviations, respectively, of the hyperparameter posteriors from the QP GP fit to the light curve data, as reported in \autoref{tab:lc_gp_posteriors}.}
    \tablenotetext{d}{$P_\mathrm{rot}$ refers to the literature stellar period as reported in \autoref{tab:systems}.}
\end{deluxetable*}

\subsection{Model 2: Keplerian + GP Fit}
\label{sec:gp+rv}

\subsubsection{The Quasiperiodic Kernel}
\label{sec:qp-kernel}

For the two models that employ a GP (Models 2 and 3), we adopt a quasiperiodic (QP) kernel as the covariance function. The QP kernel is one of the most commonly used kernels to model stellar activity signals because the QP hyperparameters can be mapped to physical quantities such as the stellar rotation period, spot evolution timescale, and spot lifetime \citep[e.g.,][]{Aigrain2012, Haywood2014, Angus2018, Kosiarek2020, Nicholson2022}. This choice reflects how single GP models are regularly employed in the literature to investigate their impact on the accuracy and precision of planetary measurements.

The traditional QP kernel used for astronomical time series combines an exponential sine-squared term and a squared exponential term. We use the functional form of the QP kernel as implemented in \texttt{pyaneti}:
\begin{equation} \label{eq:quasi_per_kernel}
    k_\mathrm{QP}(t,t') = A^2 \: \mathrm{exp}\left( -\frac{(t - t')^2}{2l_e^2} - \frac{\: \mathrm{sin}^2\left(\frac{\pi (t - t')}{P_\mathrm{GP}}\right)}{2l_p^2} \right),
\end{equation}
where $t$ and $t'$ represent the pairs of observations in time, $A$ is the variance amplitude, $l_e$ encodes the evolutionary timescale, $P_\mathrm{GP}$ is the recurrence timescale (or period) of the activity signal, and $l_p$ is the smoothness and complexity of the periodic component (or the periodic lengthscale).

\subsubsection{Uninformed GP Applied to RVs}
\label{sec:gp_on_rvs}

Using \texttt{pyaneti}'s built-in QP kernel (\autoref{eq:quasi_per_kernel}), we jointly fit the RV data using both a Keplerian model and a single red-noise GP model. During this process, we adopt the same priors for the Keplerian parameters ($T_{0}$, $P$, $\sqrt{e} \sin \omega$, $\sqrt{e} \cos \omega$, and $K$) and instrumental velocity and jitter parameters as in Model 1 (see \autoref{tab:rv_priors}). For the GP hyperparameters, we adopt uninformative, broad uniform priors. These priors are detailed in \autoref{tab:gp_priors}. The model parameter posteriors are sampled using the same MCMC algorithm, number of walkers, steps, and burn-in period as Model 1 (see \autoref{sec:rv-only}). Similarly, the GR diagnostic, with a threshold of 1.02, is used to test for convergence.

\begin{figure}
    \centering
    \includegraphics[width=\linewidth]{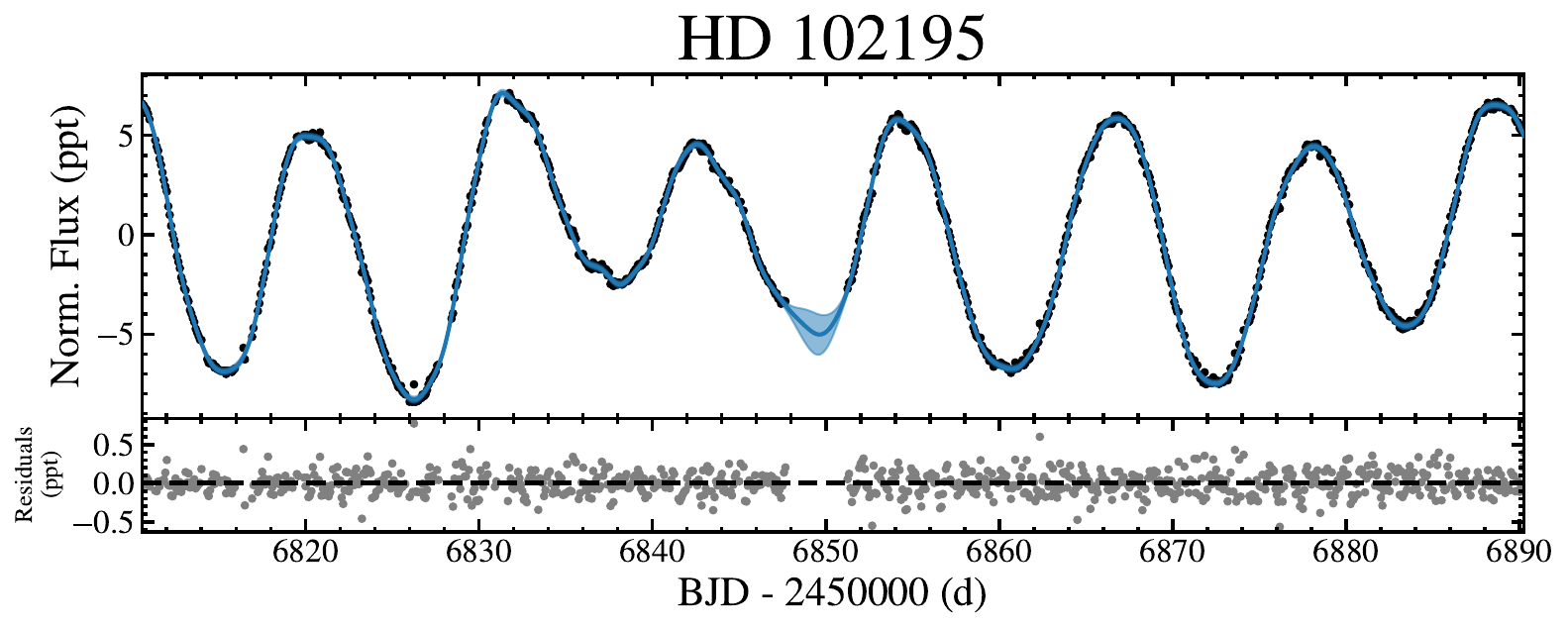}
    \includegraphics[width=\linewidth]{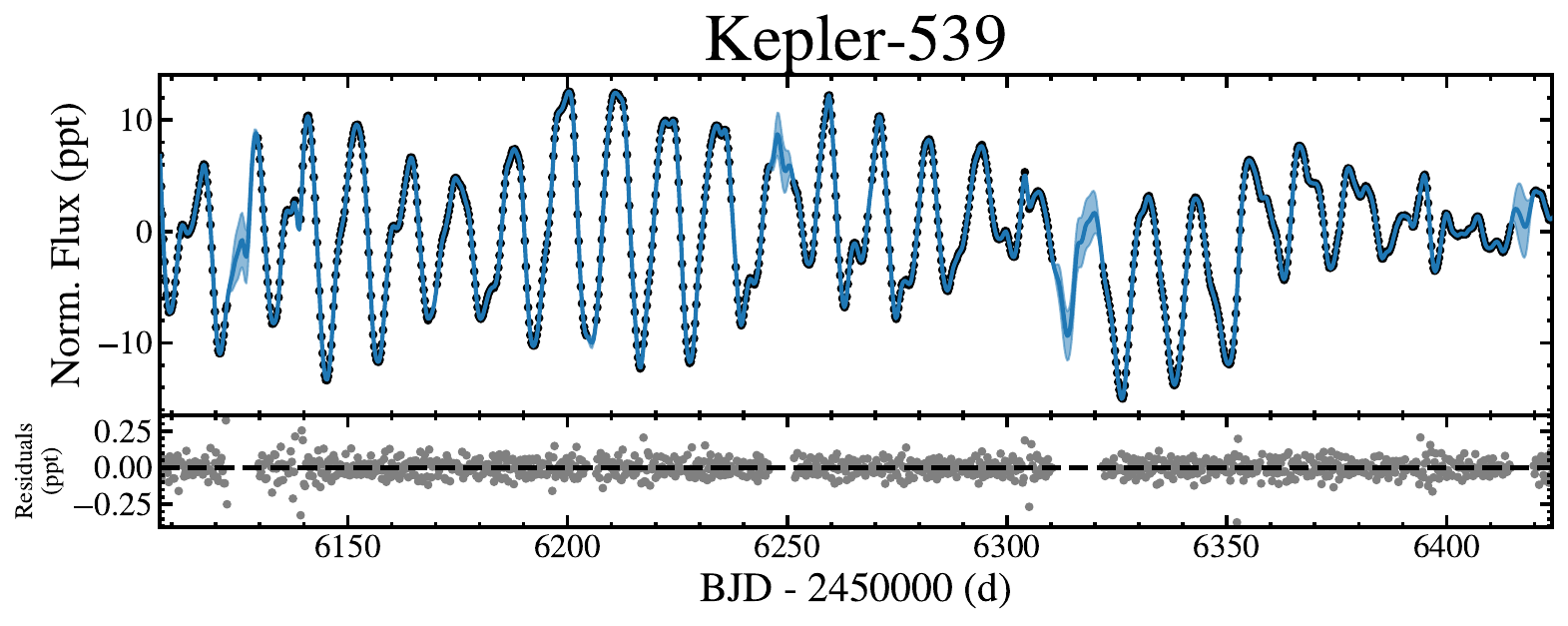}
    \includegraphics[width=\linewidth]{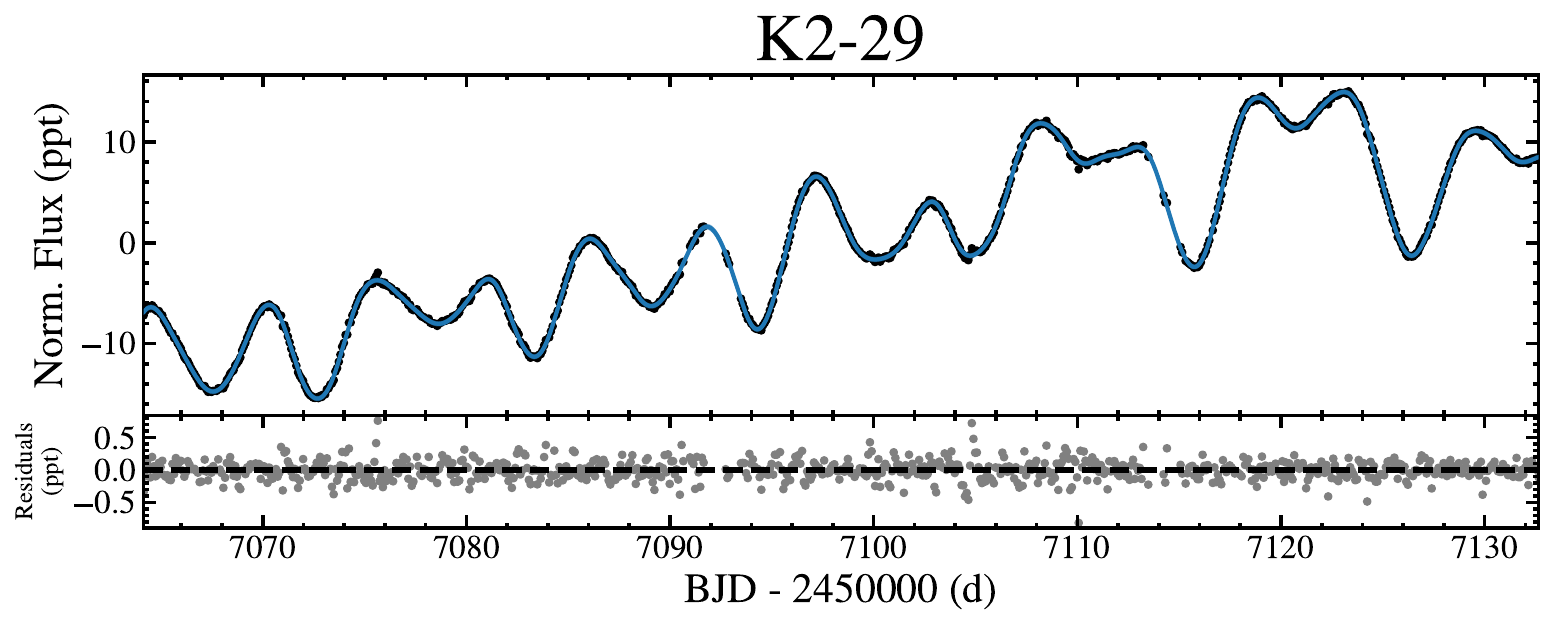}
    \includegraphics[width=\linewidth]{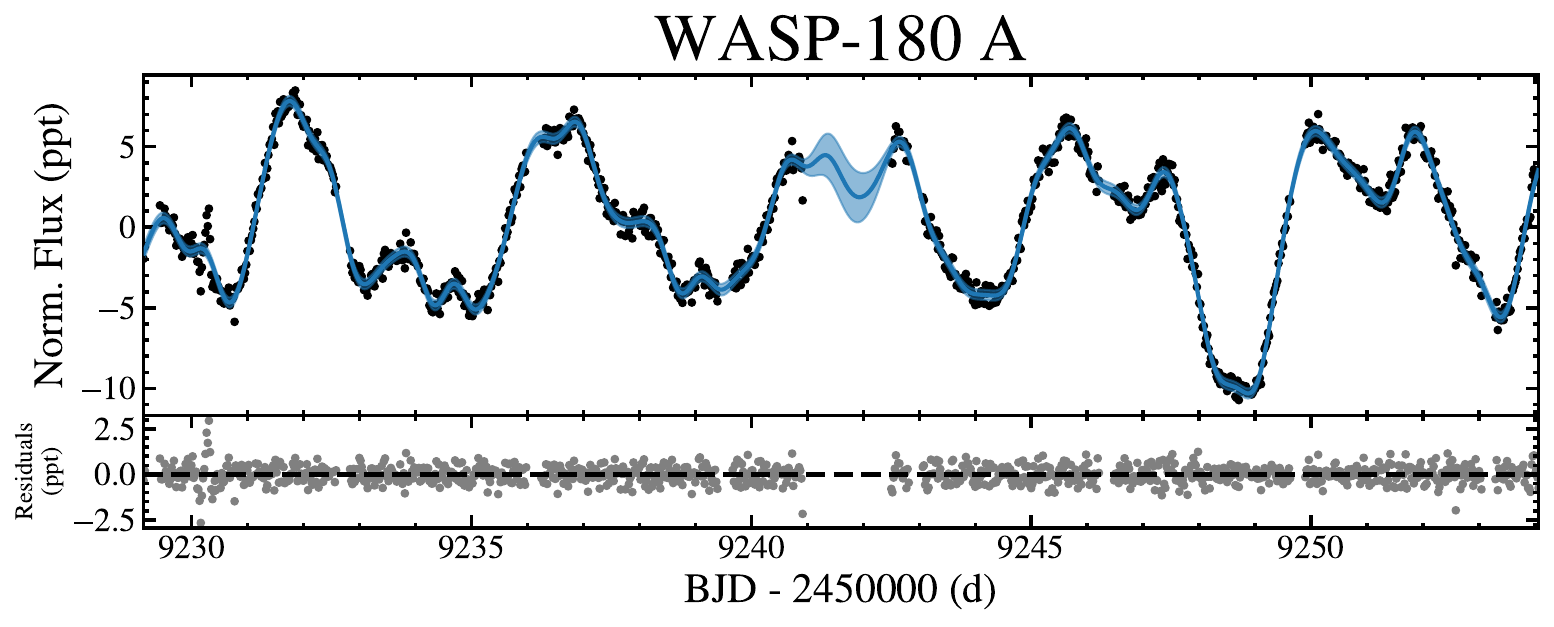}
    \caption{Representative examples of best-fit GP regression models of K2, Kepler, K2, and TESS light curves for HD 102195, Kepler-539, K2-29, and WASP-180 A, respectively. These light curves illustrate the variety of datasets available for host stars in our sample, spanning a range of periodicities, long-term trends, and evolving starspot patterns. Black circles display the binned light curve. The solid blue line and shaded region represent the mean and covariance of the GP prediction. The bottom panels display the residuals between the data and the best-fit GP model.}
    \label{fig:lc_gp_ex}
\end{figure}

\subsection{Model 3: Keplerian + pretrained GP Fit}
\label{sec:gp+lc}

\subsubsection{Pretraining GP on Photometry}
\label{sec:train_gp_on_lc}

For Model 3, we first apply a GP regression using the QP kernel defined in \autoref{eq:quasi_per_kernel} to portions of the light curves of each target in our sample. Whenever possible, we restrict the photometry to sections that are simultaneous with the RVs. If the two time series overlap, then we select for an observational window 10 days before and after the RV observations\footnote{Kepler-75 requires 50 days before and after to account for its short RV observational window.} to capture any persistent activity signals in the light curves. If the photometry and RVs are not contemporaneous, we select a segment of the light curve closest in time to the RVs that appears to cover at least one cycle of stellar modulation. Details about the relative photometry used in the modeling are listed in \autoref{tab:systems}.

As each light curve typically contains thousands of photometric measurements, we further process each light curve segment to reduce computational costs. As described in \autoref{sec:sample_and_data}, we apply a Savitzky-Golay fiter to remove transits, flares, and instrumental effects. Each photometry window is then normalized by the average flux value and converted into units of ppt. Typical photometric observations from Kepler, K2, and TESS are sampled on timescales of 30--60 min, with the number of brightness measurements reaching $10^5$ data points or more for Kepler light curves. As the computational complexity of GP regression scales as $\mathcal{O}(N^3)$ \citep{Rasmussen2006}, where $N$ is the number of data points, modeling light curves can quickly become computationally expensive. Thus, light curves are binned in time such that the number of photometric data points used in the modeling is reduced to approximately $10^3$. Light curve binning sizes are reported in \autoref{tab:systems}. We find that this binning procedure does not remove large rotationally-modulated activity signals, which operate on timescales much larger than the bin sizes (days as compared to minutes or hours).

Using this processed light curve segment, we fit for 5 parameters: the photometric noise, $\sigma_\mathrm{LC}$, and the four GP hyperparameters: $A$, $l_p$, $l_e$, and $P_\mathrm{GP}$. We use the open Python package \texttt{emcee} \citep{Foreman-Mackey2013, Foreman-Mackey2019} to sample posteriors for each GP hyperparameter and the photometric noise term. Note that we do not fit for a mean function as transit events have been removed and the light curves are normalized, allowing us to assume a zero-mean function. Broad uniform priors are adopted for each hyperparameter (see \autoref{tab:gp_priors}). We sample the posteriors using 30 walkers for $10^5$ steps and visually confirm for convergence of the chains. \autoref{tab:lc_gp_posteriors} reports the maximum a posteriori (MAP) value, median value, and 68.3\% highest density interval (HDI)\footnote{The highest density interval is the minimum width Bayesian credible interval, or the minimum range encompassing a given probability of a posterior distribution.} of the GP hyperparameter posteriors of each system. \autoref{fig:lc_gp_ex} displays four examples of the typical GP regression to the light curves using the MAP hyperparameter values.

\subsubsection{Pretrained GP Applied to RVs} \label{sec:gp+rv+lc}

For Model 3, we jointly fit the RVs with a Keplerian model and a GP model that has been pretrained on the light curve as described in \autoref{sec:train_gp_on_lc}. This procedure for fitting the RVs is similar to Model 2 detailed in \autoref{sec:gp+rv}. However, instead of uniform hyperpriors, the GP hyperparameters are now taken to be Gaussian distributions with the mean and standard deviation set to the median and standard deviation of the GP kernel hyperparameter posteriors recovered from the light curve (see \autoref{tab:gp_priors}). This choice is motivated by the approximately normally distributed marginalized posteriors of the hyperparameters. The priors for the Keplerian parameters (\autoref{tab:rv_priors}) are defined and parameter posterior distributions are sampled as in Models 1 and 2.

\section{Results}
\label{sec:results}

Here, we report and compare the posteriors for each of the three models. \autoref{tab:posterior-values} summarizes the median and 1$\sigma$ HDI constraints of all fitted parameters, planetary properties, and QP GP kernel hyperparameters for each system in our sample. The best-fit phased RV curves for each approach are presented in \autoref{sec:app_results}.

\subsection{Planet Properties via Keplerian-only Model}\label{sec:rv_only_model_results}

Whether intentional or not, previous studies presenting the discovery or follow-up RV observations of planets make specific decisions in their analysis that affect the inferred planetary properties---for instance, the choice of priors, if fitting in a Bayesian framework, inflating instrumental uncertainties to account for jitter, or adopting circular orbits and not treating eccentricity as a free parameter. As a result, each model fit is an attempt to carry out parameter inference under varying assumptions and model complexities. These heterogeneous choices make it difficult to uniformly compare the impact of the various model fits in this work to published planet orbits and minimum mass measurements of these systems.

Instead, our systematic and homogeneous treatment of each dataset means that each fit retains the same priors for the Keplerian component of the model ($T_0$, $P$, $K$, $\sqrt{e} \sin \omega$, $\sqrt{e} \cos \omega$, and $\sigma_\mathrm{jit}$), allowing us to use the posterior parameters from the Keplerian-only (Model 1) fit as a baseline comparison to assess the impact of the GP models. Overall our Model 1 results are in good agreement with fits from the literature. Slight differences between our results and published values for a handful of systems originate from alternative choices of priors or, in some instances where the planet transits, from previous fits jointly modeling the light curves together with the RVs, which we do not consider here. Furthermore, the original discoveries may use a different number of RV measurements than in this analysis because in some cases additional RVs have become available.

In particular, for five systems, the planetary eccentricity and uncertainty that we find in this work are somewhat higher compared to the original studies: HATS-29 b \citep{Espinoza2016}; HAT-P-54 b \citep{Bakos2015}; Kepler-539 b \citep{Mancini2016}; K2-260 b \citep{Johnson2018}; and WASP-180 A b \citep{Reinhold2020}. In the previous analyses, either circular orbits were adopted or upper limits are reported for eccentricities. Because these planets all transit their host stars, a simultaneous fit had been carried out for a Keplerian model, a transit model, and in some cases Rossiter-McLaughlin, Doppler tomography, or secondary eclipse models. Furthermore, each system has a modest number of observations ($N_\mathrm{RV} \leq 20$), which limits the constraining power of the RVs alone. We find that if we use priors similar to those imposed by these additional model constraints, our recovered parameters are consistently in good agreement with the original values.

\subsection{Comparison of System Properties among Models}
\label{sec:recover_params}

We are particularly interested in differences between inferred planetary properties of each model. The three parameters with the most significant changes relative to their inferred uncertainties are $K$, $e$, and $\omega$. These differences are likely driven by the GP model identifying stellar activity signals in Models 2 and 3 that were attributed to the Keplerian orbit in Model 1. Another source of this discrepancy could be from the GP model overfitting the data and attributing part of the real Keplerian signal to stellar activity. The other modeled parameters, $P$ and $T_{0}$, are consistent at the $\sim$1\% level between Models 1, 2, and 3 for all of the 17 systems.

\begin{figure*}[!t]
    \centering
    \hspace{-5mm}
    \includegraphics[width=0.93\linewidth]{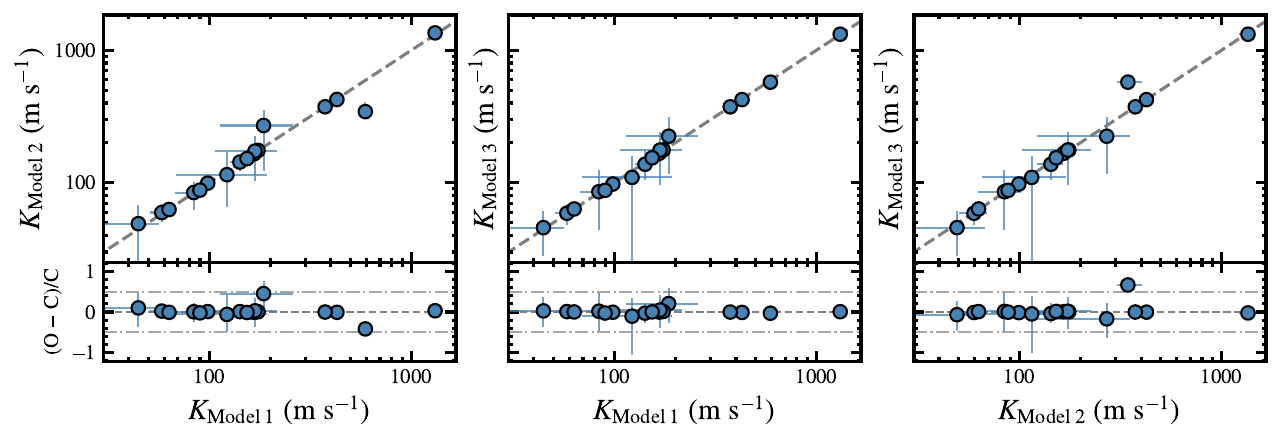}
    \caption{Relationship between the inferred Keplerian semi-amplitudes for different pairs of model fits. The fractional residuals are displayed in bottom panel, with horizontal dot-dashed lines denoting $\pm50$\% differences between the two models. The RMS of the percentage differences between pairs of models is approximately 15\%, 6\%, and 17\%, for Models 1 and 2, 1 and 3, and 2 and 3, respectively, indicating that the choice of adopted activity model can significantly affect the recovered planet semi-amplitude signal, minimum mass, and interpretation.}
    \label{fig:one-to-one_k}
\end{figure*}

\begin{figure*}%[!t]
    \centering
    \hspace{-1.4mm}
    \includegraphics[width=0.915\linewidth]{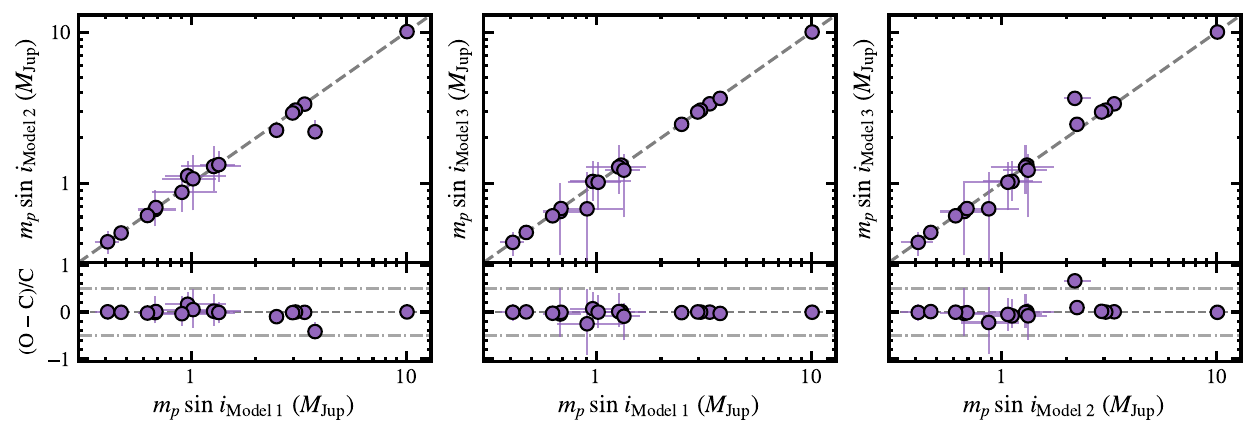}
    \caption{Similar to \autoref{fig:one-to-one_k} but for inferred minimum planetary mass. The RMS of the percentage differences between pairs of models is approximately 11\%, 7\%, and 17\%, for Models 1 and 2, 1 and 3, and 2 and 3, respectively.}
    \label{fig:one-to-one_mp}
\end{figure*}

\autoref{fig:one-to-one_k} displays 1:1 relationships and fractional residuals of the RV semi-amplitudes ($K$) for Models 1, 2, and 3. The RMS of the percentage differences between model semi-amplitudes are 15\%, 6\%, and 17\% for Models 1 and 2, 1 and 3, and 2 and 3, respectively. This is comparable to or greater than the median percentage uncertainty in $K$, 5\%, 8\%, and 7\% for Models 1, 2, and 3, respectively. The RMS of the percentage differences in RV semi-amplitude for all model comparisons is between 1\% and 67\%. Using the literature stellar mass and inferred physical parameters, we calculate the corresponding planet minimum mass, $m_p \sin  i$, for each model and find a similar characteristic variation of $\sim$12\% in $m_p\sin i$ (\autoref{fig:one-to-one_mp}). Altogether, this indicates that differences due to the choice of the model used to fit the RVs can equal or exceed the average measurement uncertainty.

\begin{rotatetable*}
\begin{deluxetable*}{lccccccccccccccc}
    \tabletypesize{\footnotesize}
    \setlength{\tabcolsep}{4pt}
    \centerwidetable
    \tablecaption{GP Hyperparameter Posteriors from Light Curve Fits \label{tab:lc_gp_posteriors}}
    \tablehead{
    \colhead{System} & 
    \multicolumn{3}{c}{ln$ \; \sigma_\mathrm{LC}$} & \multicolumn{3}{c}{$A$} & \multicolumn{3}{c}{$l_e$} & 
    \multicolumn{3}{c}{$l_p$} &
    \multicolumn{3}{c}{$P_\mathrm{GP}$} \\
    \cmidrule(lr){2-4} \cmidrule(lr){5-7} \cmidrule(lr){8-10} \cmidrule(lr){11-13} \cmidrule(lr){14-16}
    \colhead{} & \colhead{MAP\tablenotemark{a}} & \colhead{Median} & \colhead{68.3\% HDI} & \colhead{MAP} & \colhead{Median} & \colhead{68.3\% HDI} & \colhead{MAP} & \colhead{Median} & \colhead{68.3\% HDI} & \colhead{MAP} & \colhead{Median} & \colhead{68.3\% HDI} & \colhead{MAP} & \colhead{Median} & \colhead{68.3\% HDI}
    }
    \startdata
    GJ 3021 & -6.8 & -6.8 & -6.9$-$-6.6 & 2.7 & 3.1 & 2.7$-$3.5 & 7.9 & 5.0 & 4.1$-$5.6 & 0.2 & 0.3 & 0.2$-$0.3 & 7.0 & 5.6 & 5.5$-$5.6 \\ 
    HATS-29 & -27.6 & -21.9 & -27.6$-$-16.8 & 7.1 & 7.1 & 6.7$-$7.6 & 0.7 & 0.7 & 0.7$-$0.7 & 0.5 & 0.5 & 0.4$-$0.5 & 0.6 & 0.6 & 0.6$-$0.6 \\ 
    HAT-P-54 & -4.0 & -4.0 & -4.0$-$-3.9 & 5.2 & 6.0 & 4.2$-$7.2 & 21.2 & 22.0 & 15.9$-$29.1 & 0.4 & 0.5 & 0.4$-$0.5 & 15.1 & 15.1 & 14.9$-$15.3 \\
    HD 12484 & -26.2 & -19.3 & -27.6$-$-15.9 & 3.2 & 3.3 & 2.8$-$3.7 & 5.2 & 5.1 & 4.8$-$5.6 & 0.4 & 0.4 & 0.4$-$0.4 & 4.3 & 4.3 & 4.3$-$4.3 \\
    HD 102195 & -3.9 & -3.9 & -4.0$-$-3.8 & 3.7 & 3.8 & 3.3$-$4.2 & 12.7 & 9.2 & 7.5$-$10.3 & 0.4 & 0.5 & 0.5$-$0.6 & 11.7 & 10.2 & 9.8$-$10.4 \\
    HD 103720 & -4.4 & -4.4 & -4.5$-$-4.3 & 7.4 & 7.9 & 6.7$-$9.0 & 20.6 & 20.8 & 19.2$-$22.5 & 0.4 & 0.4 & 0.4$-$0.4 & 17.9 & 18.0 & 17.8$-$18.1 \\
    Kepler-17 & -1.8 & -1.8 & -1.9$-$-1.8 & 2.4 & 2.6 & 2.2$-$2.9 & 11.2 & 10.9 & 9.4$-$12.6 & 0.3 & 0.3 & 0.2$-$0.3 & 12.0 & 11.9 & 11.8$-$12.1 \\
    Kepler-43 & -4.7 & -4.6 & -4.8$-$-4.5 & 0.9 & 0.9 & 0.8$-$1.0 & 11.1 & 10.5 & 9.4$-$11.7 & 0.2 & 0.3 & 0.2$-$0.3 & 13.4 & 13.3 & 13.2$-$13.5 \\
    Kepler-75 & -13.4 & -13.4 & -13.5$-$-13.3 & 0.5 & 0.5 & 0.4$-$0.6 & 20.7 & 20.3 & 19.0$-$21.7 & 7.5 & 8.0 & 6.5$-$9.3 & 13.8 & 13.9 & 13.6$-$14.1 \\
    Kepler-77 & -5.3 & -5.3 & -5.4$-$-5.2 & 0.5 & 0.5 & 0.5$-$0.5 & 9.1 & 8.9 & 8.2$-$9.6 & 0.3 & 0.3 & 0.3$-$0.3 & 16.4 & 16.3 & 16.1$-$16.6 \\
    Kepler-447 & -4.7 & -4.7 & -4.8$-$-4.6 & 8.0 & 8.0 & 7.0$-$8.7 & 7.5 & 7.3 & 6.5$-$7.8 & 0.4 & 0.4 & 0.4$-$0.5 & 6.4 & 6.4 & 5.9$-$6.5 \\
    Kepler-539 & -4.8 & -4.8 & -4.9$-$-4.7 & 4.4 & 4.4 & 4.1$-$4.7 & 10.5 & 10.5 & 10.1$-$10.9 & 0.4 & 0.4 & 0.4$-$0.4 & 11.9 & 11.9 & 11.8$-$12.0 \\
    K2-29 & -3.7 & -3.7 & -3.7$-$-3.6 & 5.6 & 5.7 & 5.0$-$6.3 & 10.8 & 10.8 & 10.2$-$11.4 & 0.4 & 0.4 & 0.4$-$0.4 & 10.4 & 10.3 & 10.2$-$10.5 \\
    K2-237 & -2.3 & -2.3 & -2.3$-$-2.2 & 2.7 & 2.8 & 2.5$-$3.1 & 5.2 & 5.3 & 4.8$-$5.7 & 0.5 & 0.5 & 0.5$-$0.6 & 4.8 & 4.8 & 4.7$-$4.9 \\
    K2-260 & -3.0 & -3.0 & -3.0$-$-2.9 & 5.6 & 5.9 & 4.9$-$6.6 & 3.0 & 3.0 & 2.9$-$3.1 & 15.6 & 16.1 & 12.5$-$19.4 & 2.3 & 2.3 & 2.3$-$2.4 \\
    Qatar-2 & -1.8 & -1.8 & -1.8$-$-1.7 & 4.8 & 5.2 & 4.1$-$6.0 & 22.9 & 22.6 & 19.8$-$25.2 & 0.5 & 0.5 & 0.4$-$0.5 & 19.6 & 19.6 & 19.3$-$19.9 \\
    WASP-180 A & -2.5 & -2.5 & -2.6$-$-2.3 & 3.5 & 3.7 & 3.2$-$4.1 & 4.4 & 4.4 & 4.0$-$4.8 & 0.3 & 0.3 & 0.3$-$0.3 & 4.4 & 4.4 & 4.4$-$4.5 \\
    \enddata
    \tablenotetext{a}{MAP refers to the maximum a posteriori probability.}
    %\tablenotetext{b}{HDI refers to the highest density interval, the minimum range encompassing a given probability of a distribution.}
\end{deluxetable*}
\end{rotatetable*}

\begin{longrotatetable}
\begin{deluxetable*}{lccccccccccc}
\tabletypesize{\footnotesize}
\setlength{\tabcolsep}{4pt}
\centerwidetable
\tablecaption{Keplerian Parameter and QP GP Hyperparameter Posteriors of Each System. \label{tab:posterior-values}}
\tablehead{
\colhead{Model} & \colhead{$T_{0}$} & \colhead{$P$} & \colhead{$\sqrt{e}\;\mathrm{sin}\;\omega$} & \colhead{$\sqrt{e}\;\mathrm{sin}\;\omega$} & \colhead{$e$} & \colhead{$\omega$\tablenotemark{a}} & \colhead{$K$} & \colhead{$A$} & \colhead{$l_e$} & \colhead{$l_p$} & \colhead{$P_\mathrm{GP}$} \\
\colhead{} & \colhead{(BJD - 2450000)} & \colhead{(d)} & \colhead{} & \colhead{} & \colhead{} & \colhead{$(\degree)$} & \colhead{(m s$^{-1}$)} & \colhead{(m s$^{-1}$)} & \colhead{} & \colhead{} & \colhead{(d)}
}
\startdata
\multicolumn{12}{c}{\textbf{GJ 3021}} \\
Model 1 & $1592.84_{-2.06}^{+2.01}$ & $133.702_{-0.194}^{+0.207}$ & $-0.67_{-0.02}^{+0.02}$ & $0.26_{-0.03}^{+0.04}$ & $0.51_{-0.02}^{+0.02}$ & $291.15_{-3.07}^{+2.98}$ & $167.1_{-3.9}^{+3.8}$ & $\cdots$ & $\cdots$ & $\cdots$ & $\cdots$ \\
Model 2 & $1594.03_{-2.91}^{+2.84}$ & $133.665_{-0.228}^{+0.258}$ & $-0.67_{-0.02}^{+0.02}$ & $0.24_{-0.04}^{+0.05}$ & $0.51_{-0.02}^{+0.02}$ & $289.61_{-3.88}^{+3.93}$ & $165.6_{-5.4}^{+5.6}$ & $17.6_{-4.7}^{+5.9}$ & $20.79_{-8.59}^{+7.43}$ & $0.39_{-0.20}^{+0.16}$ & $7.18_{-0.25}^{+0.19}$ \\
Model 3 & $1592.81_{-2.42}^{+2.18}$ & $133.708_{-0.241}^{+0.219}$ & $-0.67_{-0.02}^{+0.02}$ & $0.26_{-0.04}^{+0.04}$ & $0.51_{-0.02}^{+0.02}$ & $291.22_{-3.26}^{+3.26}$ & $166.9_{-4.4}^{+4.4}$ & $10.8_{-8.3}^{+5.3}$ & $4.83_{-0.78}^{+0.79}$ & $0.25_{-0.02}^{+0.02}$ & $5.59_{-0.06}^{+0.06}$ \\
\hline
\multicolumn{12}{c}{\textbf{HATS-29}} \\
Model 1 & $7032.01_{-0.18}^{+0.44}$ & $4.607_{-1.128}^{+0.031}$ & $-0.18_{-0.32}^{+0.30}$ & $0.06_{-0.27}^{+0.26}$ & $0.15_{-0.15}^{+0.08}$ & $255.72_{-65.51}^{+104.28}$ & $83.8_{-16.3}^{+12.9}$ & $\cdots$ & $\cdots$ & $\cdots$ & $\cdots$ \\
Model 2 & $7031.96_{-0.29}^{+0.36}$ & $4.611_{-0.718}^{+1.080}$ & $-0.14_{-0.37}^{+0.34}$ & $0.04_{-0.30}^{+0.29}$ & $0.17_{-0.17}^{+0.09}$ & $238.39_{-85.71}^{+121.61}$ & $84.3_{-21.9}^{+17.7}$ & $243.8_{-242.8}^{+134.1}$ & $31.11_{-8.58}^{+18.89}$ & $27.00_{-9.58}^{+21.76}$ & $25.49_{-9.87}^{+24.09}$ \\
Model 3 & $7031.99_{-0.19}^{+0.47}$ & $4.831_{-1.806}^{+1.937}$ & $-0.09_{-0.48}^{+0.38}$ & $0.08_{-0.41}^{+0.43}$ & $0.27_{-0.27}^{+0.19}$ & $216.17_{-87.98}^{+143.83}$ & $85.5_{-41.7}^{+38.3}$ & $30.4_{-29.4}^{+17.9}$ & $0.68_{-0.02}^{+0.02}$ & $0.45_{-0.02}^{+0.02}$ & $0.641_{-0.004}^{+0.004}$ \\
\hline
\multicolumn{12}{c}{\textbf{HAT-P-54}} \\
Model 1 & $6299.31_{-0.38}^{+0.35}$ & $3.797_{-0.003}^{+0.003}$ & $0.47_{-0.18}^{+0.39}$ & $0.28_{-0.23}^{+0.33}$ & $0.40_{-0.40}^{+0.15}$ & $63.44_{-34.18}^{+33.47}$ & $185.7_{-72.1}^{+74.1}$ & $\cdots$ & $\cdots$ & $\cdots$ & $\cdots$ \\
Model 2 & $6299.32_{-0.33}^{+0.40}$ & $3.796_{-0.004}^{+0.004}$ & $0.69_{-0.12}^{+0.23}$ & $0.24_{-0.23}^{+0.33}$ & $0.65_{-0.15}^{+0.24}$ & $71.58_{-27.76}^{+23.83}$ & $270.3_{-147.5}^{+83.4}$ & $59.9_{-29.4}^{+20.7}$ & $37.62_{-7.06}^{+12.38}$ & $25.51_{-8.98}^{+24.48}$ & $25.07_{-11.21}^{+22.76}$ \\
Model 3 & $6299.25_{-0.38}^{+0.37}$ & $3.798_{-0.003}^{+0.004}$ & $0.63_{-0.15}^{+0.28}$ & $0.21_{-0.25}^{+0.36}$ & $0.56_{-0.19}^{+0.31}$ & $73.29_{-32.91}^{+28.95}$ & $224.4_{-108.9}^{+88.4}$ & $52.2_{-18.1}^{+14.9}$ & $30.67_{-7.04}^{+6.83}$ & $0.51_{-0.05}^{+0.05}$ & $15.16_{-0.13}^{+0.13}$ \\
\hline
\multicolumn{12}{c}{\textbf{HD 12484}} \\
Model 1 & $6698.49_{-0.71}^{+0.76}$ & $58.853_{-0.074}^{+0.074}$ & $-0.07_{-0.14}^{+0.12}$ & $0.17_{-0.08}^{+0.11}$ & $0.05_{-0.04}^{+0.03}$ & $303.48_{-130.84}^{+56.52}$ & $153.9_{-5.2}^{+5.2}$ & $\cdots$ & $\cdots$ & $\cdots$ & $\cdots$ \\
Model 2 & $6698.17_{-0.80}^{+0.82}$ & $58.840_{-0.075}^{+0.080}$ & $0.00_{-0.15}^{+0.15}$ & $0.20_{-0.08}^{+0.10}$ & $0.06_{-0.04}^{+0.03}$ & $144.20_{-144.20}^{+175.79}$ & $151.9_{-5.8}^{+5.8}$ & $12.7_{-9.1}^{+6.0}$ & $29.35_{-8.84}^{+20.65}$ & $24.26_{-24.26}^{+9.22}$ & $24.36_{-24.31}^{+9.30}$ \\
Model 3 & $6698.52_{-0.76}^{+0.81}$ & $58.836_{-0.081}^{+0.082}$ & $-0.04_{-0.16}^{+0.13}$ & $0.17_{-0.09}^{+0.12}$ & $0.05_{-0.04}^{+0.03}$ & $286.91_{-219.53}^{+73.09}$ & $154.1_{-6.5}^{+6.5}$ & $24.9_{-5.0}^{+6.6}$ & $5.11_{-0.40}^{+0.40}$ & $0.39_{-0.02}^{+0.02}$ & $4.28_{-0.02}^{+0.02}$ \\
\hline
\multicolumn{12}{c}{\textbf{HD 102195}} \\
Model 1 & $3732.51_{-0.04}^{+0.04}$ & $4.1133_{-0.0001}^{+0.0001}$ & $0.19_{-0.06}^{+0.07}$ & $-0.12_{-0.11}^{+0.08}$ & $0.06_{-0.03}^{+0.03}$ & $122.59_{-25.27}^{+27.80}$ & $63.3_{-1.6}^{+1.6}$ & $\cdots$ & $\cdots$ & $\cdots$ & $\cdots$ \\
Model 2 & $3732.48_{-0.04}^{+0.03}$ & $4.1133_{-0.0001}^{+0.0001}$ & $0.07_{-0.09}^{+0.11}$ & $-0.08_{-0.12}^{+0.10}$ & $0.02_{-0.02}^{+0.01}$ & $146.63_{-68.54}^{+49.99}$ & $62.8_{-2.2}^{+1.8}$ & $85.3_{-26.5}^{+19.5}$ & $43.95_{-3.07}^{+6.05}$ & $20.39_{-17.36}^{+11.11}$ & $10.30_{-10.30}^{+12.88}$ \\
Model 3 & $3732.52_{-0.04}^{+0.03}$ & $4.1133_{-0.0001}^{+0.0001}$ & $0.15_{-0.04}^{+0.05}$ & $-0.15_{-0.09}^{+0.07}$ & $0.05_{-0.02}^{+0.02}$ & $135.96_{-20.09}^{+23.40}$ & $63.4_{-1.5}^{+1.5}$ & $9.5_{-1.9}^{+1.8}$ & $9.01_{-1.34}^{+1.36}$ & $0.51_{-0.05}^{+0.05}$ & $10.19_{-0.31}^{+0.30}$ \\
\hline
\multicolumn{12}{c}{\textbf{HD 103720}} \\
Model 1 & $5384.24_{-0.08}^{+4.51}$ & $4.5556_{-0.0001}^{+0.0001}$ & $-0.27_{-0.05}^{+0.04}$ & $-0.03_{-0.08}^{+0.07}$ & $0.08_{-0.02}^{+0.02}$ & $263.52_{-17.36}^{+15.89}$ & $90.0_{-2.0}^{+2.0}$ & $\cdots$ & $\cdots$ & $\cdots$ & $\cdots$ \\
Model 2 & $5384.21_{-0.09}^{+4.46}$ & $4.5556_{-0.0001}^{+0.0001}$ & $-0.29_{-0.03}^{+0.02}$ & $-0.01_{-0.05}^{+0.05}$ & $0.08_{-0.01}^{+0.01}$ & $267.34_{-9.35}^{+9.33}$ & $88.0_{-1.5}^{+2.1}$ & $12.4_{-1.7}^{+1.8}$ & $1.65_{-1.35}^{+28.98}$ & $9.65_{-9.65}^{+15.69}$ & $18.17_{-13.85}^{+13.31}$ \\
Model 3 & $5388.74_{-4.54}^{+0.04}$ & $4.5556_{-0.0001}^{+0.0001}$ & $-0.29_{-0.02}^{+0.02}$ & $-0.02_{-0.04}^{+0.03}$ & $0.09_{-0.01}^{+0.01}$ & $266.12_{-6.89}^{+6.49}$ & $87.6_{-1.4}^{+1.4}$ & $15.9_{-2.1}^{+2.5}$ & $21.19_{-1.64}^{+1.64}$ & $0.37_{-0.02}^{+0.02}$ & $17.92_{-0.18}^{+0.18}$ \\
\hline
\multicolumn{12}{c}{\textbf{K2-29}} \\
Model 1 & $7383.70_{-0.06}^{+0.05}$ & $3.254_{-0.004}^{+0.004}$ & $-0.05_{-0.14}^{+0.13}$ & $0.15_{-0.14}^{+0.18}$ & $0.05_{-0.05}^{+0.02}$ & $258.01_{-163.15}^{+101.99}$ & $98.3_{-3.4}^{+3.5}$ & $\cdots$ & $\cdots$ & $\cdots$ & $\cdots$ \\
Model 2 & $7383.68_{-0.07}^{+0.06}$ & $3.254_{-0.004}^{+0.004}$ & $-0.02_{-0.14}^{+0.14}$ & $0.19_{-0.14}^{+0.18}$ & $0.06_{-0.06}^{+0.03}$ & $236.45_{-236.45}^{+76.41}$ & $99.3_{-4.1}^{+4.1}$ & $10.7_{-9.7}^{+7.0}$ & $31.26_{-8.76}^{+18.74}$ & $25.32_{-17.69}^{+15.72}$ & $24.34_{-24.34}^{+9.33}$ \\
Model 3 & $7383.70_{-0.07}^{+0.06}$ & $3.254_{-0.004}^{+0.004}$ & $-0.04_{-0.15}^{+0.14}$ & $0.16_{-0.14}^{+0.18}$ & $0.06_{-0.06}^{+0.02}$ & $254.10_{-175.44}^{+105.90}$ & $97.8_{-3.9}^{+4.1}$ & $6.4_{-5.4}^{+2.4}$ & $10.81_{-0.61}^{+0.62}$ & $0.43_{-0.02}^{+0.02}$ & $10.35_{-0.12}^{+0.12}$ \\
\hline
\multicolumn{12}{c}{\textbf{K2-237}} \\
Model 1 & $7684.63_{-0.39}^{+0.34}$ & $2.182_{-0.003}^{+0.002}$ & $0.11_{-0.12}^{+0.14}$ & $0.05_{-0.18}^{+0.18}$ & $0.05_{-0.05}^{+0.02}$ & $92.64_{-92.32}^{+51.98}$ & $174.3_{-9.6}^{+8.7}$ & $\cdots$ & $\cdots$ & $\cdots$ & $\cdots$ \\
Model 2 & $7684.57_{-0.36}^{+0.33}$ & $2.182_{-0.003}^{+0.002}$ & $0.10_{-0.15}^{+0.17}$ & $0.08_{-0.22}^{+0.21}$ & $0.06_{-0.06}^{+0.03}$ & $101.65_{-101.65}^{+63.38}$ & $175.5_{-10.8}^{+9.2}$ & $81.6_{-80.6}^{+80.7}$ & $32.87_{-8.78}^{+17.13}$ & $25.55_{-9.25}^{+24.45}$ & $24.48_{-16.75}^{+16.71}$ \\
Model 3 & $7684.61_{-0.50}^{+0.37}$ & $2.183_{-0.004}^{+0.003}$ & $0.11_{-0.19}^{+0.17}$ & $0.11_{-0.22}^{+0.22}$ & $0.08_{-0.08}^{+0.04}$ & $91.07_{-91.07}^{+69.54}$ & $176.6_{-12.1}^{+10.5}$ & $22.1_{-21.1}^{+13.8}$ & $5.24_{-0.46}^{+0.47}$ & $0.53_{-0.06}^{+0.06}$ & $4.83_{-0.12}^{+0.12}$ \\
\hline
\multicolumn{12}{c}{\textbf{K2-260}} \\
Model 1 & $7820.77_{-1.43}^{+1.31}$ & $2.532_{-0.890}^{+1.131}$ & $0.11_{-0.46}^{+0.56}$ & $0.26_{-0.32}^{+0.49}$ & $0.39_{-0.39}^{+0.17}$ & $124.29_{-124.29}^{+130.64}$ & $168.3_{-61.2}^{+48.8}$ & $\cdots$ & $\cdots$ & $\cdots$ & $\cdots$ \\
Model 2 & $7820.78_{-0.81}^{+1.92}$ & $2.544_{-0.902}^{+1.135}$ & $0.14_{-0.43}^{+0.61}$ & $0.22_{-0.34}^{+0.52}$ & $0.41_{-0.41}^{+0.18}$ & $121.11_{-121.11}^{+121.40}$ & $173.3_{-70.2}^{+53.7}$ & $69.7_{-68.7}^{+44.8}$ & $29.85_{-9.48}^{+20.15}$ & $25.53_{-9.35}^{+24.47}$ & $25.33_{-9.45}^{+24.65}$ \\
Model 3 & $7820.76_{-0.44}^{+0.90}$ & $2.626_{-1.017}^{+1.252}$ & $0.13_{-0.40}^{+0.71}$ & $0.21_{-0.38}^{+0.54}$ & $0.47_{-0.47}^{+0.18}$ & $128.68_{-128.67}^{+112.22}$ & $176.3_{-79.4}^{+64.2}$ & $46.9_{-45.9}^{+25.6}$ & $3.04_{-0.10}^{+0.11}$ & $16.15_{-3.36}^{+3.60}$ & $2.35_{-0.66}^{+0.68}$ \\
\hline
\multicolumn{12}{c}{\textbf{Kepler-17}} \\
Model 1 & $5185.69_{-0.12}^{+0.13}$ & $1.486_{-0.001}^{+0.001}$ & $-0.14_{-0.18}^{+0.15}$ & $0.00_{-0.15}^{+0.15}$ & $0.05_{-0.05}^{+0.03}$ & $252.37_{-56.71}^{+85.38}$ & $428.4_{-21.9}^{+21.2}$ & $\cdots$ & $\cdots$ & $\cdots$ & $\cdots$ \\
Model 2 & $5185.65_{-0.19}^{+0.11}$ & $1.486_{-0.001}^{+0.001}$ & $-0.10_{-0.18}^{+0.19}$ & $-0.01_{-0.18}^{+0.16}$ & $0.06_{-0.06}^{+0.03}$ & $234.74_{-70.40}^{+106.46}$ & $425.0_{-27.7}^{+25.3}$ & $43.2_{-42.2}^{+32.0}$ & $30.44_{-10.04}^{+19.56}$ & $26.18_{-17.43}^{+15.78}$ & $25.98_{-25.86}^{+7.50}$ \\
Model 3 & $5185.67_{-0.16}^{+0.16}$ & $1.486_{-0.001}^{+0.001}$ & $-0.13_{-0.20}^{+0.18}$ & $-0.01_{-0.18}^{+0.18}$ & $0.06_{-0.06}^{+0.03}$ & $243.97_{-58.78}^{+96.02}$ & $423.6_{-27.9}^{+29.0}$ & $29.5_{-28.5}^{+14.8}$ & $10.92_{-1.62}^{+1.64}$ & $0.26_{-0.02}^{+0.02}$ & $11.94_{-0.18}^{+0.18}$ \\
\hline
\multicolumn{12}{c}{\textbf{Kepler-43}} \\
Model 1 & $4965.46_{-2.99}^{+0.06}$ & $3.0239_{-0.0002}^{+0.0002}$ & $0.09_{-0.07}^{+0.09}$ & $-0.19_{-0.06}^{+0.05}$ & $0.05_{-0.02}^{+0.02}$ & $155.41_{-25.04}^{+22.21}$ & $374.4_{-6.5}^{+6.4}$ & $\cdots$ & $\cdots$ & $\cdots$ & $\cdots$ \\
Model 2 & $4965.49_{-0.04}^{+0.03}$ & $3.0239_{-0.0002}^{+0.0002}$ & $0.11_{-0.06}^{+0.08}$ & $-0.19_{-0.06}^{+0.05}$ & $0.05_{-0.02}^{+0.02}$ & $151.07_{-22.72}^{+19.99}$ & $374.3_{-6.3}^{+6.1}$ & $15.7_{-14.7}^{+7.0}$ & $29.65_{-10.12}^{+20.35}$ & $25.11_{-9.19}^{+24.85}$ & $25.05_{-15.40}^{+18.65}$ \\
Model 3 & $4965.49_{-0.04}^{+0.04}$ & $3.0239_{-0.0002}^{+0.0002}$ & $0.10_{-0.07}^{+0.08}$ & $-0.20_{-0.06}^{+0.05}$ & $0.05_{-0.02}^{+0.02}$ & $153.84_{-22.20}^{+20.07}$ & $373.9_{-6.9}^{+6.8}$ & $13.6_{-10.1}^{+6.2}$ & $10.46_{-1.15}^{+1.15}$ & $0.28_{-0.04}^{+0.04}$ & $13.29_{-0.15}^{+0.15}$ \\
\hline
\multicolumn{12}{c}{\textbf{Kepler-75}} \\
Model 1 & $5739.68_{-11.35}^{+8.22}$ & $8.913_{-0.058}^{+0.080}$ & $0.67_{-0.03}^{+0.03}$ & $0.33_{-0.06}^{+0.06}$ & $0.56_{-0.02}^{+0.02}$ & $63.48_{-5.41}^{+4.88}$ & $1293.2_{-70.1}^{+60.4}$ & $\cdots$ & $\cdots$ & $\cdots$ & $\cdots$ \\
Model 2 & $5740.27_{-10.71}^{+9.62}$ & $8.921_{-0.124}^{+0.100}$ & $0.67_{-0.03}^{+0.04}$ & $0.33_{-0.08}^{+0.07}$ & $0.56_{-0.04}^{+0.03}$ & $63.73_{-5.36}^{+5.82}$ & $1294.1_{-77.7}^{+67.2}$ & $603.1_{-602.1}^{+441.1}$ & $123.42_{-39.11}^{+62.34}$ & $27.31_{-13.49}^{+17.20}$ & $26.29_{-9.08}^{+23.71}$ \\
Model 3 & $5739.96_{-9.52}^{+3.70}$ & $8.895_{-0.099}^{+0.133}$ & $0.67_{-0.03}^{+0.03}$ & $0.31_{-0.07}^{+0.07}$ & $0.56_{-0.04}^{+0.03}$ & $65.09_{-5.97}^{+5.23}$ & $1284.6_{-73.4}^{+59.7}$ & $166.4_{-165.4}^{+110.5}$ & $20.41_{-1.24}^{+1.48}$ & $8.20_{-1.49}^{+1.44}$ & $13.90_{-0.27}^{+0.26}$ \\
\hline
\multicolumn{12}{c}{\textbf{Kepler-77}} \\
Model 1 & $5096.51_{-1.04}^{+0.84}$ & $3.576_{-0.004}^{+0.005}$ & $0.20_{-0.25}^{+0.36}$ & $-0.01_{-0.25}^{+0.27}$ & $0.15_{-0.15}^{+0.07}$ & $118.42_{-112.03}^{+53.82}$ & $58.2_{-7.4}^{+7.4}$ & $\cdots$ & $\cdots$ & $\cdots$ & $\cdots$ \\
Model 2 & $5096.01_{-0.90}^{+1.57}$ & $3.575_{-0.029}^{+0.007}$ & $0.16_{-0.31}^{+0.38}$ & $-0.00_{-0.29}^{+0.29}$ & $0.17_{-0.17}^{+0.09}$ & $126.04_{-126.04}^{+62.83}$ & $59.6_{-9.4}^{+9.9}$ & $19.9_{-18.9}^{+14.3}$ & $31.22_{-8.78}^{+18.78}$ & $25.86_{-14.97}^{+17.91}$ & $25.45_{-23.65}^{+9.93}$ \\
Model 3 & $5096.08_{-0.83}^{+1.56}$ & $3.576_{-0.029}^{+0.006}$ & $0.14_{-0.32}^{+0.40}$ & $-0.00_{-0.30}^{+0.33}$ & $0.18_{-0.18}^{+0.10}$ & $131.38_{-131.38}^{+72.41}$ & $58.8_{-10.7}^{+9.9}$ & $10.0_{-9.0}^{+5.6}$ & $8.94_{-0.69}^{+0.68}$ & $0.30_{-0.01}^{+0.01}$ & $16.32_{-0.30}^{+0.28}$ \\
\hline
\multicolumn{12}{c}{\textbf{Kepler-447}} \\
Model 1 & $4970.05_{-0.59}^{+0.67}$ & $7.795_{-0.005}^{+0.003}$ & $0.27_{-0.17}^{+0.21}$ & $0.11_{-0.24}^{+0.28}$ & $0.15_{-0.15}^{+0.06}$ & $76.82_{-61.02}^{+42.79}$ & $141.9_{-15.6}^{+15.9}$ & $\cdots$ & $\cdots$ & $\cdots$ & $\cdots$ \\
Model 2 & $4970.26_{-0.84}^{+0.79}$ & $7.955_{-0.166}^{+0.009}$ & $0.26_{-0.23}^{+0.24}$ & $0.07_{-0.31}^{+0.34}$ & $0.17_{-0.17}^{+0.08}$ & $92.23_{-79.84}^{+44.68}$ & $143.2_{-20.5}^{+21.3}$ & $73.4_{-72.4}^{+35.3}$ & $32.20_{-9.07}^{+17.80}$ & $24.40_{-14.32}^{+18.39}$ & $26.11_{-9.18}^{+23.89}$ \\
Model 3 & $4970.15_{-1.04}^{+0.97}$ & $7.797_{-0.269}^{+0.538}$ & $0.21_{-0.32}^{+0.32}$ & $0.07_{-0.35}^{+0.41}$ & $0.22_{-0.22}^{+0.13}$ & $109.41_{-109.34}^{+61.48}$ & $137.9_{-32.5}^{+36.1}$ & $32.4_{-31.4}^{+36.4}$ & $7.25_{-0.61}^{+0.62}$ & $0.41_{-0.06}^{+0.05}$ & $6.44_{-0.29}^{+0.28}$ \\
\hline
\multicolumn{12}{c}{\textbf{Kepler-539}} \\
Model 1 & $5621.59_{-19.59}^{+23.01}$ & $122.499_{-1.976}^{+2.112}$ & $-0.28_{-0.41}^{+0.32}$ & $-0.06_{-0.44}^{+0.43}$ & $0.31_{-0.31}^{+0.14}$ & $237.41_{-50.70}^{+107.04}$ & $44.6_{-14.0}^{+11.9}$ & $\cdots$ & $\cdots$ & $\cdots$ & $\cdots$ \\
Model 2 & $5618.83_{-26.31}^{+27.13}$ & $121.728_{-2.857}^{+3.466}$ & $-0.27_{-0.44}^{+0.31}$ & $-0.13_{-0.54}^{+0.48}$ & $0.39_{-0.39}^{+0.16}$ & $225.29_{-56.64}^{+104.70}$ & $49.1_{-23.3}^{+18.4}$ & $17.7_{-16.7}^{+8.2}$ & $25.30_{-15.48}^{+16.32}$ & $25.30_{-15.81}^{+17.56}$ & $25.08_{-9.26}^{+24.92}$ \\
Model 3 & $5620.39_{-23.27}^{+25.76}$ & $122.254_{-2.419}^{+2.664}$ & $-0.25_{-0.45}^{+0.33}$ & $-0.08_{-0.48}^{+0.49}$ & $0.36_{-0.36}^{+0.16}$ & $226.88_{-54.97}^{+114.68}$ & $45.8_{-18.1}^{+15.3}$ & $12.5_{-11.5}^{+5.5}$ & $10.51_{-0.38}^{+0.37}$ & $0.43_{-0.01}^{+0.01}$ & $11.89_{-0.10}^{+0.09}$ \\
\hline
\multicolumn{12}{c}{\textbf{Qatar 2}} \\
Model 1 & $5624.87_{-0.05}^{+0.04}$ & $3.931_{-0.002}^{+0.002}$ & $0.00_{-0.13}^{+0.13}$ & $-0.06_{-0.14}^{+0.13}$ & $0.03_{-0.03}^{+0.01}$ & $179.15_{-107.78}^{+86.80}$ & $592.0_{-23.4}^{+22.8}$ & $\cdots$ & $\cdots$ & $\cdots$ & $\cdots$ \\
Model 2 & $5624.97_{-0.06}^{+0.05}$ & $3.930_{-0.003}^{+0.004}$ & $0.03_{-0.11}^{+0.11}$ & $-0.05_{-0.13}^{+0.11}$ & $0.02_{-0.02}^{+0.01}$ & $158.70_{-105.92}^{+72.13}$ & $344.0_{-39.9}^{+60.4}$ & $330.9_{-172.7}^{+111.5}$ & $47.54_{-1.56}^{+2.46}$ & $1.48_{-0.75}^{+0.50}$ & $0.571_{-0.208}^{+0.001}$ \\
Model 3 & $5624.91_{-0.05}^{+0.04}$ & $3.930_{-0.002}^{+0.002}$ & $-0.08_{-0.13}^{+0.10}$ & $-0.11_{-0.13}^{+0.11}$ & $0.03_{-0.03}^{+0.02}$ & $211.38_{-55.70}^{+65.49}$ & $574.1_{-19.6}^{+19.3}$ & $71.0_{-24.8}^{+20.3}$ & $22.77_{-2.73}^{+2.69}$ & $0.51_{-0.05}^{+0.05}$ & $19.54_{-0.29}^{+0.29}$ \\
\hline
\multicolumn{12}{c}{\textbf{WASP-180 A}} \\
Model 1 & $7763.23_{-0.50}^{+1.43}$ & $3.786_{-0.410}^{+0.583}$ & $0.37_{-0.35}^{+0.52}$ & $-0.25_{-0.49}^{+0.41}$ & $0.52_{-0.20}^{+0.45}$ & $130.06_{-77.77}^{+63.44}$ & $122.4_{-53.5}^{+69.8}$ & $\cdots$ & $\cdots$ & $\cdots$ & $\cdots$ \\
Model 2 & $7763.14_{-1.12}^{+0.64}$ & $3.791_{-0.503}^{+0.408}$ & $0.43_{-0.28}^{+0.54}$ & $-0.15_{-0.50}^{+0.43}$ & $0.53_{-0.21}^{+0.44}$ & $119.63_{-78.84}^{+58.59}$ & $115.0_{-49.8}^{+54.8}$ & $109.4_{-70.7}^{+53.7}$ & $28.72_{-7.81}^{+21.28}$ & $26.75_{-9.48}^{+23.21}$ & $24.58_{-21.49}^{+12.78}$ \\
Model 3 & $7763.24_{-1.20}^{+0.74}$ & $3.814_{-1.531}^{+2.099}$ & $0.03_{-0.51}^{+0.74}$ & $-0.17_{-0.78}^{+0.38}$ & $0.67_{-0.21}^{+0.33}$ & $175.08_{-110.75}^{+105.59}$ & $109.7_{-106.5}^{+49.0}$ & $83.7_{-44.4}^{+38.4}$ & $4.40_{-0.38}^{+0.38}$ & $0.30_{-0.02}^{+0.02}$ & $4.44_{-0.03}^{+0.04}$
\enddata
\tablenotetext{a}{$\omega$ refers to the  argument of periastron of the star, $\omega_*$.}
\end{deluxetable*}
\end{longrotatetable}

\begin{figure*}[!t]
    \centering
    \hspace*{-1.9mm}
    \includegraphics[width=0.9268\linewidth]{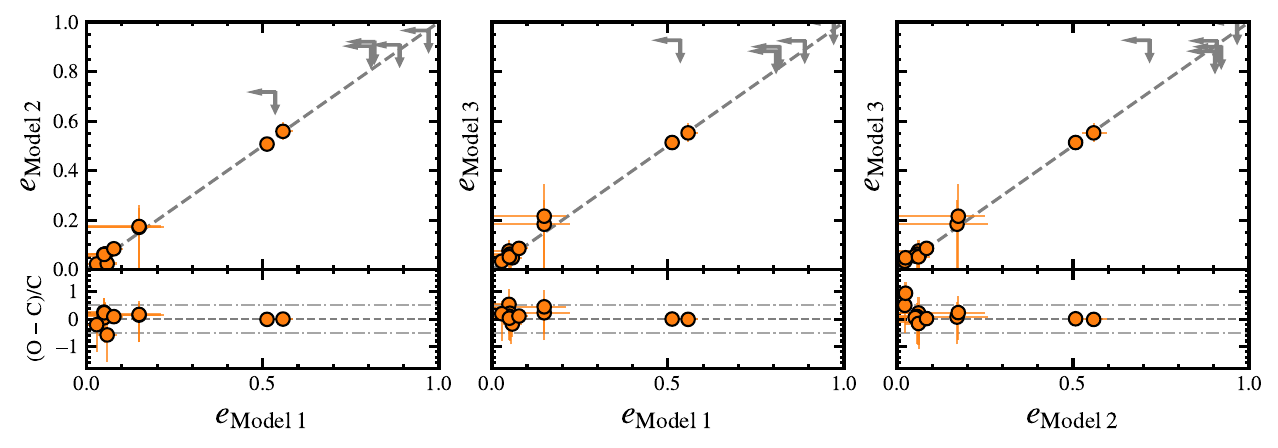}
    \caption{Similar to \autoref{fig:one-to-one_k} but for inferred eccentricities. gray arrows represent the 95\% upper limits for systems with the poorest eccentricity constraints listed in \autoref{sec:rv_only_model_results}. The RMS of the percentage differences in eccentricities between each model pair, excluding the systems with unconstrained eccentricities, is approximately 22\%, 24\%, and 33\% for Models 1 and 2, 1 and 3, and 2 and 3, respectively.}
    \label{fig:one-to-one_e}
\end{figure*}

\autoref{fig:one-to-one_e} shows the 1:1 relationship and residuals of the eccentricities ($e$) between each of the models. For systems that are unconstrained (see \autoref{sec:rv_only_model_results}), 95\% upper limits are plotted as gray arrows. Excluding the five systems with the poorest eccentricity constraints, the RMS in percentage differences of $e$ is 22\%, 24\%, and 33\% between Models 1 and 2, 1 and 3, and 2 and 3, respectively. The median absolute difference in $e$ across all model pairs is 0.01. The median measurement uncertainty in eccentricity for the same subsample of 12 systems is 0.03, 0.03, and 0.04 for Models 1, 2, and 3, respectively. The range of the percentage and absolute differences in eccentricity between all model comparisons is $<$1\%--95\% and $<$0.01--0.25, respectively.

\begin{figure*}%[!t]
    \centering
    \hspace*{-3.5mm}
    \includegraphics[width=0.928\linewidth]{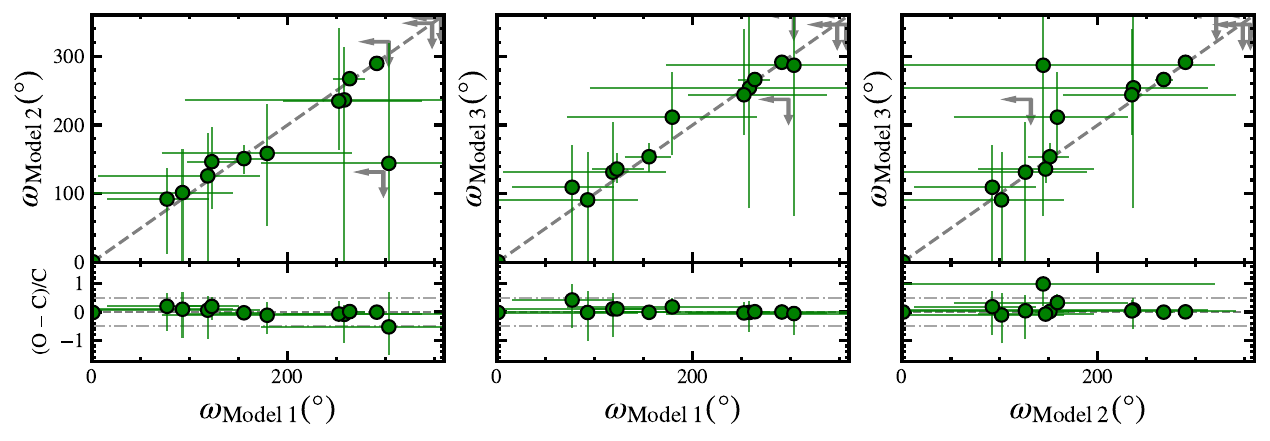}
    \caption{Similar to \autoref{fig:one-to-one_k} but for inferred arguments of periastron. Gray arrows represent the 95\% upper limits for systems with poorest constraints listed in \autoref{sec:rv_only_model_results}. The RMS of the percentage differences in the arguments of periastron between each model pair, excluding the outliers, is approximately 18\%, 14\%, and 31\%, for Models 1 and 2, 1 and 3, and 2 and 3, respectively.}
    \label{fig:one-to-one_omega}
\end{figure*}

\autoref{fig:one-to-one_omega} displays the 1:1 relationship and fractional residuals of inferred arguments of periastron ($\omega$) between each of the pairs of models. $e$ and $\omega$ are highly correlated, so similar to our analysis of $e$, we remove the five systems with poorly constrained eccentricities and only plot their 95\% upper limits as gray arrows. For the subsample of 12 well-constrained systems, the RMS of the percentage differences in $\omega$ is 18\%, 14\%, and 31\% between Models 1 and 2, 1 and 3, and 2 and 3, respectively. The median uncertainties of the arguments of periastron are 28\%, 45\%, and 26\% for Models 1, 2, and 3, respectively. The range of the absolute percentage differences in $\omega$ is between 1\% and 99\%.

We conclude that the variations among models that treat activity as a source of white noise,  those that implement GPs to describe stellar activity imprinted in RVs, and those that pretrain the GP kernel parameters can be as large as the typical measurement precision of $K$, and therefore $m_p \sin i$. On the other hand, deviations in inferred values for $e$ and $\omega$ are typically below measurement errors. Unless planet mass can independently be constrained (e.g., with transit timing variations), the underlying truth is unknown and these observed differences ($\sim$13\% in $K$ and $\sim$26\% in $e$) suggest that the inferred physical and orbital properties of planets can be highly sensitive to model choices---at least for giant planets orbiting active stars.

\section{Discussion}
\label{sec:discuss}

\subsection{Comparison between GP Kernel Hyperparameters from Photometry and RVs}

The primary assumption of Model 3 is that photospheric activity signals should manifest similarly in both photometry and RVs. This correlation is expected to improve for datasets taken closer in time. Independent GP model fits to photometric and RV time series should therefore yield similar kernel parameters. While not the primary focus of this study, here we briefly compare the GP hyperparameters inferred from the GP model trained on the photometry (\autoref{sec:train_gp_on_lc}) and Model 2 (the untrained GP model directly applied to RVs; \autoref{sec:gp_on_rvs}).

\subsubsection{Rotation Periods from Photometry and RVs} \label{sec:gp_params_from_lv_v_rc}

When inferred from photometry, the $P_\mathrm{GP}$ hyperparameter of the QP kernel can be interpreted as the stellar rotation period \citep{Angus2018, Gordon2021}. $P_\mathrm{GP}$ quantifies the recurrence timescale of the underlying process, which is equivalent to the rotation period for activity signals resulting from the modulation of starspots. Measuring the stellar rotation period through this method has been shown to be more accurate and precise than traditional non-inference-based methods such as the Lomb-Scargle periodogram \citep{Lomb1976, Scargle1982} and the autocorrelation function \citep{McQuillan2013}. This is particularly true for systems with evolving stellar features, such as the active stars considered in this work \citep{Gillen2020}.

For these reasons, we choose to adopt the inferred $P_\mathrm{GP}$ from the GP model fit to the photometric data as the stellar rotation period for each system. This choice also ensures that stellar rotation periods are determined homogeneously. Our rotation periods inferred from the photometry ($P_\mathrm{GP}$) can be found in \autoref{tab:lc_gp_posteriors}. This systematic treatment allows us to compare $P_\mathrm{GP}$ values inferred from GP model fits to the photometry and the RVs.\footnote{Note that if differential rotation is present, the rotation period inferred from light curves may overestimate the true rotation period if spots are located at mid-latitudes. On the Sun, differential rotation varies from $\sim$25~days at the equator to $\sim$34~days at the poles. However, if the nature of the periodicity imprinted in RVs and light curves are both predominately caused by spots, then these periodic signals should be similar in both datasets.} By contrast, the values reported in \autoref{tab:systems} are estimated with a variety of methods.

All 17 systems yield $P_\mathrm{GP}$ with uncertainties at the $\sim$20\% level or below when fit to the photometry, but only GJ 3021 has a well-constrained $P_\mathrm{GP}$ estimate from the simultaneous Keplerian and GP fit of Model 2. For the 16 other systems, $P_\mathrm{GP}$ is effectively unconstrained. These poor constraints are likely a result of modest RV sampling at both high and low frequencies. Conversely, GJ 3021 has a comparably greater number of RV observations ($N_\mathrm{RV} > 60$) from a single spectrograph over a modest observational window ($\sim$1.3 yrs).

For GJ 3021, $P_\mathrm{GP}$ inferred from the GP fit to the light curve is formally inconsistent from the Model 2 value. The $P_\mathrm{GP}$ value from training the GP on the light curve is $5.6^{+0.1}_{-0.1}$ d (\autoref{tab:lc_gp_posteriors}). From the Model 2 fit, we determine $P_\mathrm{GP} = 7.2_{-0.3}^{+0.2}$ d (\autoref{tab:posterior-values}). This difference is likely due to the considerable gap between the light curve and RVs. The observations are separated by $\sim$18 yrs, or nearly 1100 stellar rotations. It is likely that both signals are ``correct'' (in that they are accurately reflecting the astrophysical signal) as this break is substantial enough that the starspots (and possibly even the broader activity cycle) have evolved significantly in that time frame. Likewise, if differential rotation is present, the average latitude of starspots may shifted during this time as part of the overall activity cycle. In this context, the general agreement (7.2 d versus 5.6 d) is encouraging.

This agreement supports findings that, with sufficient sampling, an uninformed GP model can accurately identify and infer stellar activity signals in RVs \citep[e.g.,][]{Faria2016, Faria2020, Kosiarek2020, Langellier2021, Nicholson2022}. Although demonstrated for only one system, activity signals seem to share similar periodic features, and this correlation decreases as a function of the separation in time between the two time series, in line with the assumptions of Model 3.

\subsubsection{Comparison of Other Kernel Parameters Inferred from Photometry and RVs} \label{sec:comp_kernel_photo_v_rv}

Comparisons between the GP model fit to the light curves and the Model 2 values for other kernel parameters, $l_e$ (spot evolutionary time) and $l_p$ (scale of local variations), are more challenging to carry out and interpret. For all 17 systems, $l_p$ from Model 2 is unconstrained. Conversely, the Model 2 $l_e$ estimates systematically prefer larger values (although the posteriors are broad and span nearly all possible values) than $l_e$ inferred from the light curves. One system, HD 102195, yields a significantly larger $l_e$ from Model 2 ($l_{e,\mathrm{Model \; 2}} = 44.0_{-3.1}^{+6.1}$) than the photometrically inferred value ($l_{e, \mathrm{LC}} = 20.8_{-1.6}^{+1.7}$).

Both the unconstrained inference of $l_p$ and the systematic preference for larger values of $l_e$ can likely be attributed to sparse RV sampling cadence over long observational baselines. \citet{Nicholson2022} observed similar trends when comparing inferred QP kernel parameters between GP models trained on synthetic photometry and RV time series. They found that $l_p$ is often unconstrained even for synthetic data.\footnote{\citet{Nicholson2022} implement a QP kernel with a $\Gamma$ parameter, which equates to 1/$l_p$. For the discussion here, this difference in parameterization has no significant effect as the general trends remain the same.} Similarly, \citet{Nicholson2022} noted that if the RV observational cadence is much lower than the spot evolutionary timescale, then the model is prone to overestimating $l_e$ as it can only place an upper limit on this parameter. This is true for all systems in this study and particularly applicable to HD 102195, which has a large number of observations ($N_\mathrm{RV} = 98$) but over a very wide observational window (approximately 4000 days).

\subsection{Effects of Stellar Activity Models on Parameter Inference and Planet Interpretation} \label{sec:activity_models_param_infer}

\subsubsection{Young Planets Orbiting Young Stars} \label{sec:discuss_young_stars}

In this study, we selected for giant planets orbiting active stars, defined by their photometric variability, to mimic the observing conditions of young, magnetically active systems. This focus is motivated by a desire to robustly detect and characterize young planets around young stars, which serve as important laboratories for testing models of thermal evolution, orbital migration, and atmospheric erosion \citep[e.g.,][]{Owen2017, Gupta2019, Suarez2022, Tran2024}. These mechanisms are expected to operate over timescales spanning Myrs to Gyrs, so characterizing young planets, particularly at ages less than several hundred Myr when these processes have yet to terminate, is critical to determining the evolutionary pathways of planets.

However, distinguishing between competing mechanisms requires robust mass and density determinations, which for young systems can be strongly influenced by the adopted stellar activity mitigation scheme \citep[e.g.,][]{Setiawan2008, Huelamo2008, Donati2016, Yu2017a, Damasso2020, Kossakowski2022}. Even for planets that are independently detected through other methods such as transits, the choice of stellar activity model can impact mass and density inferences. For example, \citet{Suarez2022} measured densities for V1298 Tau b and e that were higher than expected for planets at these ages, suggesting that they contracted much faster than predicted by planet formation theories. However, \citet{Blunt2023} found that mass determinations for V1298 Tau b and e may suffer from overfitting due to the joint Keplerian and GP model, which would bias the resulting planet properties. Similarly, mass estimates for other young planets such as those in the AU Mic system have been shown to differ significantly depending on the choice of stellar activity model \citep{Klein2021, Cale2021, Zicher2022}.

These studies, combined with the results in this work, demonstrate that the adoption and specific choice of a stellar activity model can profoundly impact the interpretation of the underlying planet characteristics. For the 17 active stars harboring giant planets that we analyzed in this study, we find a variation in minimum planetary mass (and, for a transiting planet, density) of approximately 12\% and as high as 67\% depending on the stellar activity mitigation strategy and which specific GP framework is adopted.

Deviations in the physical or orbital parameters of planets around active stars at the levels that we found in this study can have important consequences, particularly for young systems. For instance, for a transiting giant planet, a change in mass (say, at the 30\% level) translates to a comparable impact on density. Depending on the age of the system, an adjustment at this level could significantly alter its inferred thermal contraction timescale \citep[e.g.,][]{Fortney2007, Suarez2022}. Similarly, if a population of young hot Jupiters on near-circular orbits is misinterpreted as having a modest range of eccentricities, this can impact the inferred timescale of tidal circularization, the timescale of tidal decay, and the implied range of tidal quality factors \citep[e.g.,][]{Adams2006, Fabrycky2007, Dawson2018}. To make robust parameter inference and draw correct conclusions about the evolutionary histories of young planets, these model discrepancies should be taken into account when interpreting results for planets around active stars.

\subsubsection{Connection to Low-Mass Planet and Low-Amplitude Activity Signals} \label{sec:discuss_low_mass_planets}

The analysis presented here is an internally consistent comparison between two stellar activity mitigation approaches employing a red-noise GP filter. This work is complementary to the RV fitting challenges conducted by \citet{Dumusque2017} and \citet{Zhao2022}, which compared the results of different teams that attempted to mitigate stellar activity signals and detect planets. These studies were concerned with low-mass planets and low-amplitude activity signals (both on the order of a few m s$^{-1}$) while in this work we are primarily interested in the large amplitude signals of giant planets around young, active stars (on the order of several hundreds of m s$^{-1}$). However, for both regimes there is a similarity in the ratio of the amplitude of the Keplerian signal and the scale of the stellar variability. Comparing these parallel but distinct cases offers insight into the benefits and drawbacks of stellar activity mitigation strategies.

In the RV data challenges, each of the teams is supplied with identical synthetic and real datasets and tasked with modeling astrophysical noise. For the six simulated planets studied in \citet{Dumusque2017}, the teams that employed a GP model inferred $K$ values that differed on average by 36\% and in one instance had a percentage difference greater than 80\%. Similarly, for the four systems examined in \citet[][HD 101501, HD 26965, HD 10700, and HD 34411]{Zhao2022}, the approaches of teams that employed a GP framework had residual RMS values with percentage differences ranging from 2--98\%. This variance is large enough to have consequences for the recovery of a low-mass planet or its implied physical properties, as discussed in \citet{Zhao2022}. This tendency for planet parameters to depend on the details of the GP model is similar to our results, despite our focus on a different regime of activity and planet mass.

A consistent trend emerges when considering our results together with the findings from these data challenges. Across a wide range of astrophysical noise levels and planetary masses, the choice of GP model to correct for stellar activity can yield significantly different inferred planetary properties. For both giant planets around active stars and low-mass planets around inactive stars, these deviations associated with the modeling framework are comparable to the parameter uncertainties.

\section{Summary and Conclusions} \label{sec:summary}

We have conducted a homogeneous re-analysis of radial velocity datasets for 17 photometrically active stars hosting giant planets and assessed the effects of two red-noise stellar activity GP models on inferred planetary properties compared to a fiducial Keplerian orbit fit with a white-noise only model. We summarize our primary conclusions below.

\begin{enumerate}[topsep=1ex,itemsep=1ex,partopsep=0ex,parsep=0ex]
    \item Constraints on physical and orbital planetary properties can vary dramatically: by as much as 67\% in minimum planet mass and 95\% in eccentricity between different implementations of stellar activity models. These differences are greater than the median measurement uncertainties for these parameters (7\% and 69\%, respectively).
    \item The scale of variations in inferred planetary parameters associated with the choice of stellar activity mitigation framework can impact the interpretation of planets around young stars. These young systems provide important constraints on evolutionary mechanisms impacting young planets such as planet densities and thermal evolution, circularization timescales of young hot Jupiters, and tidal quality factors, so properly characterizing signals in RV datasets requires correctly assessing all forms of uncertainties.
    \item As part of this analysis, we compare the inferred QP kernel parameters from a GP model trained on photometry with parameters determined from RVs alone. We find that high-cadence RV sampling is necessary to robustly measure the stellar rotation period with a GP model trained only on RV data. For one of our systems where this constraint was possible (GJ 3021), we found that the inferred stellar rotation period is in reasonably good agreement with the stellar rotation period derived from a GP model fit to the photometry given the gap in observations. 
    \item This study is complementary to the data challenges carried out by \citet{Dumusque2017} and \citet{Zhao2022}. In all instances, recovered planet parameters fluctuate greatly as a function of the adopted activity mitigation scheme. These results suggest that employing GPs can impact the inference of planetary properties across a range of planetary and stellar activity amplitudes. Given these observed variations, we recommend comparing results from multiple stellar activity mitigation models to assess if individual systems and datasets are robust against different model choices.
\end{enumerate}

As stellar activity signals increasingly become the dominant obstacle to precise and robust mass determinations of young planets, employing activity mitigation techniques, such as GPs, will be critical. The results of this study indicate that a large number of RVs over small timescales---ideally nightly cadence within the window of several stellar rotation cycles---will best help constrain GP frameworks designed to model starspot-driven modulations. For instances where ancillary photometry is leveraged to supplement RVs, near-simultaneous observations are especially beneficial.

A natural extension of this study would be a similar comparative analysis for other GP stellar activity models and time series observations. Of particular interest are how multidimensional GP frameworks, which leverage ancillary spectroscopic time series to constrain stellar activity signals, affect inferred planetary properties. This comparison would require selecting targets with dense RV sampling over short time periods, as the multidimensional GP framework requires relatively high observational cadence to be effective. This would also open up the inclusion of advanced model comparison tests such as cross validation algorithms, which assesses the predictive power of each model, as well as experiments with different choices of GP kernels and GP frameworks.

\section{Acknowledgements}

The authors are grateful to the referee for their helpful comments, which improved the quality of this manuscript.

Q.H.T. and B.P.B. acknowledge the support from a NASA FINESST grant (80NSSC20K1554). B.P.B. acknowledges support from the National Science Foundation grant AST-1909209, NASA Exoplanet Research Program grant 20-XRP20$\_$2-0119, and the Alfred P. Sloan Foundation.

This paper includes data collected by the Kepler mission and obtained from the MAST data archive at the Space Telescope Science Institute (STScI). Funding for the Kepler mission is provided by the NASA Science Mission Directorate. STScI is operated by the Association of Universities for Research in Astronomy, Inc., under NASA contract NAS 5–26555. This paper includes data collected by the TESS mission. Funding for the TESS mission is provided by the NASA's Science Mission Directorate.

This research has made use of the NASA Exoplanet Archive, which is operated by the California Institute of Technology, under contract with the National Aeronautics and Space Administration under the Exoplanet Exploration Program.

\facilities{Exoplanet Archive, Kepler, K2, TESS.}

\software{\texttt{emcee} \citep{Foreman-Mackey2013, Foreman-Mackey2019},
\texttt{lightkurve} \citep{Lightkurve2018},
    \texttt{matplotlib} \citep{Hunter4160265},
    \texttt{numpy} \citep{vanderWalt2011},
    \texttt{pyaenti} \citep{Barragan2019a, Barragan2022},
    \texttt{scipy} \citep{Virtanen2020}.
}

\newpage

\appendix
\section{Results of Model Fits}
\label{sec:app_results}

\begin{figure*}
    \centering
    \hspace{0 mm}
    \includegraphics[width=0.8\linewidth]{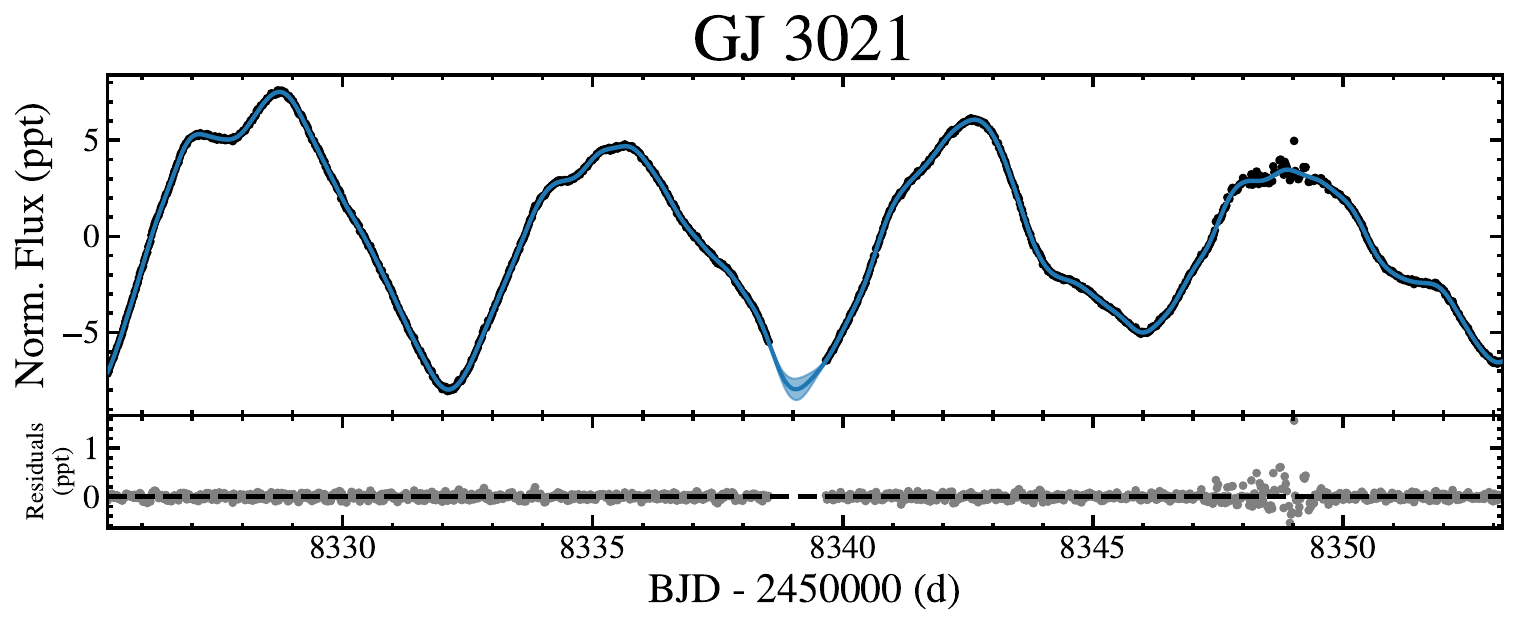}%

    \hspace{-3.25mm}
    \includegraphics[width=0.8175\linewidth]{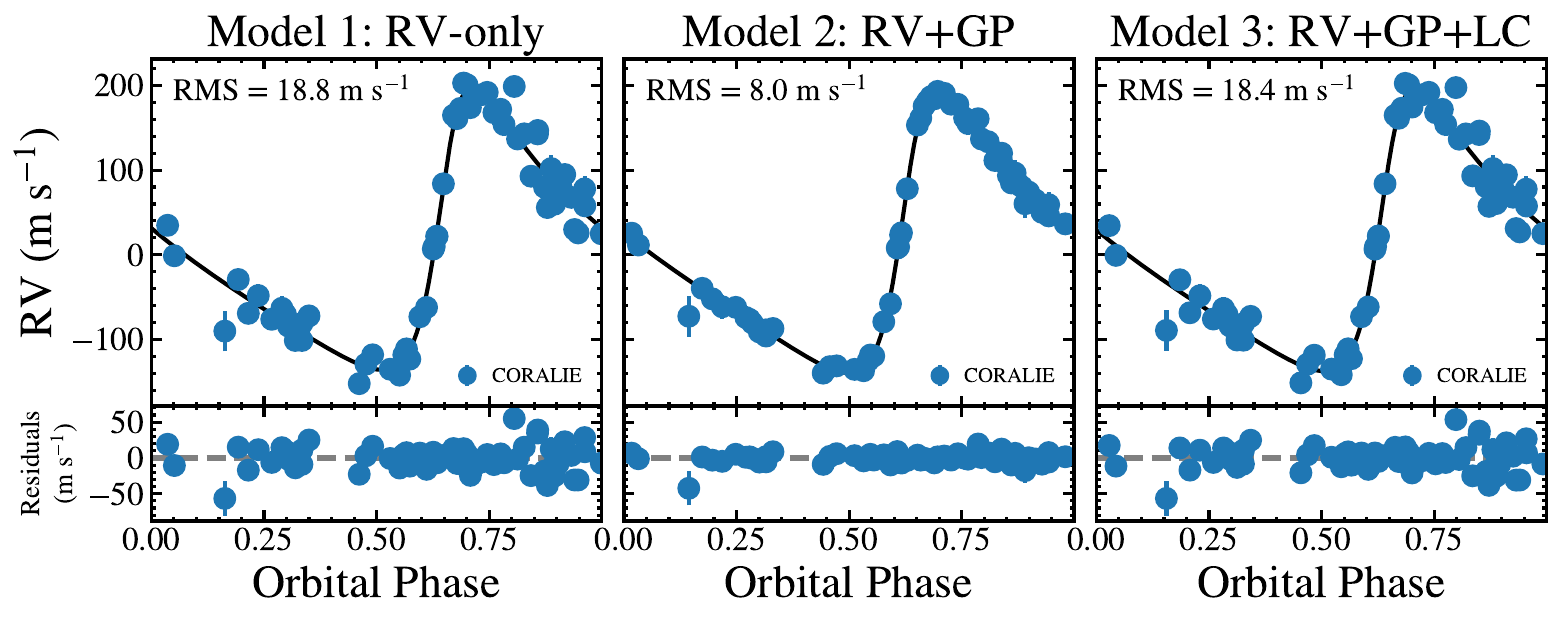}%

    \hspace{6.5mm}
    \includegraphics[width=0.785\linewidth]{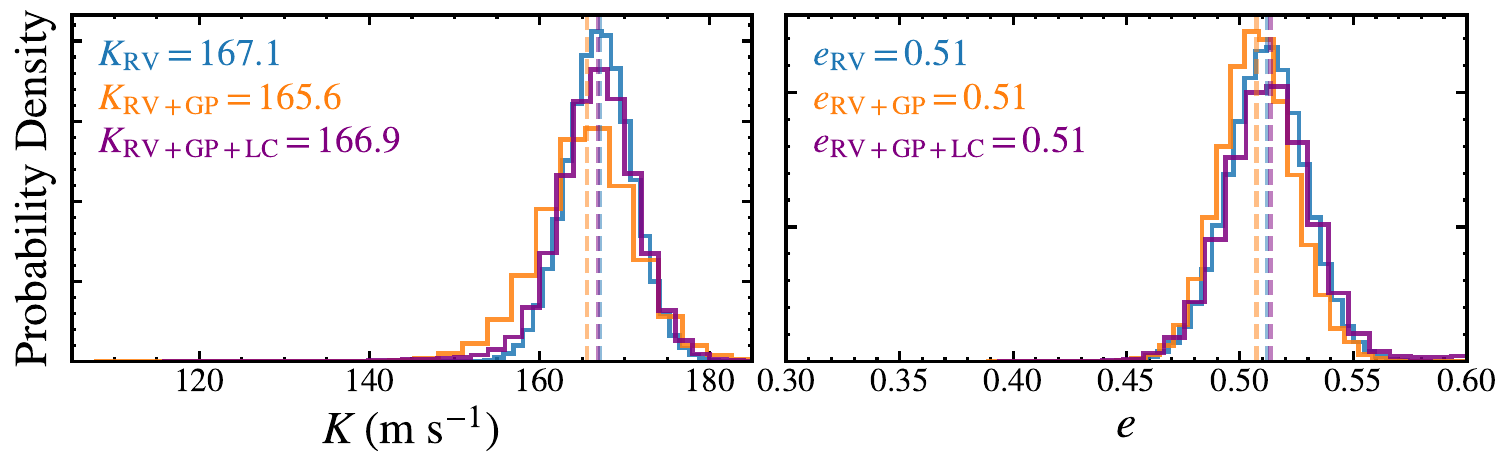}%

    \hspace{-2.5mm}
    \includegraphics[width=0.815\linewidth]{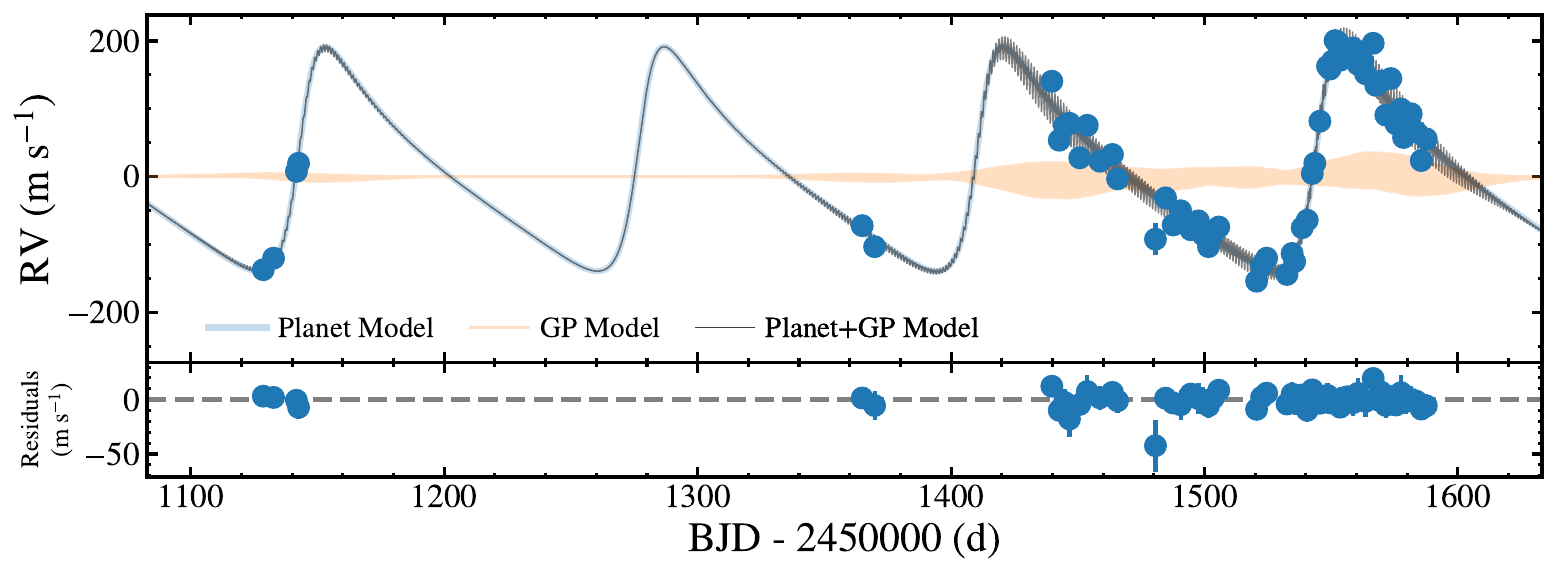}%
    
    \caption{The results of each model fit to the TESS Sector 1 photometry and CORALIE RVs from \citet{Naef2001} for GJ 3021. The top panel displays the best-fit GP model and variance (blue) and residuals. The phased RV curve and model residuals, with the stellar activity GP model removed, for (from left to right) Model 1, Model 2, and Model 3 are displayed in the second panel. The RV semi-amplitude and eccentricity distributions for each of those frameworks are presented in the third panel in blue, orange, and purple respectively. In the bottom panel, the best-fit Model 2 and residuals are shown, with the planetary component in blue, the GP component in orange, and the total model in black.}
    \label{fig:GJ_3021_results}
\end{figure*}

\begin{figure*}
    \centering
    \hspace{0 mm}
    \includegraphics[width=0.8\linewidth]{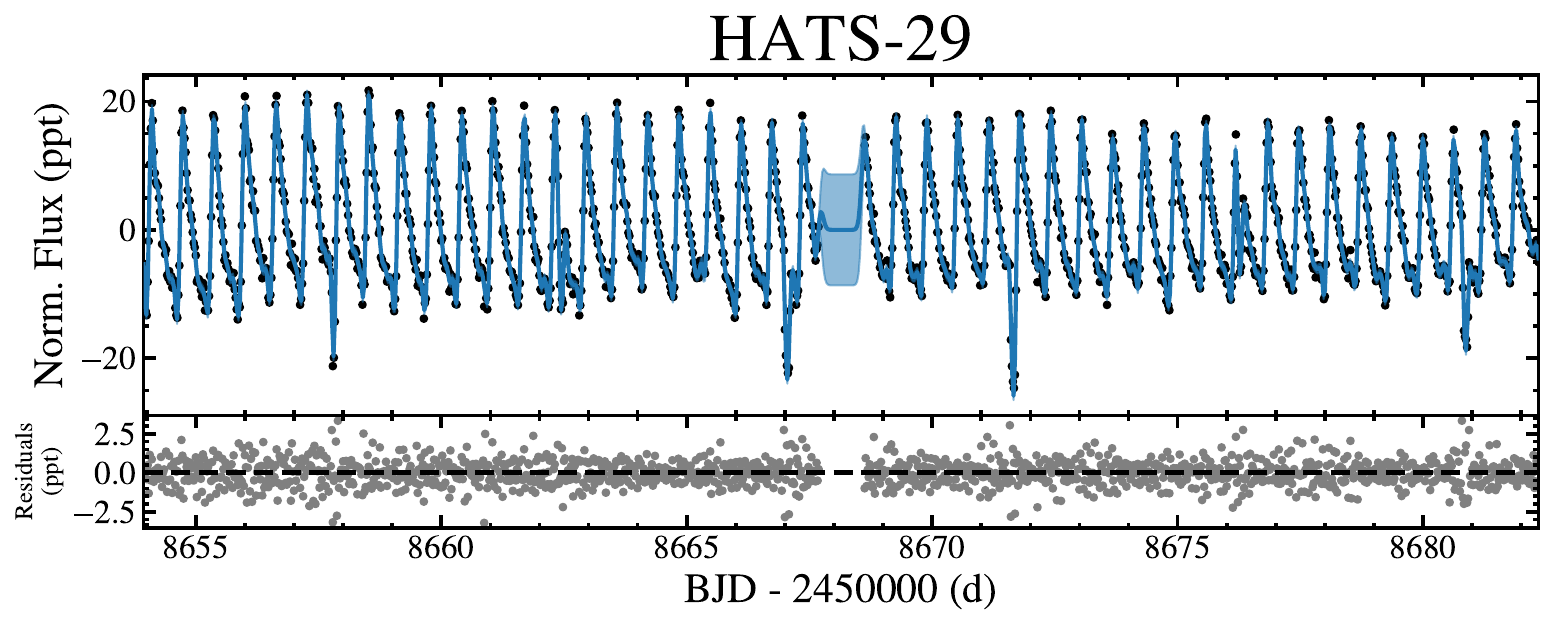}%

    \hspace{-3.25mm}
    \includegraphics[width=0.8175\linewidth]{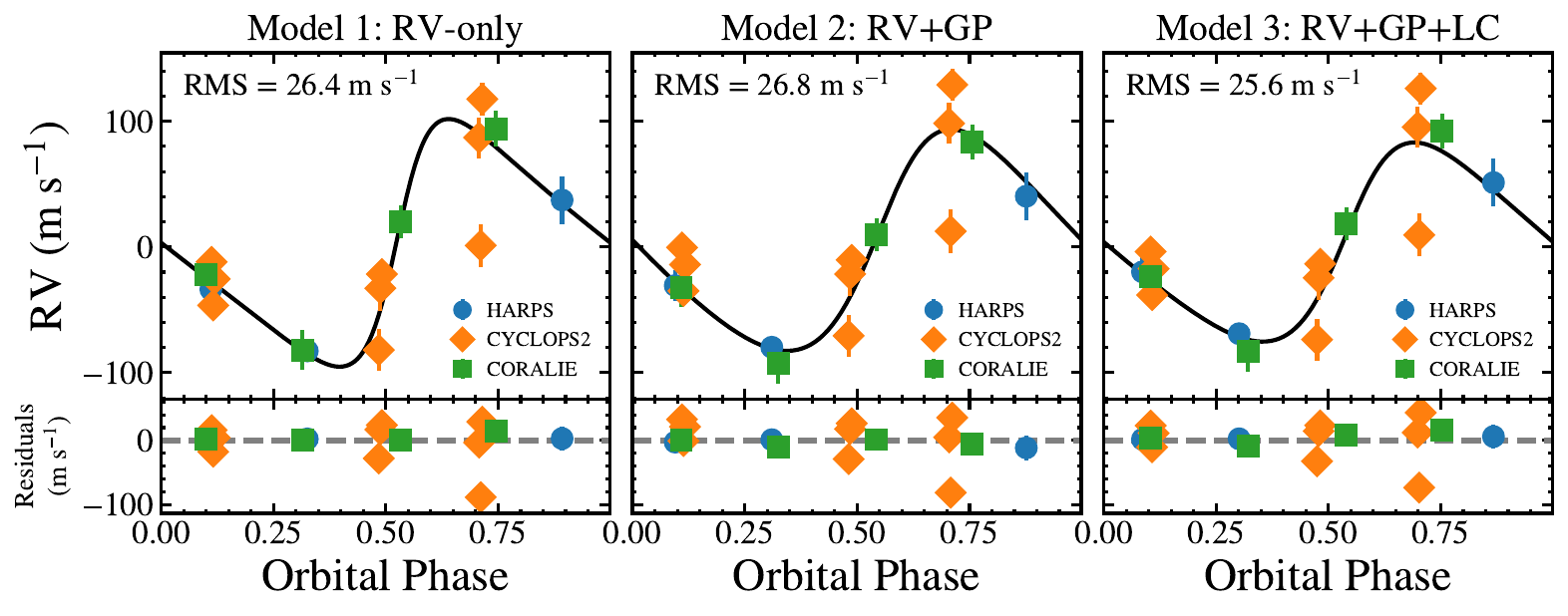}%

    \hspace{6.5mm}
    \includegraphics[width=0.785\linewidth]{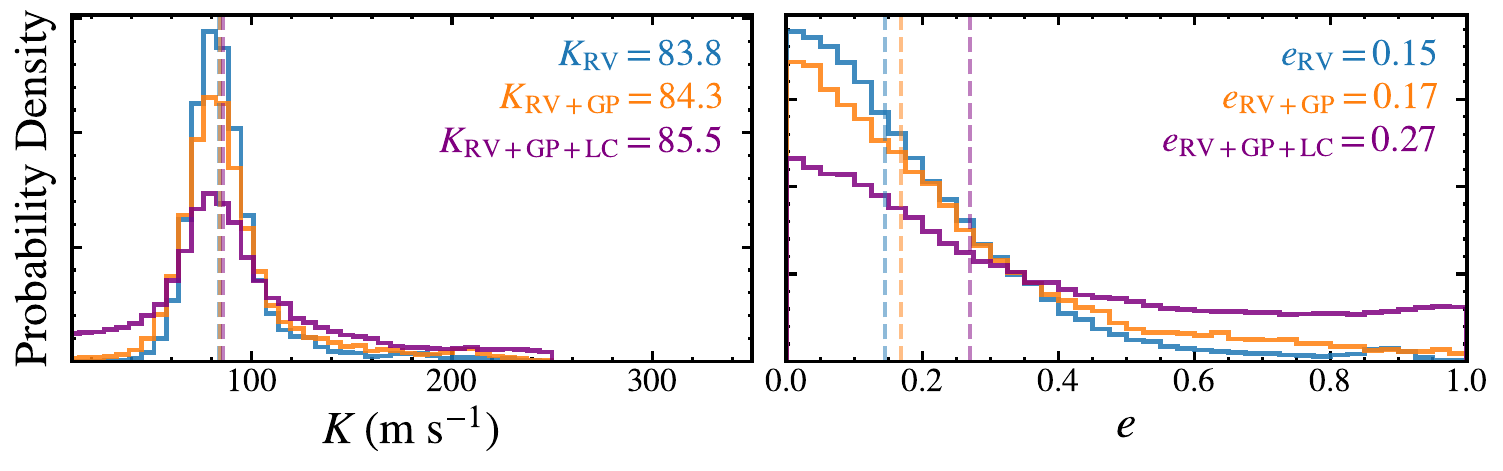}%

    \hspace{-2.5mm}
    \includegraphics[width=0.815\linewidth]{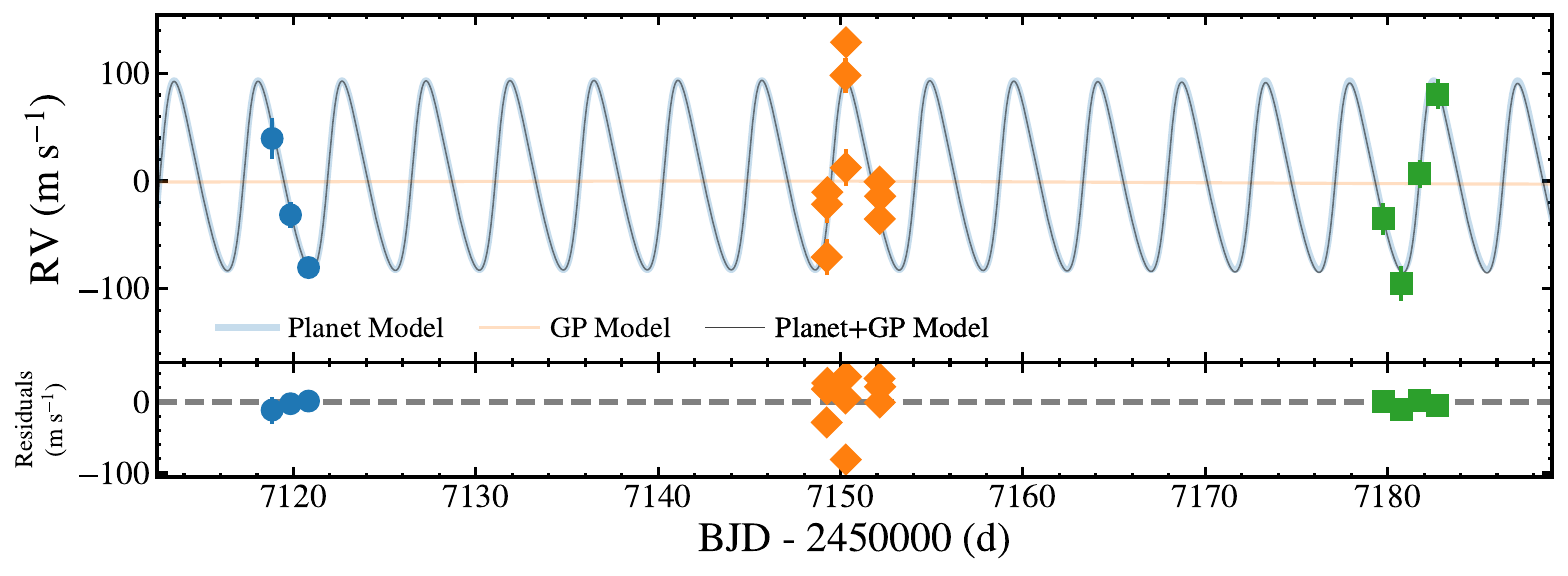}%
    
    \caption{The results of each model fit to the TESS Sector 13 photometry and HARPS, CYCLOPS2 (UCLES), and CORALIE RVs from \citet{Espinoza2016}} for HATS-29. The panels are the same as in \autoref{fig:GJ_3021_results}.
    \label{fig:HATS-29_results}
\end{figure*}

\begin{figure*}
    \centering
    \hspace{0 mm}
    \includegraphics[width=0.8\linewidth]{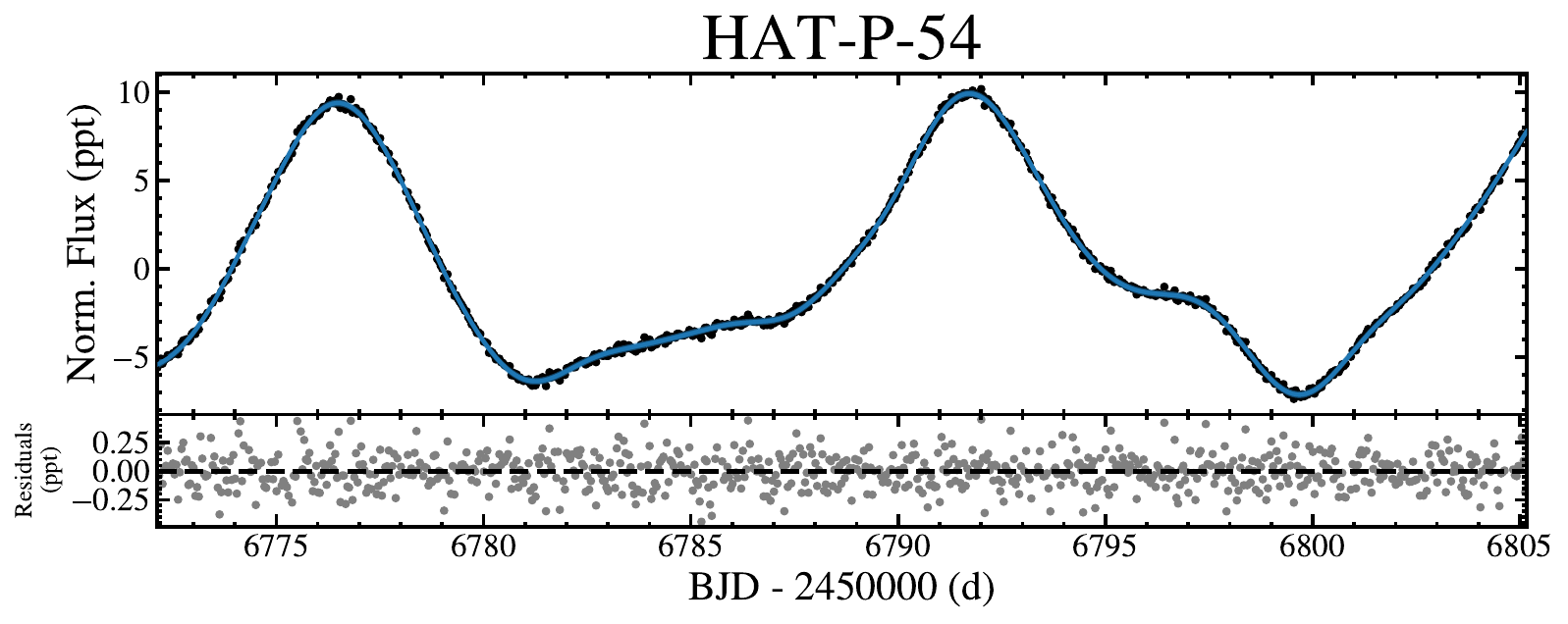}%

    \hspace{-3.25mm}
    \includegraphics[width=0.8175\linewidth]{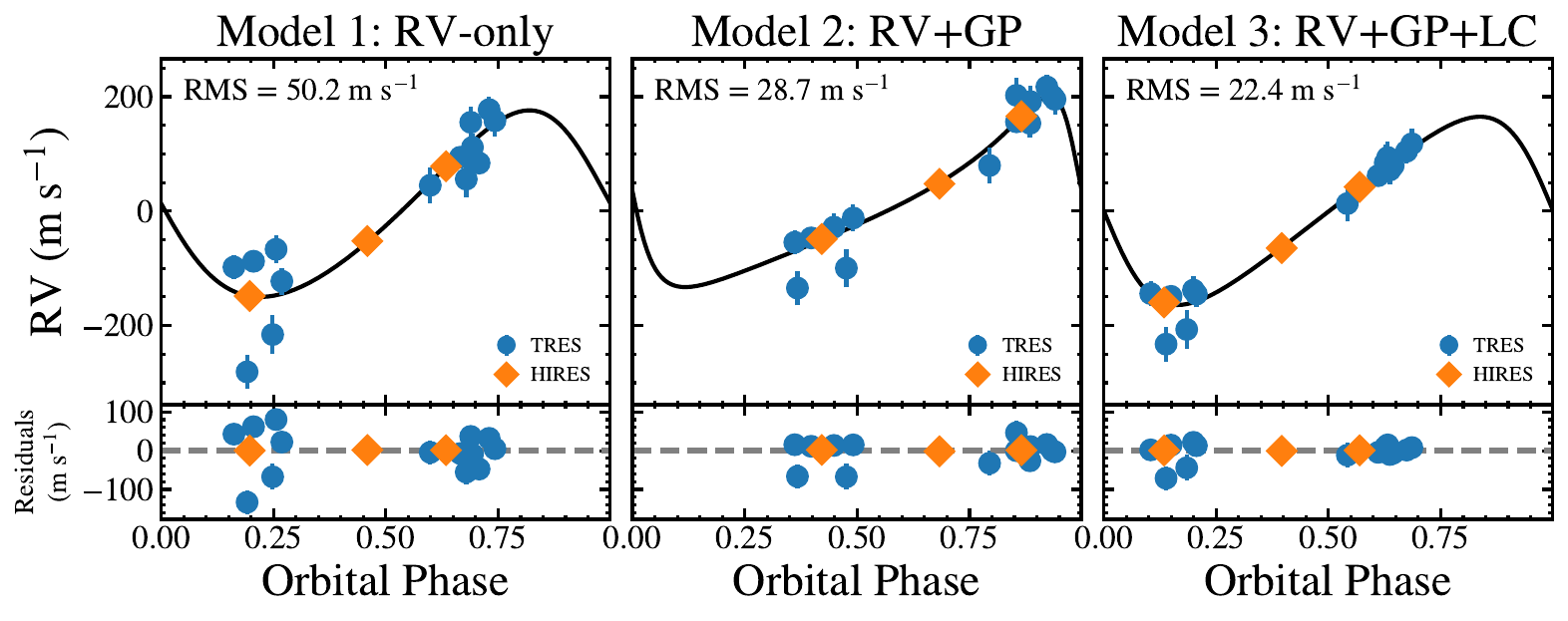}%

    \hspace{6.5mm}
    \includegraphics[width=0.785\linewidth]{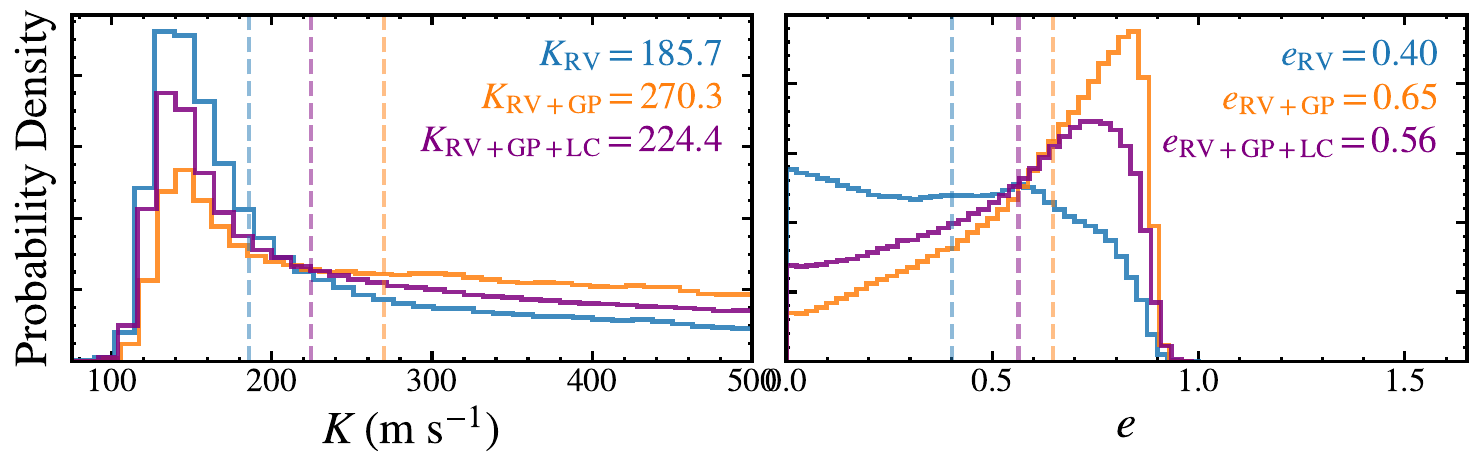}%

    \hspace{-2.5mm}
    \includegraphics[width=0.815\linewidth]{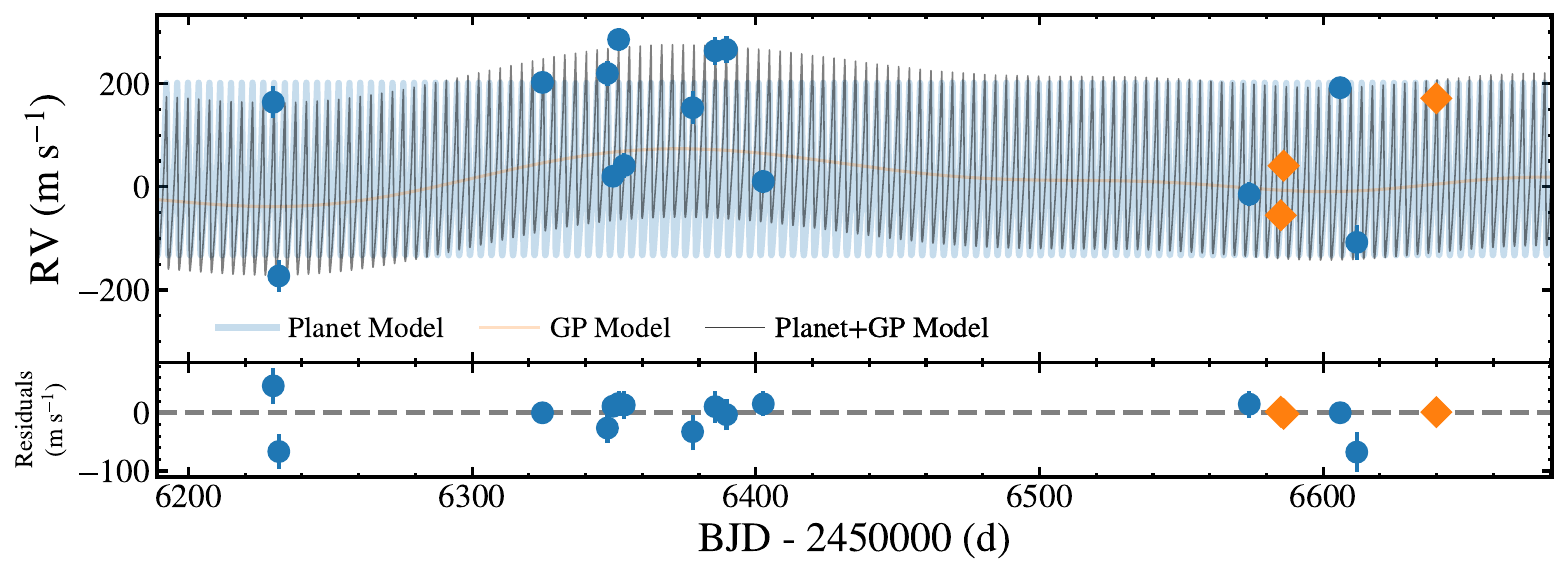}%
    
    \caption{The results of each model fit to the K2 Campaign 0 photometry and TRES and HIRES RVs from \citet{Bakos2015} for HAT-P-54. The panels are the same as in \autoref{fig:GJ_3021_results}.}
    \label{fig:HAT-P-54_results}
\end{figure*}

\begin{figure*}
    \centering
    \hspace{0 mm}
    \includegraphics[width=0.8\linewidth]{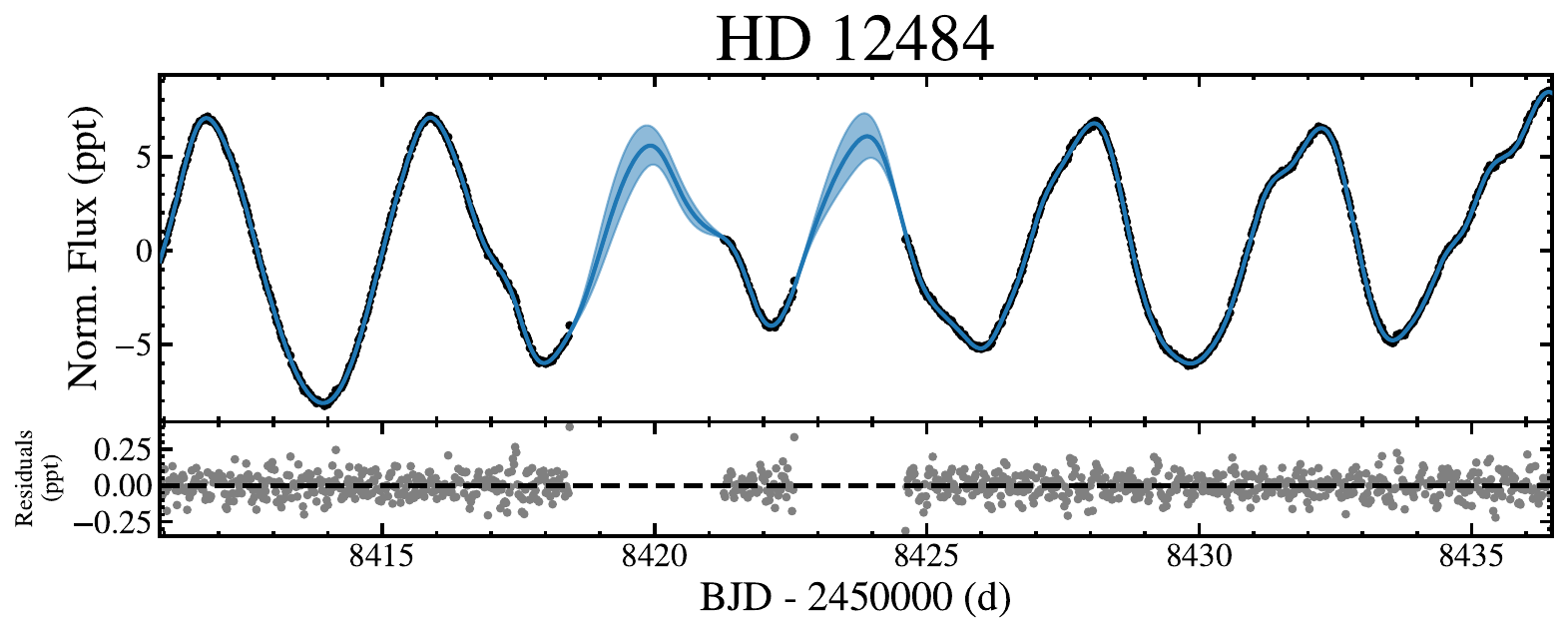}%

    \hspace{-3.25mm}
    \includegraphics[width=0.8175\linewidth]{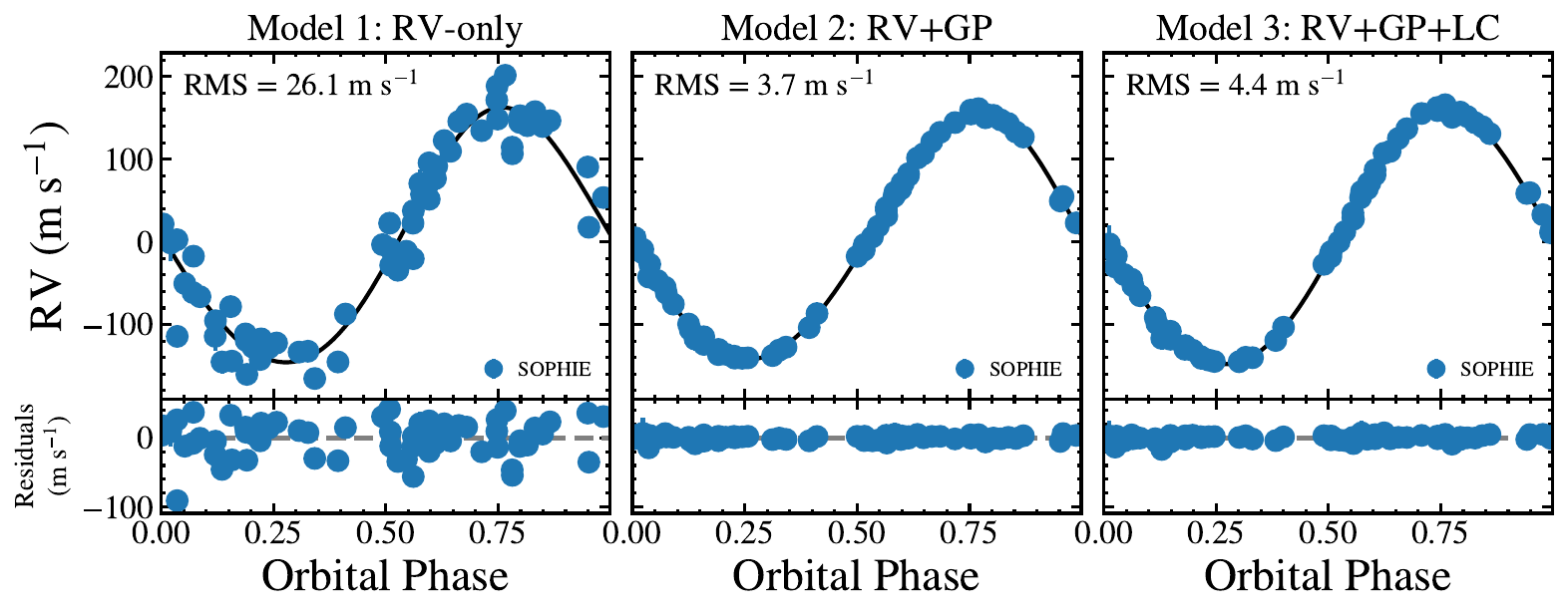}%

    \hspace{6.5mm}
    \includegraphics[width=0.785\linewidth]{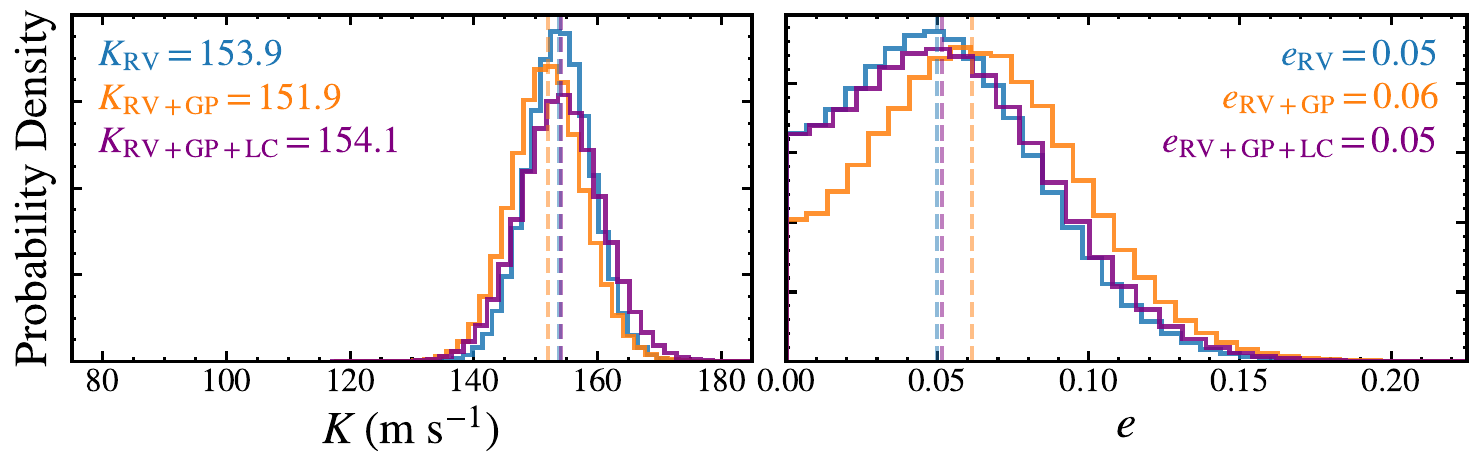}%

    \hspace{-2.5mm}
    \includegraphics[width=0.815\linewidth]{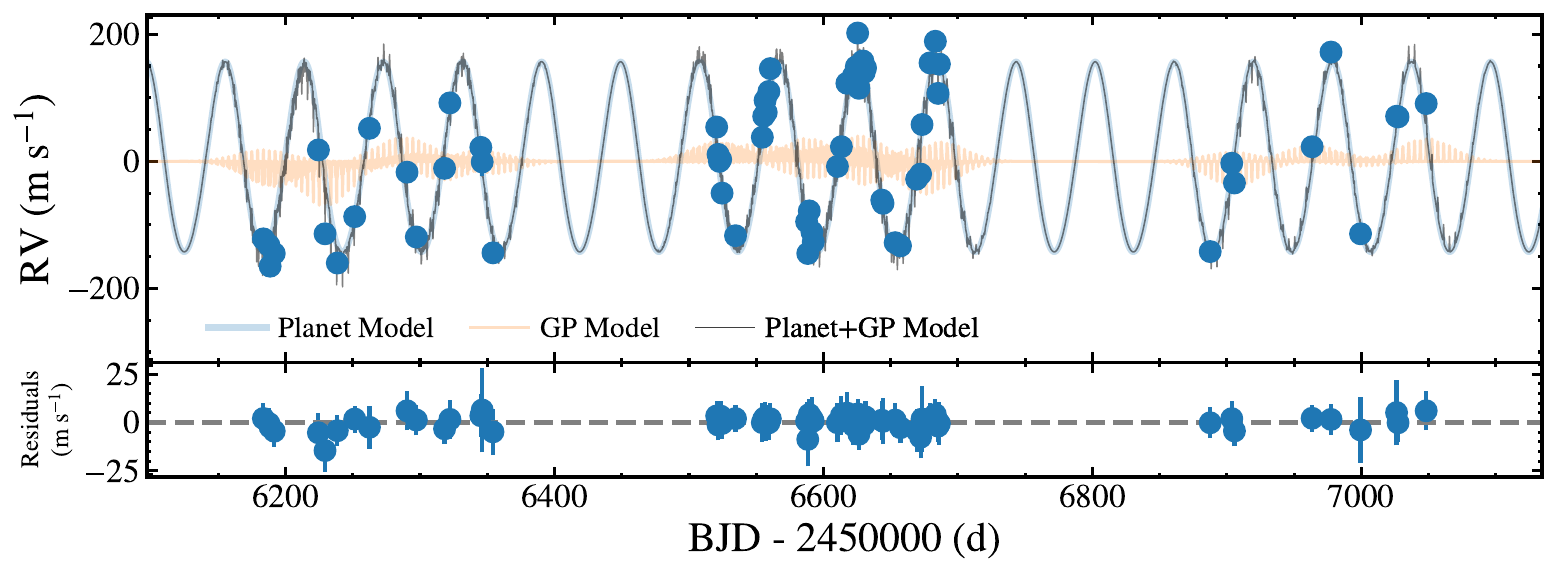}%
    
    \caption{The results of each model fit to the TESS Sector 4 photometry and SOPHIE RVs from \citet{Hebrard2016} for HD 12484. The panels are the same as in \autoref{fig:GJ_3021_results}.}
    \label{fig:HD_12484_results}
\end{figure*}

\begin{figure*}
    \centering
    \hspace{0 mm}
    \includegraphics[width=0.8\linewidth]{Figures/HD_102195_lc_gp.pdf}%

    \hspace{-3.25mm}
    \includegraphics[width=0.8175\linewidth]{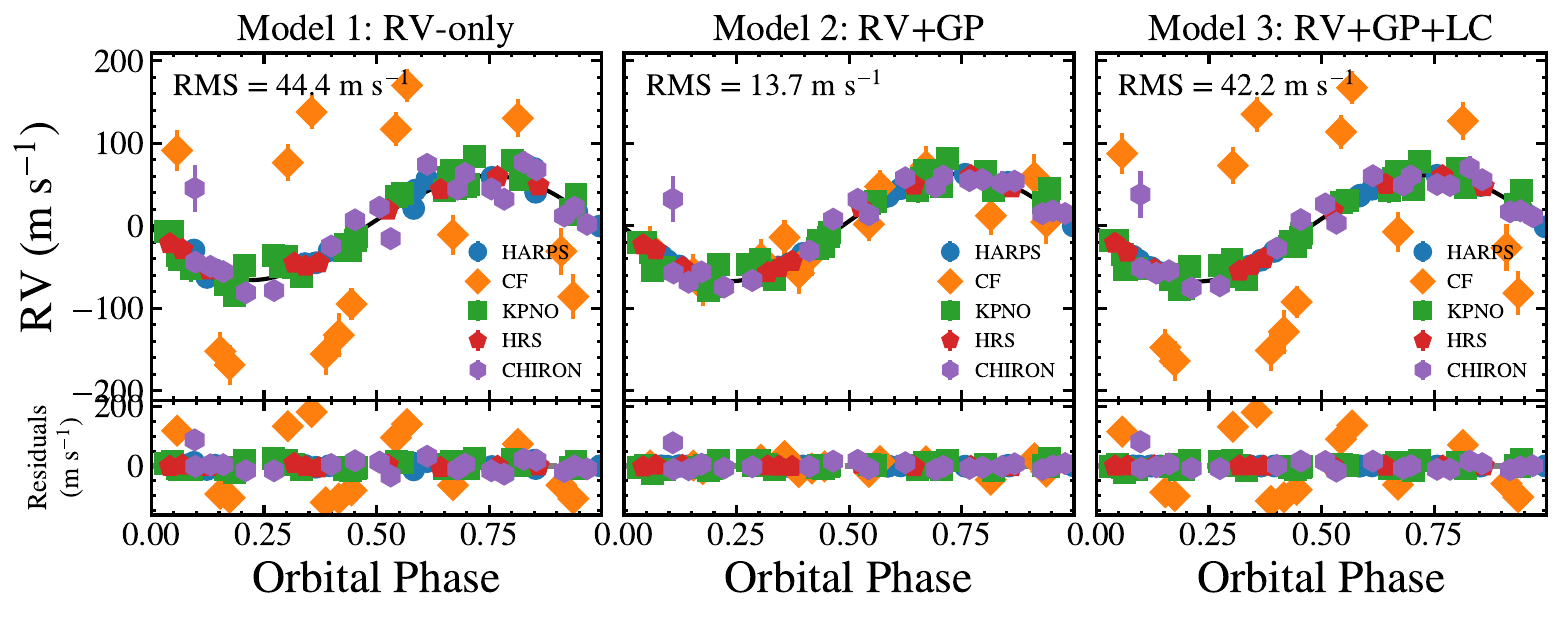}%

    \hspace{6.5mm}
    \includegraphics[width=0.785\linewidth]{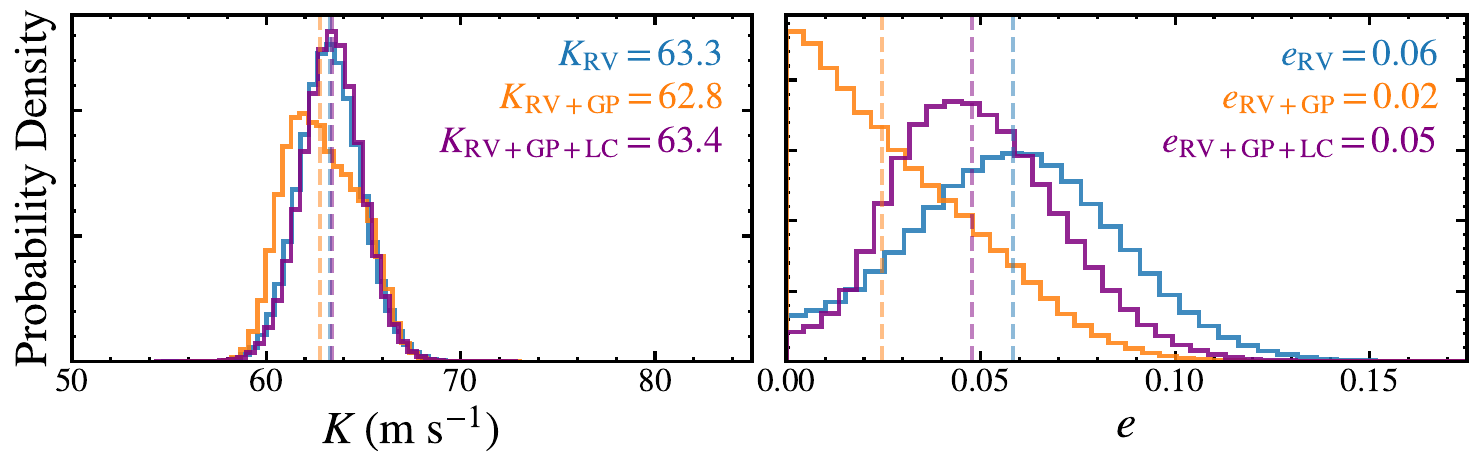}%

    \hspace{-2.5mm}
    \includegraphics[width=0.815\linewidth]{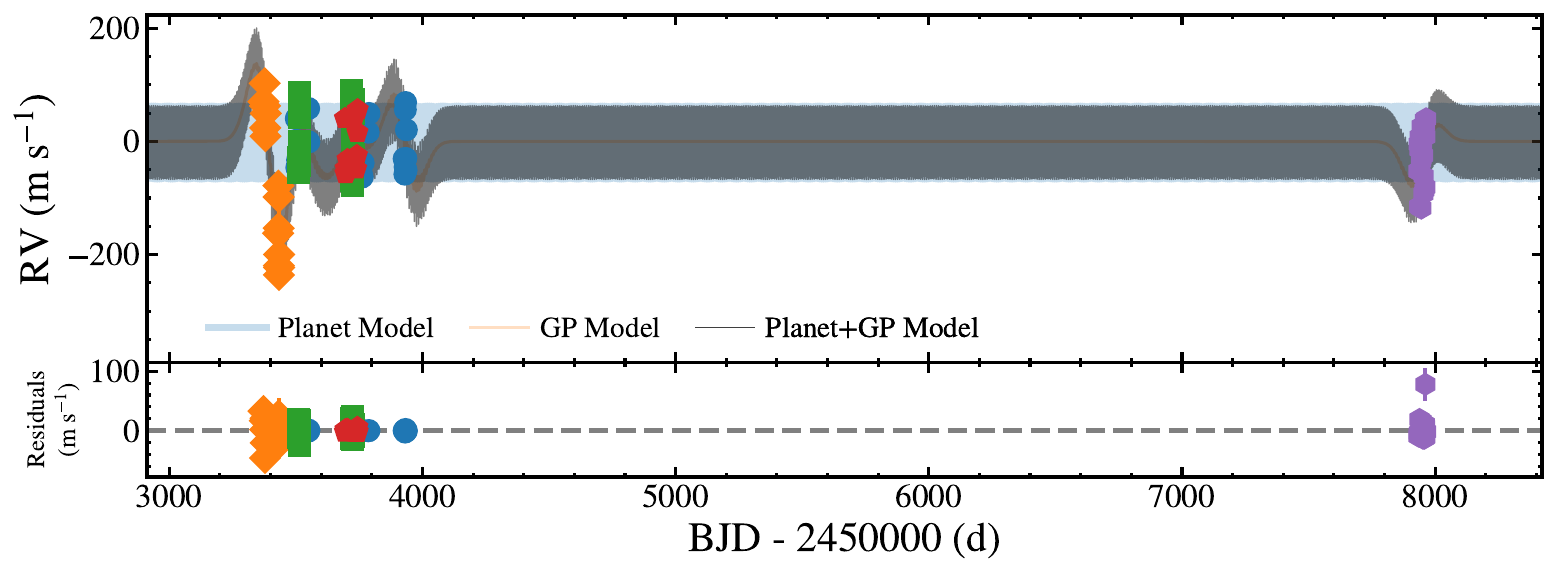}%
    
    \caption{The results of each model fit to the K2 Campaign 1 photometry and HARPS, ET on the 0.9-m telescope (CF) and 2.1-m telescope (KPNO), HRS (HET), and CHIRON RVs from \citet{Ge2006}, \citet{Melo2007}, and \citet{Paredes2021} for HD 102195. The panels are the same as in \autoref{fig:GJ_3021_results}.}
    \label{fig:HD_102195_results}
\end{figure*}

\begin{figure*}
    \centering
    \hspace{0 mm}
    \includegraphics[width=0.8\linewidth]{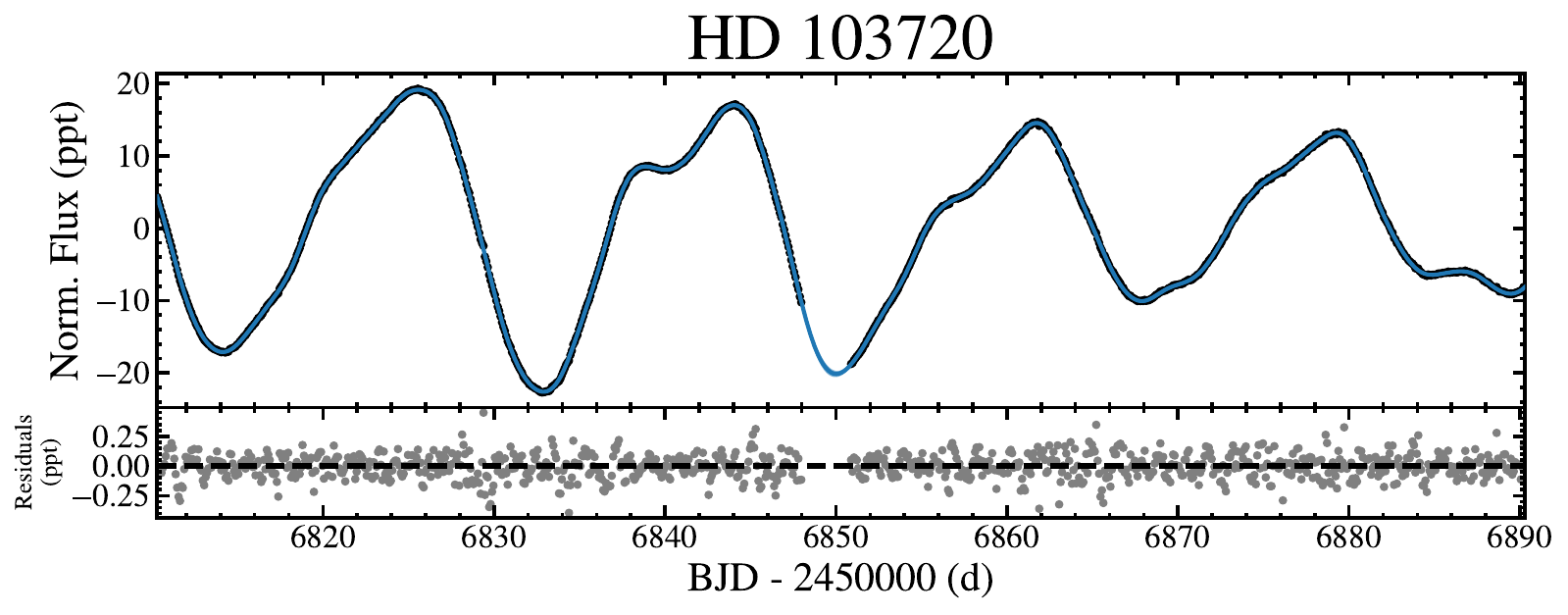}%

    \hspace{-3.25mm}
    \includegraphics[width=0.8175\linewidth]{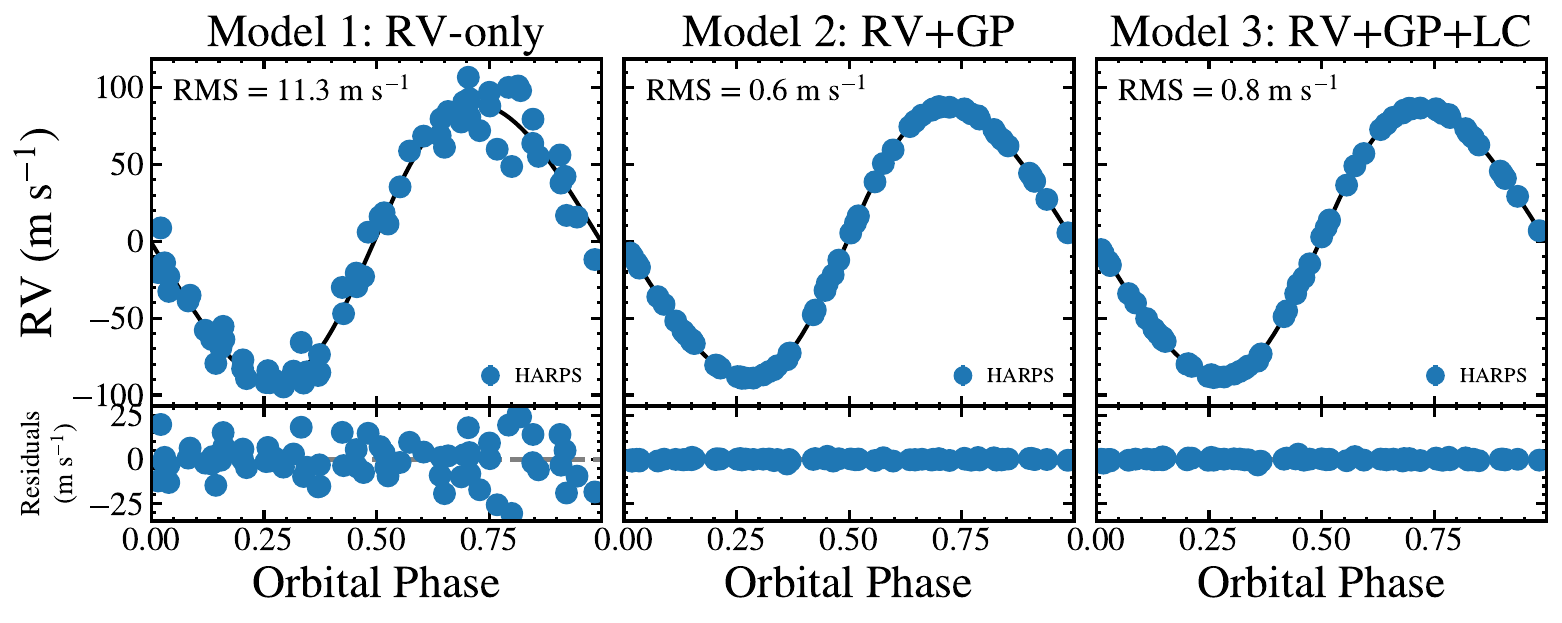}%

    \hspace{6.5mm}
    \includegraphics[width=0.785\linewidth]{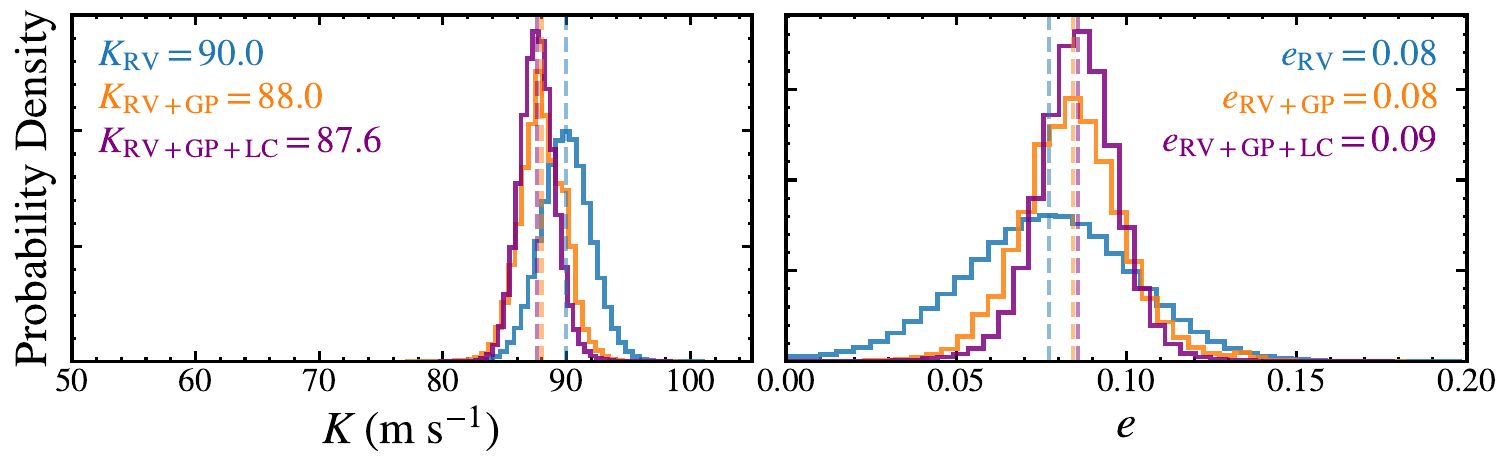}%

    \hspace{-2.5mm}
    \includegraphics[width=0.815\linewidth]{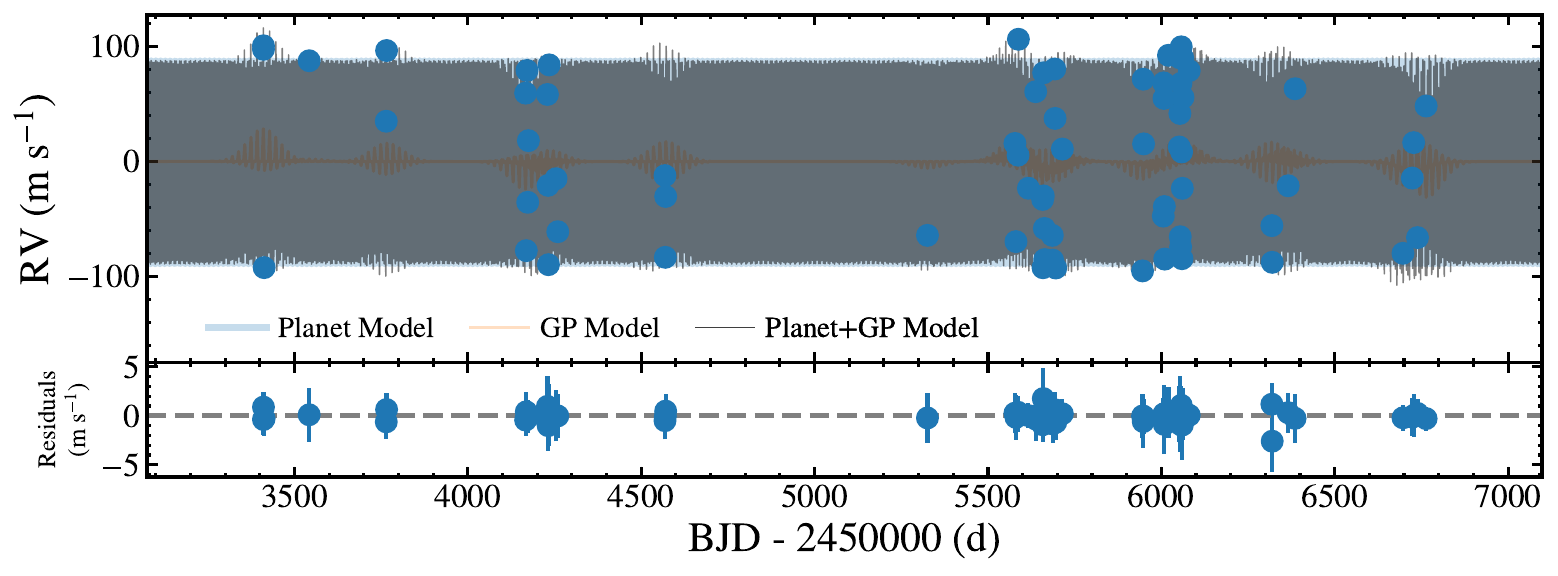}%
    
    \caption{The results of each model fit to the K2 Campaign 1 photometry and HARPS RVs from \citet{Moutou2015} for HD 103720. The panels are the same as in \autoref{fig:GJ_3021_results}.}
    \label{fig:HD_103720_results}
\end{figure*}

\begin{figure*}
    \centering
    \hspace{0 mm}
    \includegraphics[width=0.8\linewidth]{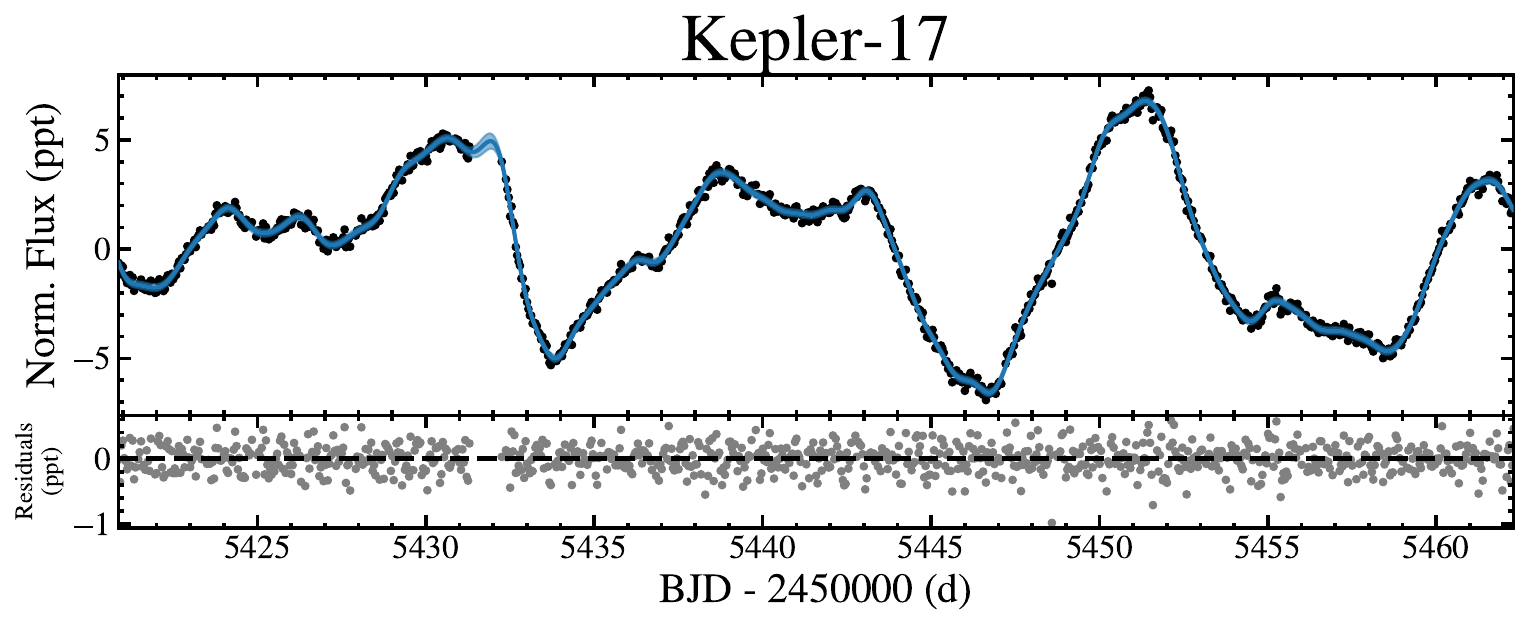}%

    \hspace{-3.25mm}
    \includegraphics[width=0.8175\linewidth]{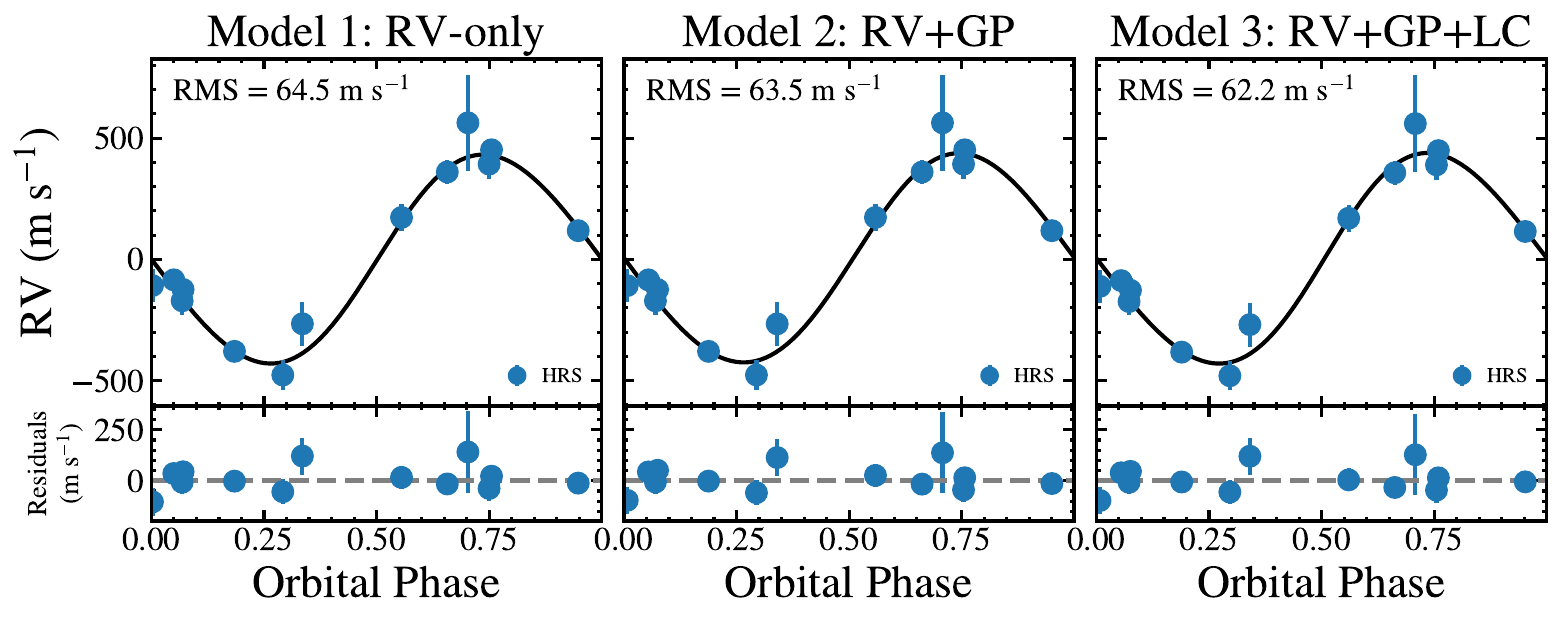}%

    \hspace{6.5mm}
    \includegraphics[width=0.785\linewidth]{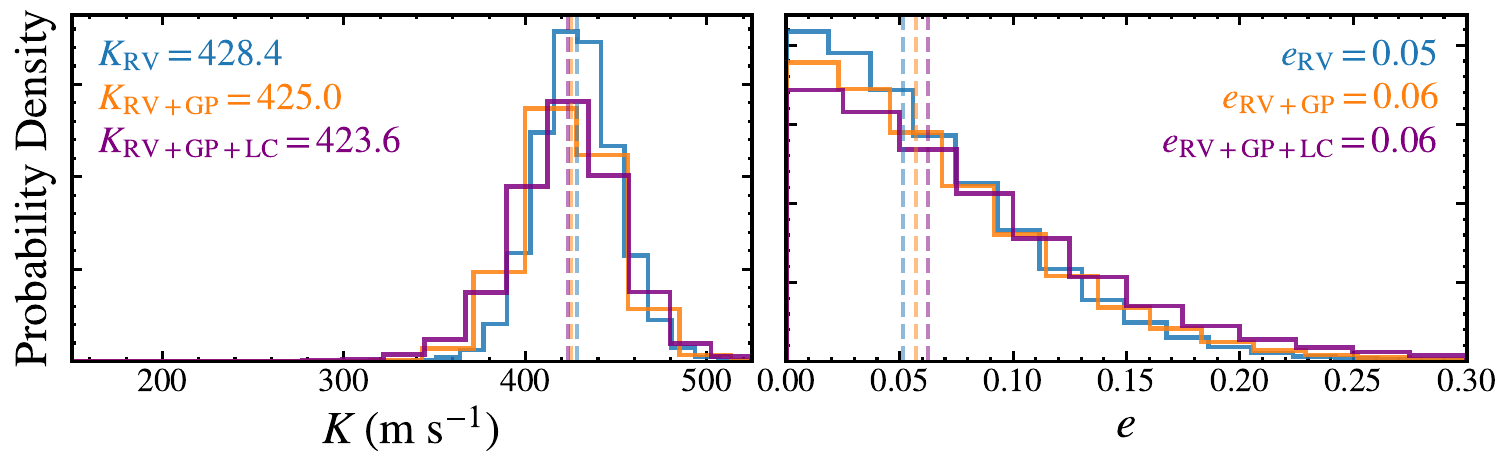}%

    \hspace{-2.5mm}
    \includegraphics[width=0.815\linewidth]{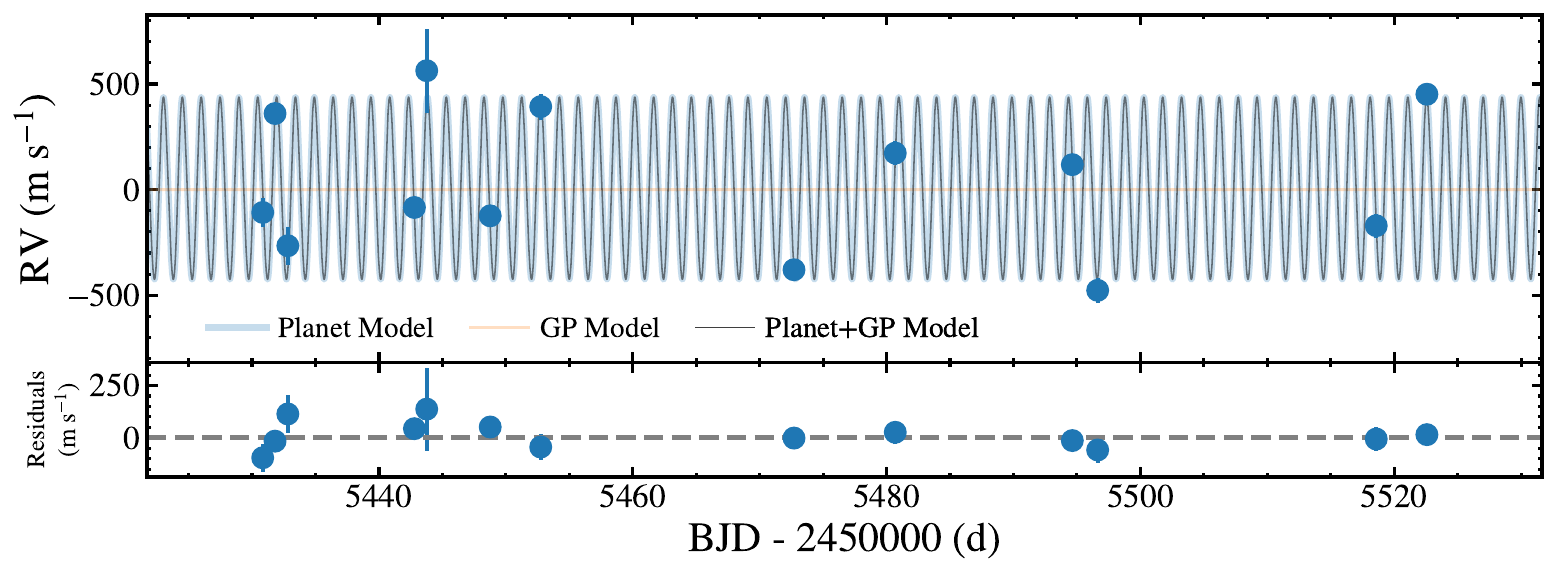}%
    
    \caption{The results of each model fit to the Kepler Quarter 6 photometry and HRS RVs from \citet{Desert2011} for Kepler-17. The panels are the same as in \autoref{fig:GJ_3021_results}.}
    \label{fig:Kepler-17_results}
\end{figure*}

\begin{figure*}
    \centering
    \hspace{0 mm}
    \includegraphics[width=0.8\linewidth]{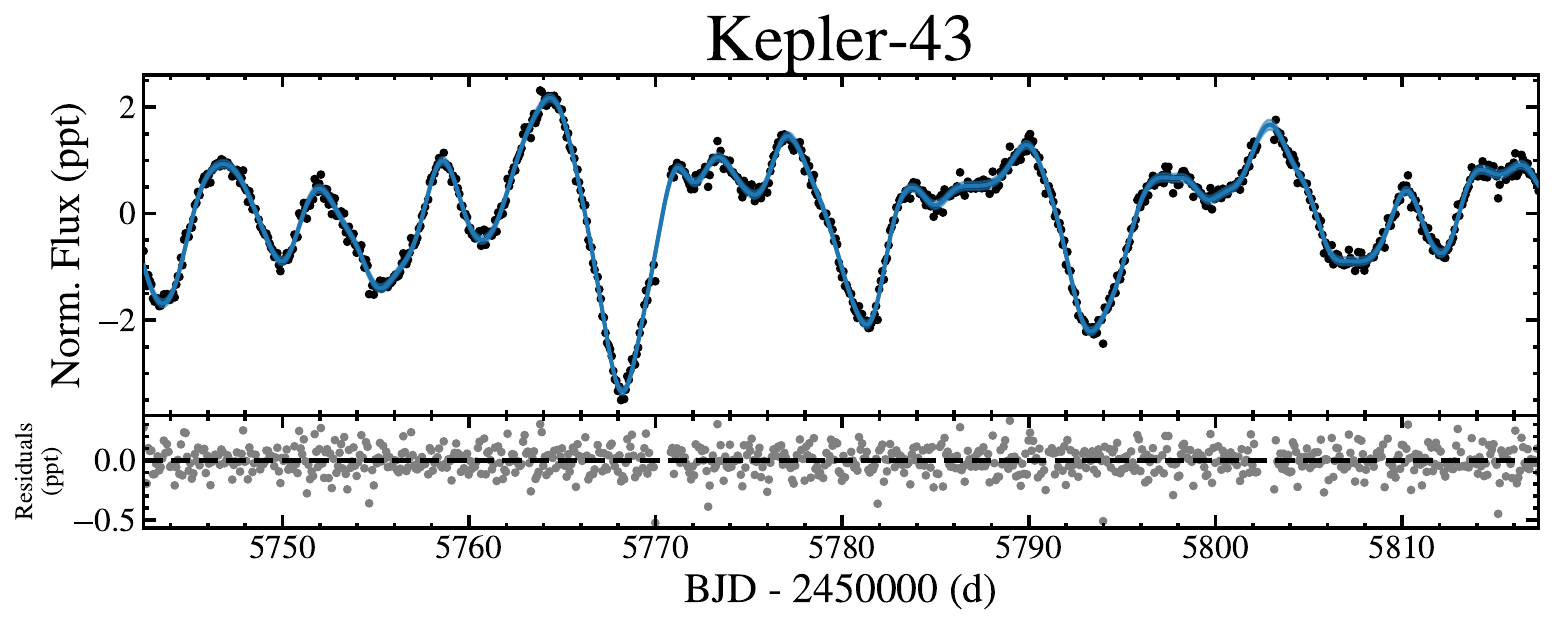}%

    \hspace{-3.25mm}
    \includegraphics[width=0.8175\linewidth]{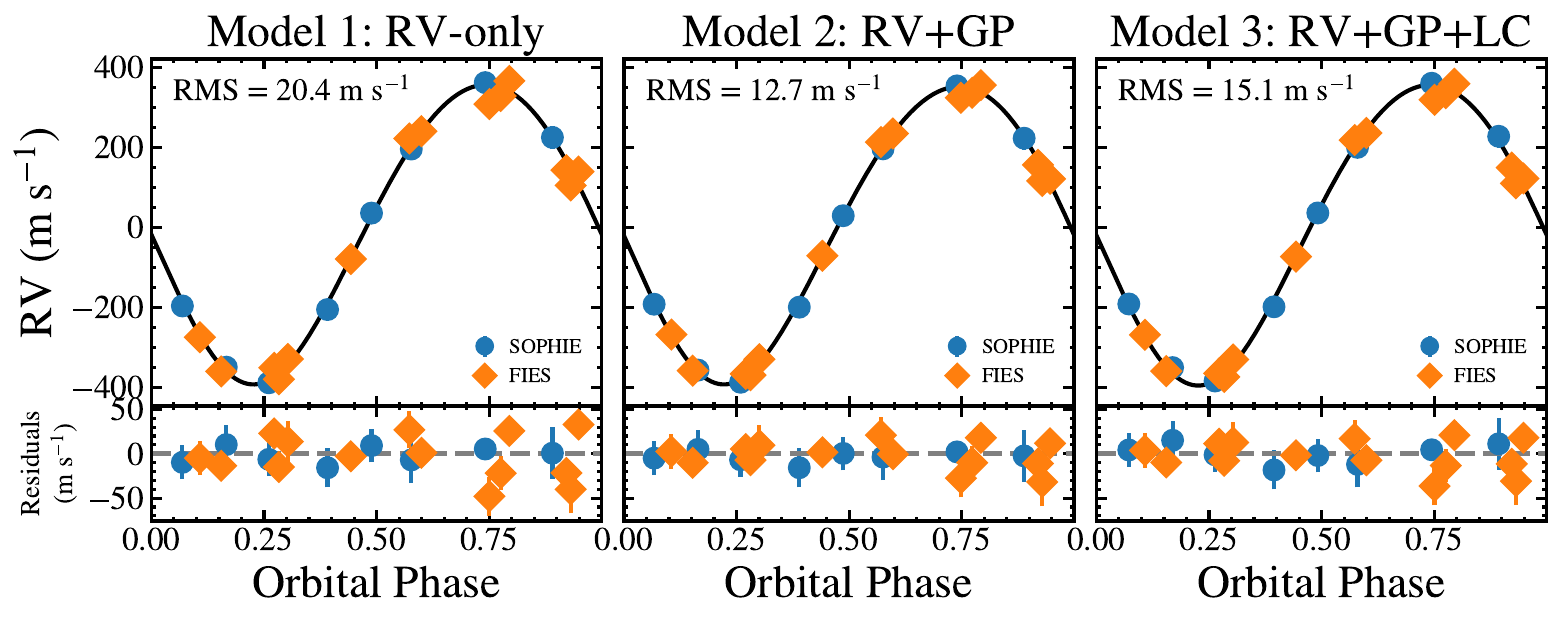}%

    \hspace{6.5mm}
    \includegraphics[width=0.785\linewidth]{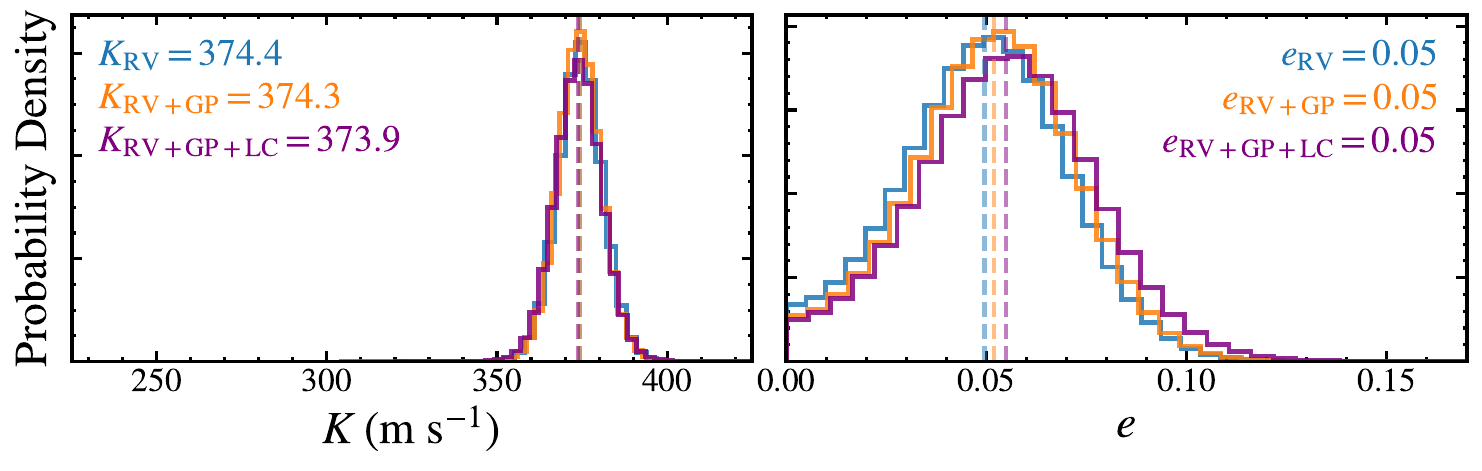}%

    \hspace{-2.5mm}
    \includegraphics[width=0.815\linewidth]{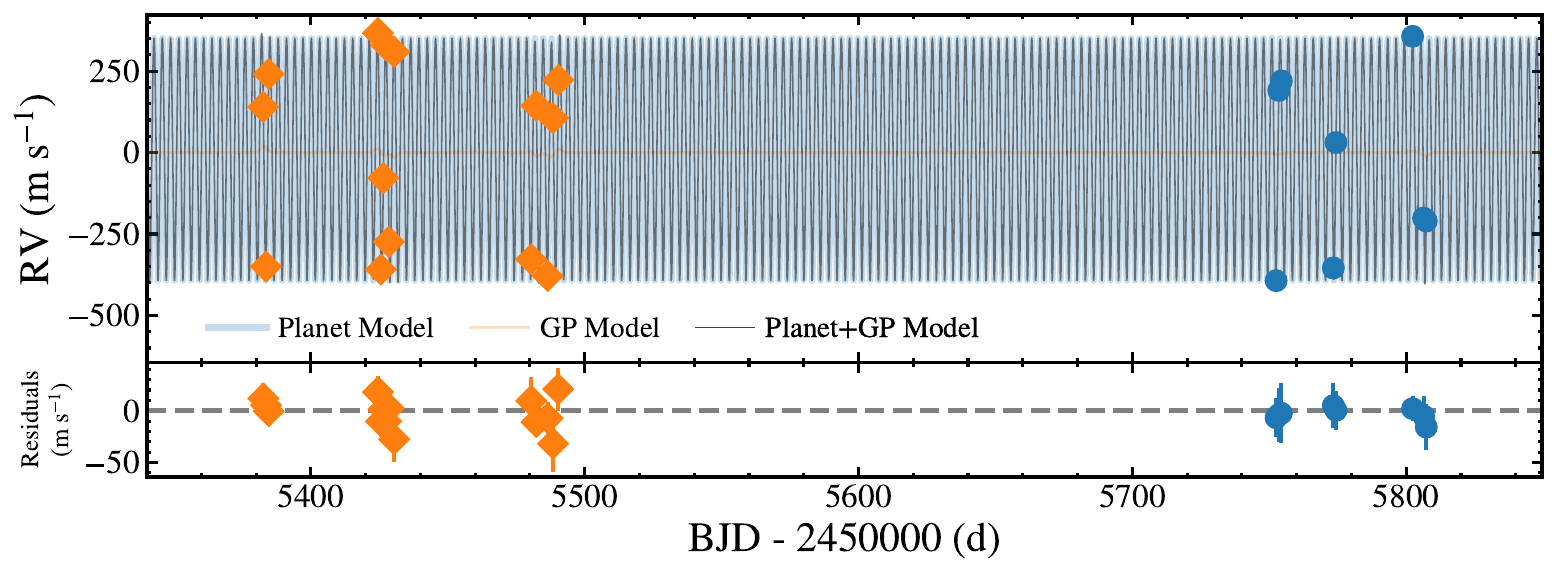}%
    
    \caption{The results of each model fit to the Kepler Quarter 10 photometry and SOPHIE and FIES RVs from \citet{Bonomo2012} and \citet{Endl2014} for Kepler-43. The panels are the same as in \autoref{fig:GJ_3021_results}.}
    \label{fig:Kepler-43_results}
\end{figure*}

\begin{figure*}
    \centering
    \hspace{0 mm}
    \includegraphics[width=0.8\linewidth]{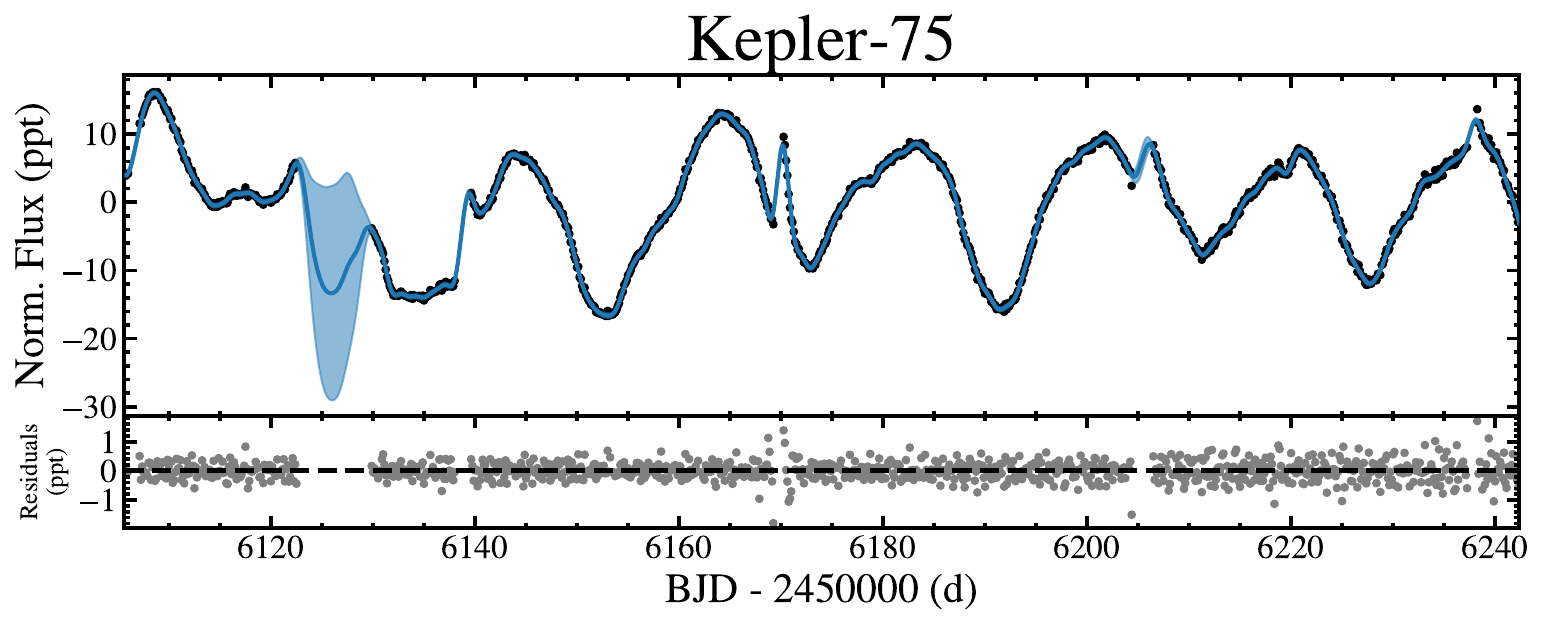}%

    \hspace{-3.25mm}
    \includegraphics[width=0.8175\linewidth]{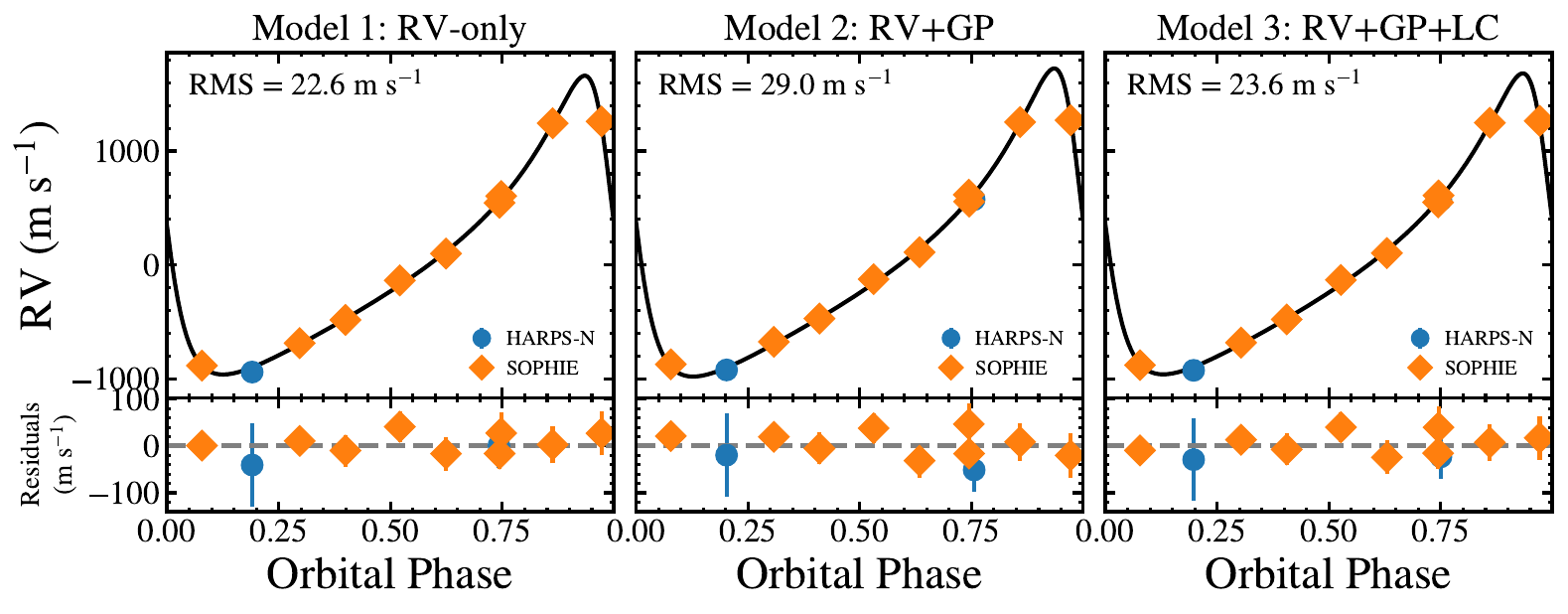}%

    \hspace{6.5mm}
    \includegraphics[width=0.785\linewidth]{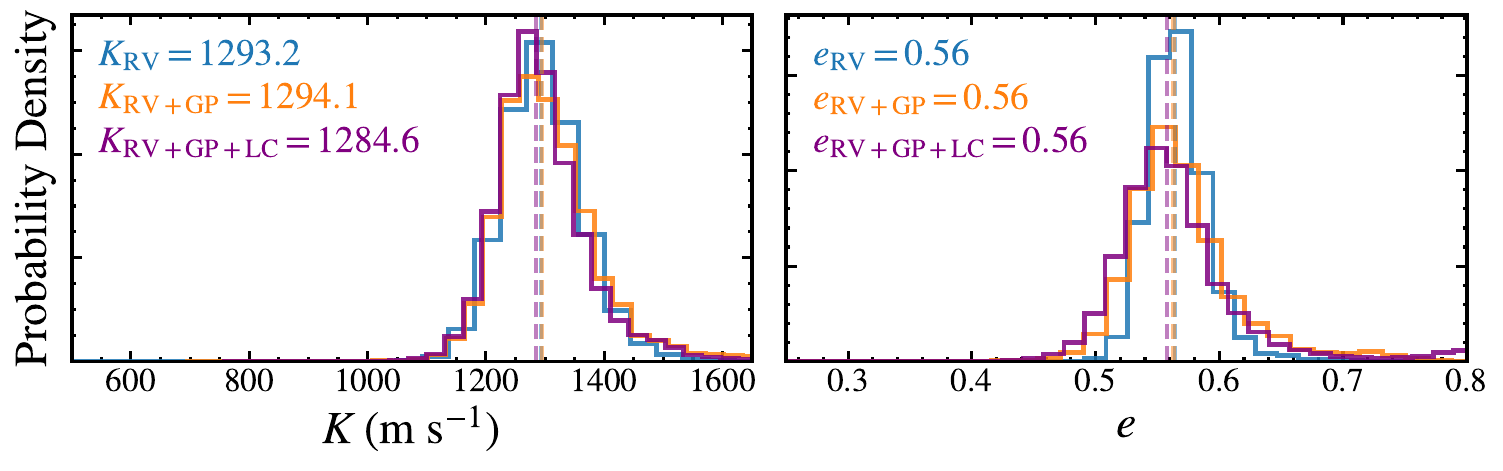}%

    \hspace{-2.5mm}
    \includegraphics[width=0.815\linewidth]{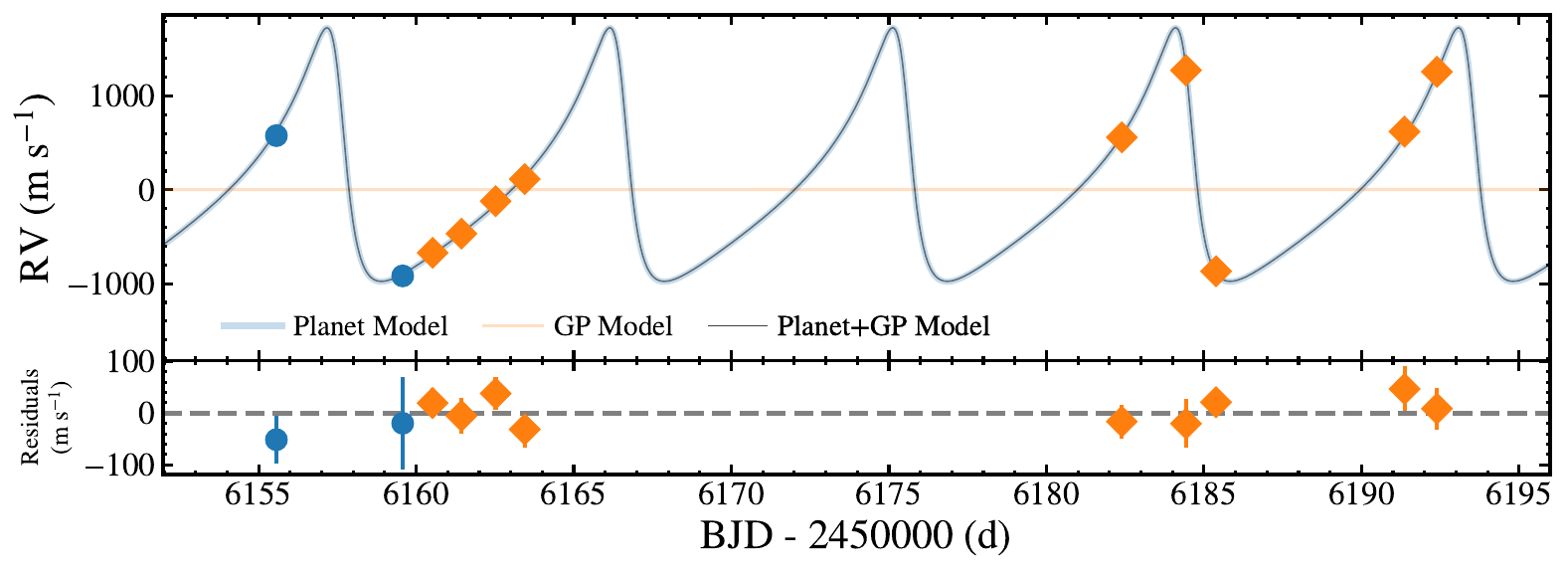}%
    
    \caption{The results of each model fit to the Kepler Quarter 15 photometry and HARPS-N and SOPHIE RVs from \citet{Hebrard2013} for Kepler-75. The panels are the same as in \autoref{fig:GJ_3021_results}.}
    \label{fig:Kepler-75_results}
\end{figure*}

\begin{figure*}
    \centering
    \hspace{0 mm}
    \includegraphics[width=0.8\linewidth]{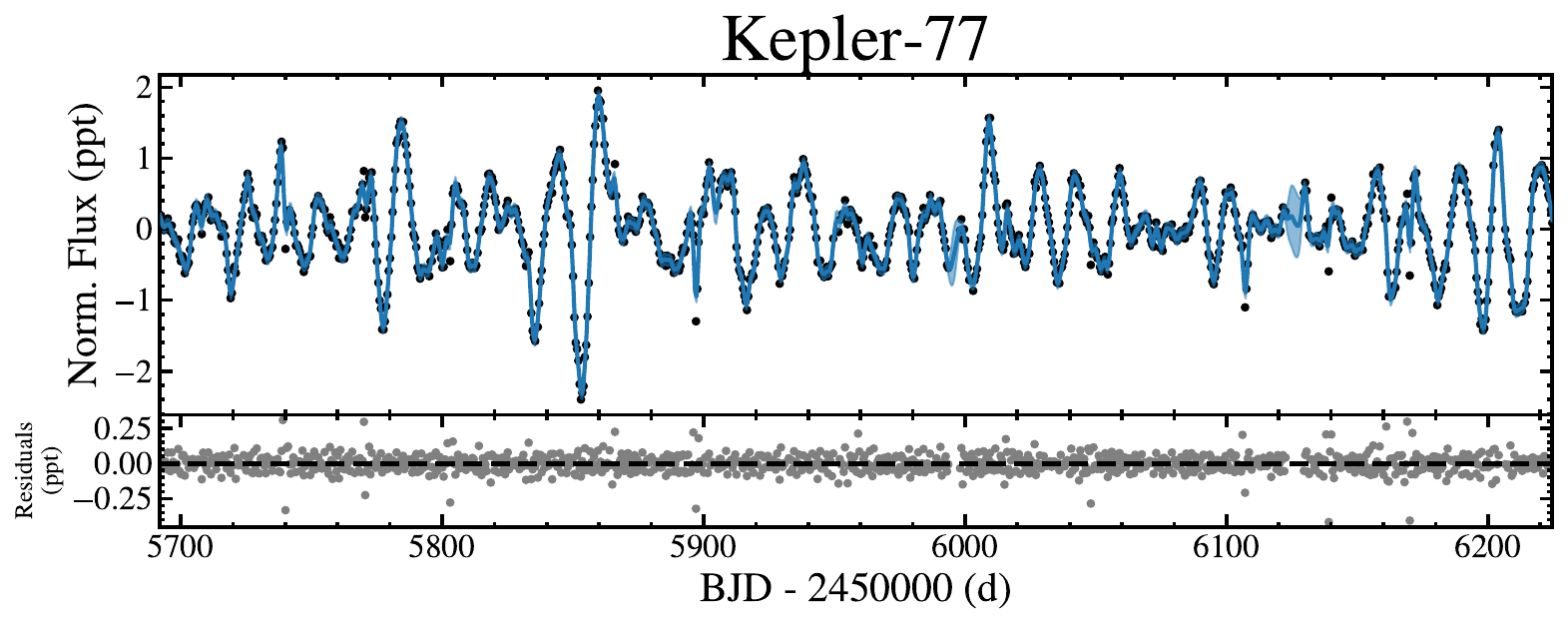}%

    \hspace{-3.25mm}
    \includegraphics[width=0.8175\linewidth]{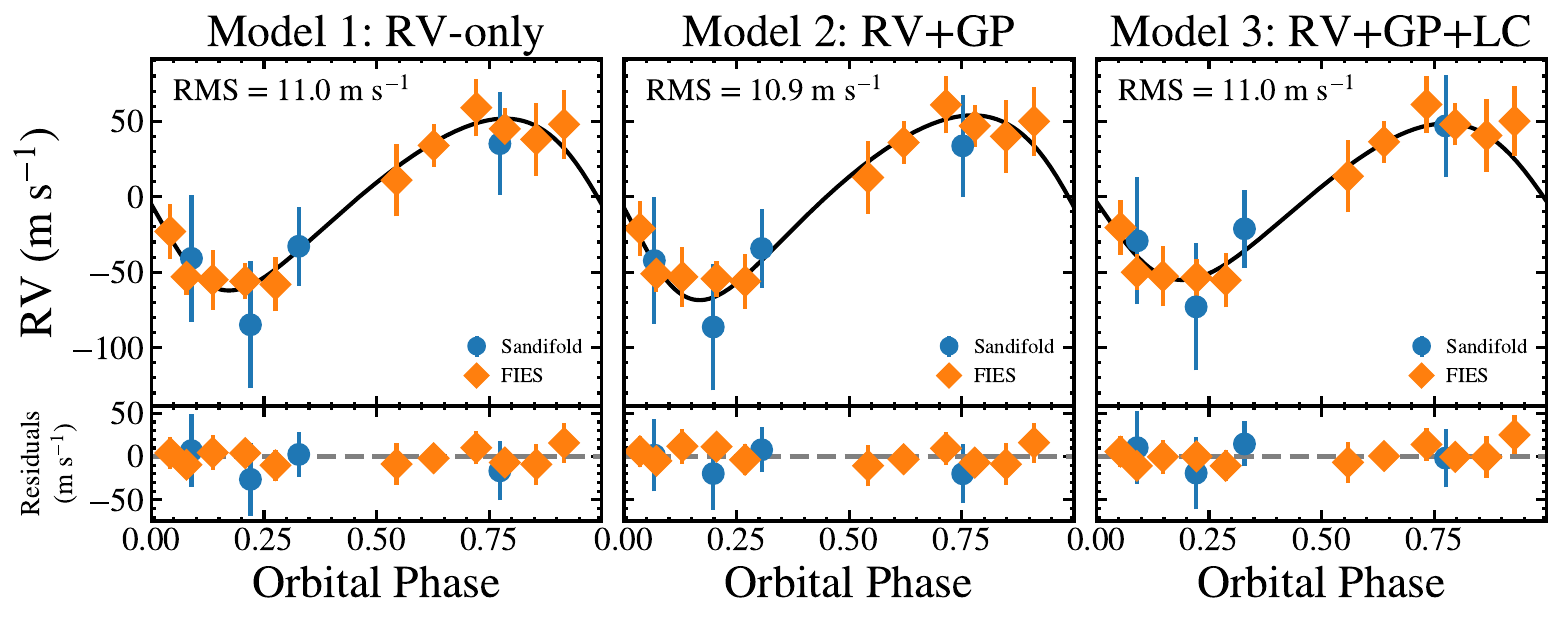}%

    \hspace{6.5mm}
    \includegraphics[width=0.785\linewidth]{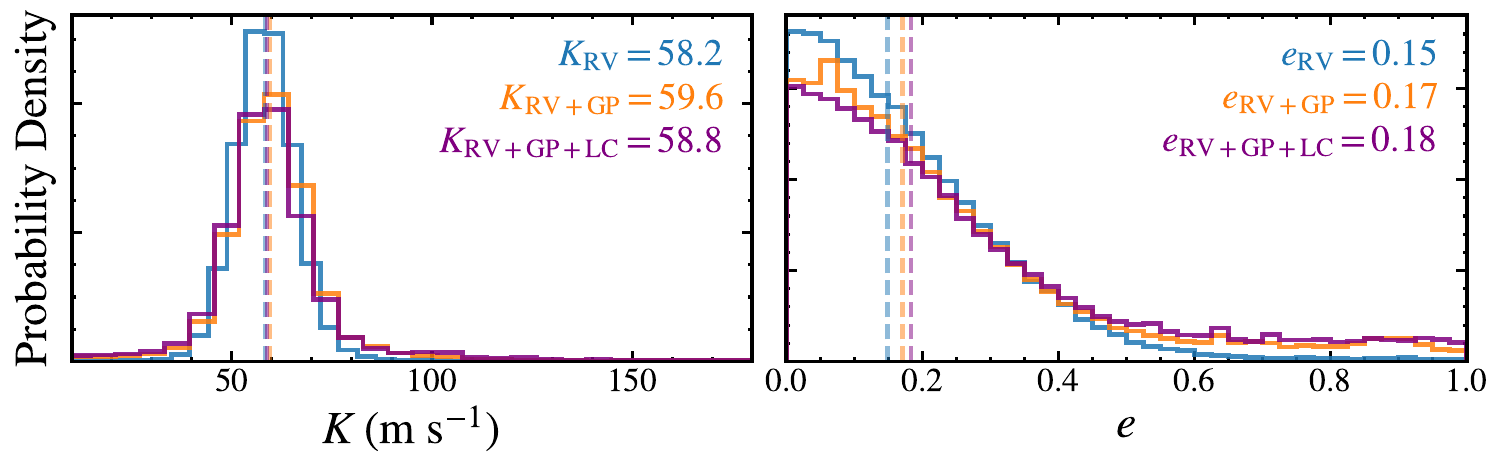}%

    \hspace{-2.5mm}
    \includegraphics[width=0.815\linewidth]{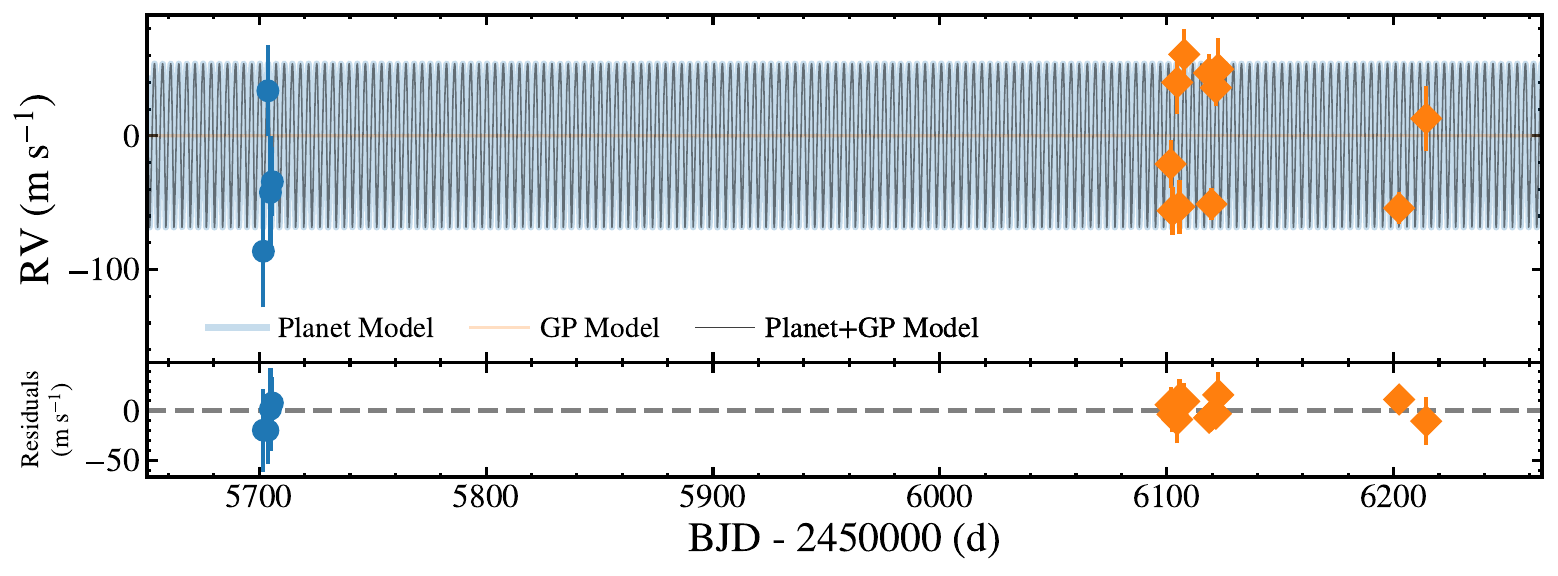}%
    
    \caption{The results of each model fit to the Kepler Quarters 9--15 photometry and Sandiford and FIES RVs from \citet{Gandolfi2013} for Kepler-77. The panels are the same as in \autoref{fig:GJ_3021_results}.}
    \label{fig:Kepler-77_results}
\end{figure*}

\begin{figure*}
    \centering
    \hspace{0 mm}
    \includegraphics[width=0.8\linewidth]{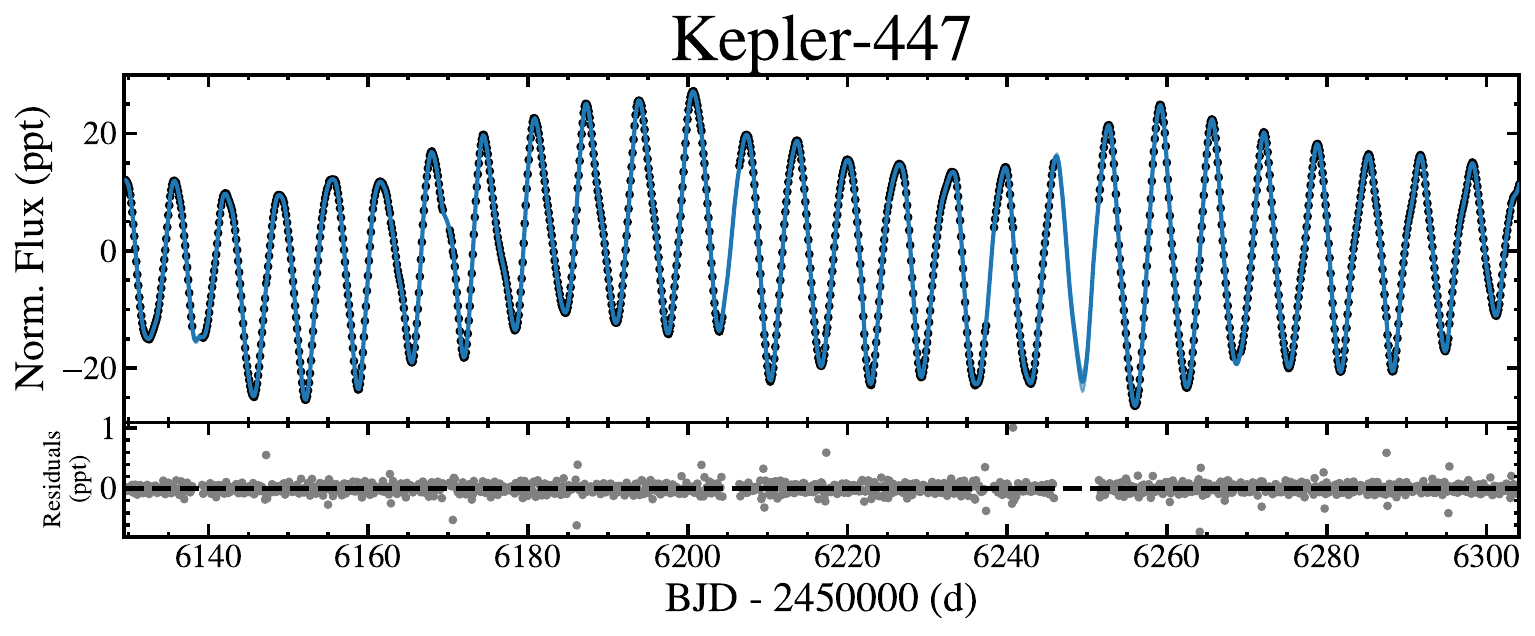}%

    \hspace{-3.25mm}
    \includegraphics[width=0.8175\linewidth]{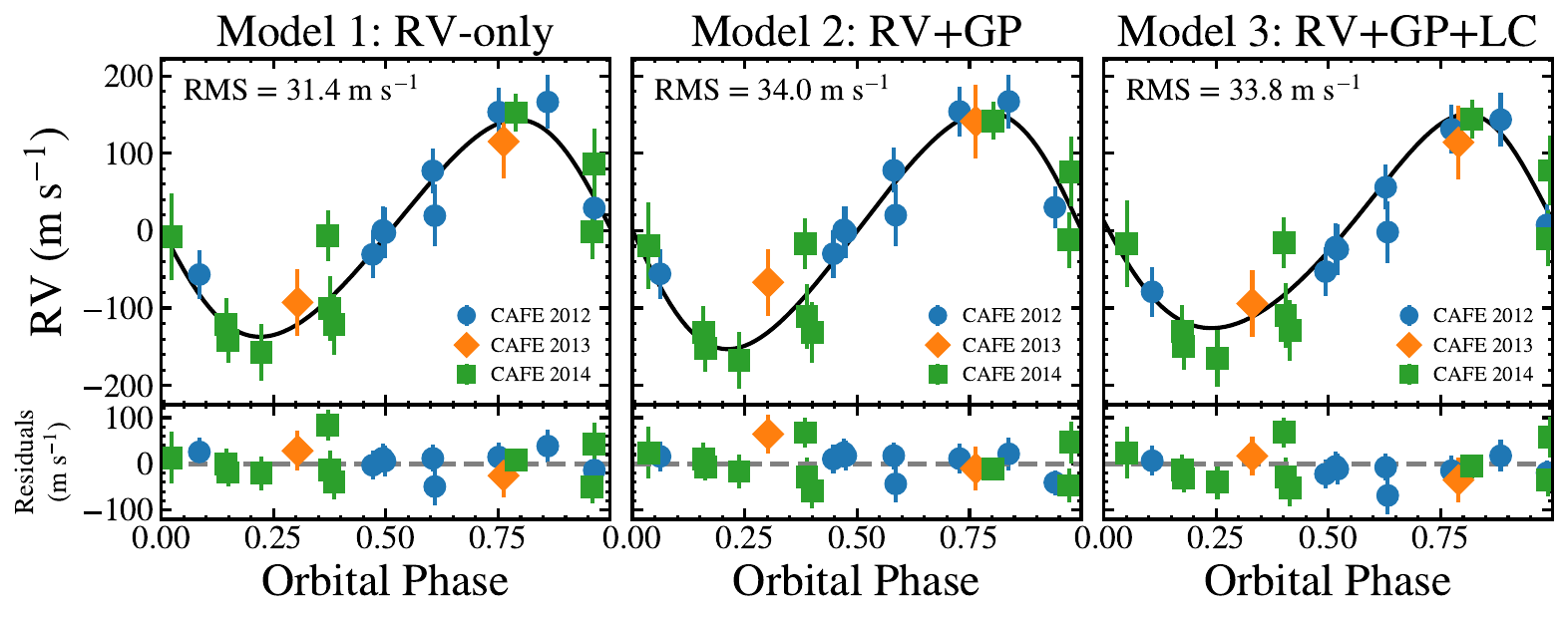}%

    \hspace{6.5mm}
    \includegraphics[width=0.785\linewidth]{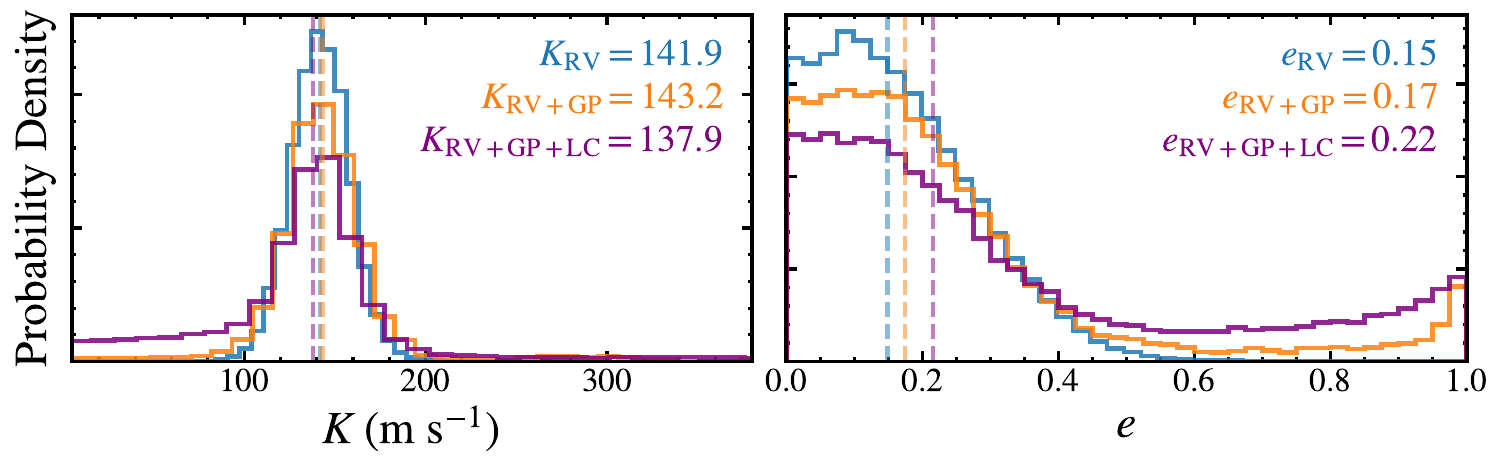}%

    \hspace{-2.5mm}
    \includegraphics[width=0.815\linewidth]{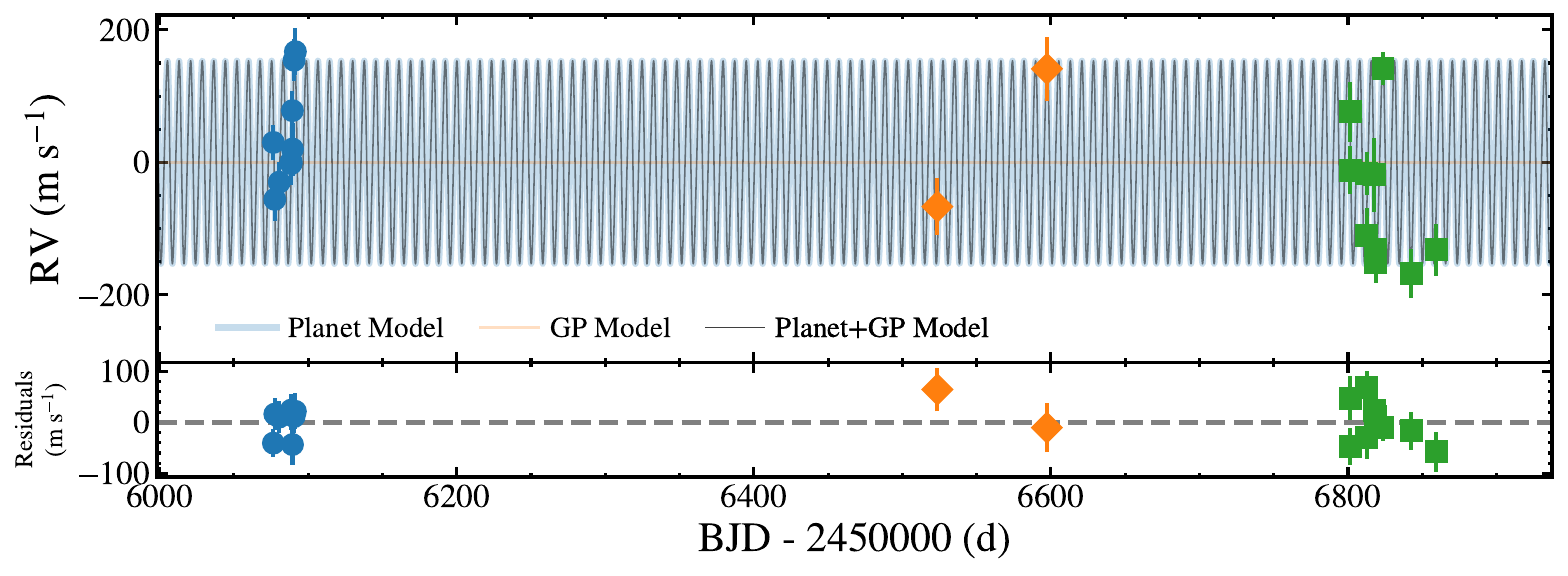}%
    
    \caption{The results of each model fit to the Kepler Quarters 14--15 photometry and CAFE RVs from \citet{Lillo-Box2015} for Kepler-447. The panels are the same as in \autoref{fig:GJ_3021_results}.}
    \label{fig:Kepler-447_results}
\end{figure*}

\begin{figure*}
    \centering
    \hspace{0 mm}
    \includegraphics[width=0.8\linewidth]{Figures/Kepler-539_lc_gp.pdf}%

    \hspace{-3.25mm}
    \includegraphics[width=0.8175\linewidth]{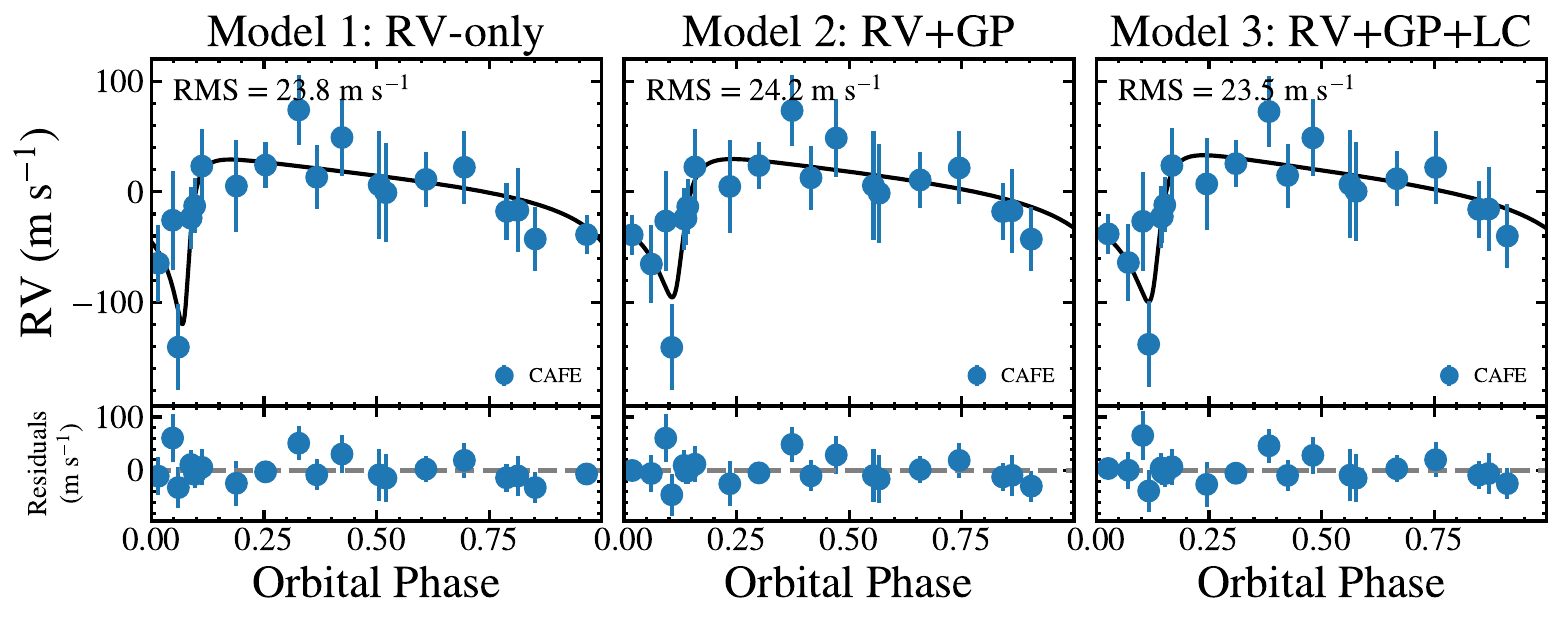}%

    \hspace{6.5mm}
    \includegraphics[width=0.785\linewidth]{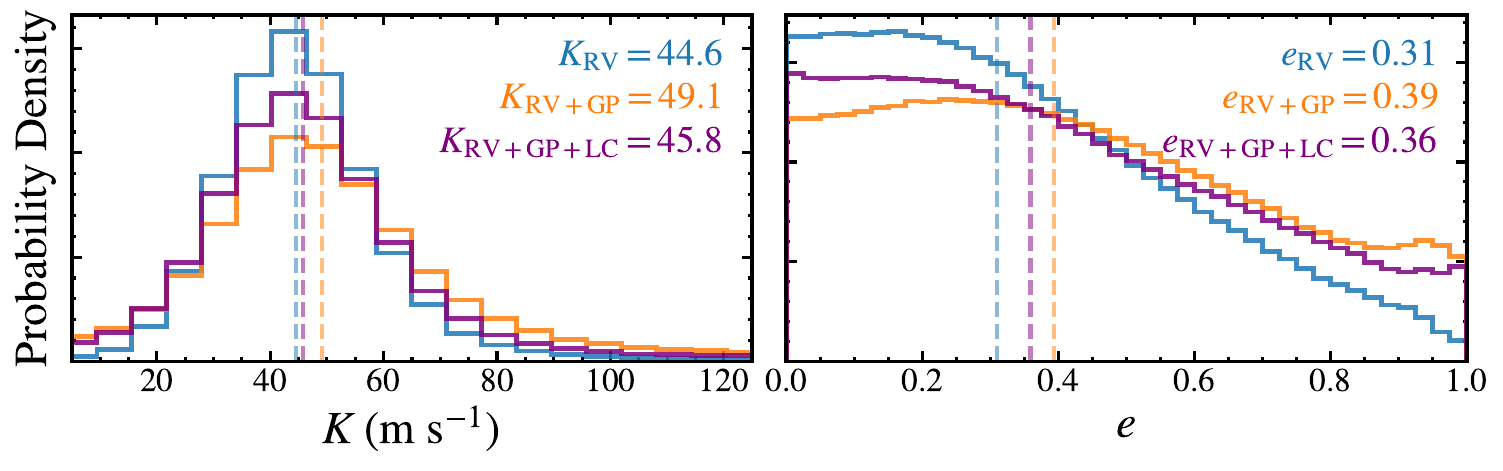}%

    \hspace{-2.5mm}
    \includegraphics[width=0.815\linewidth]{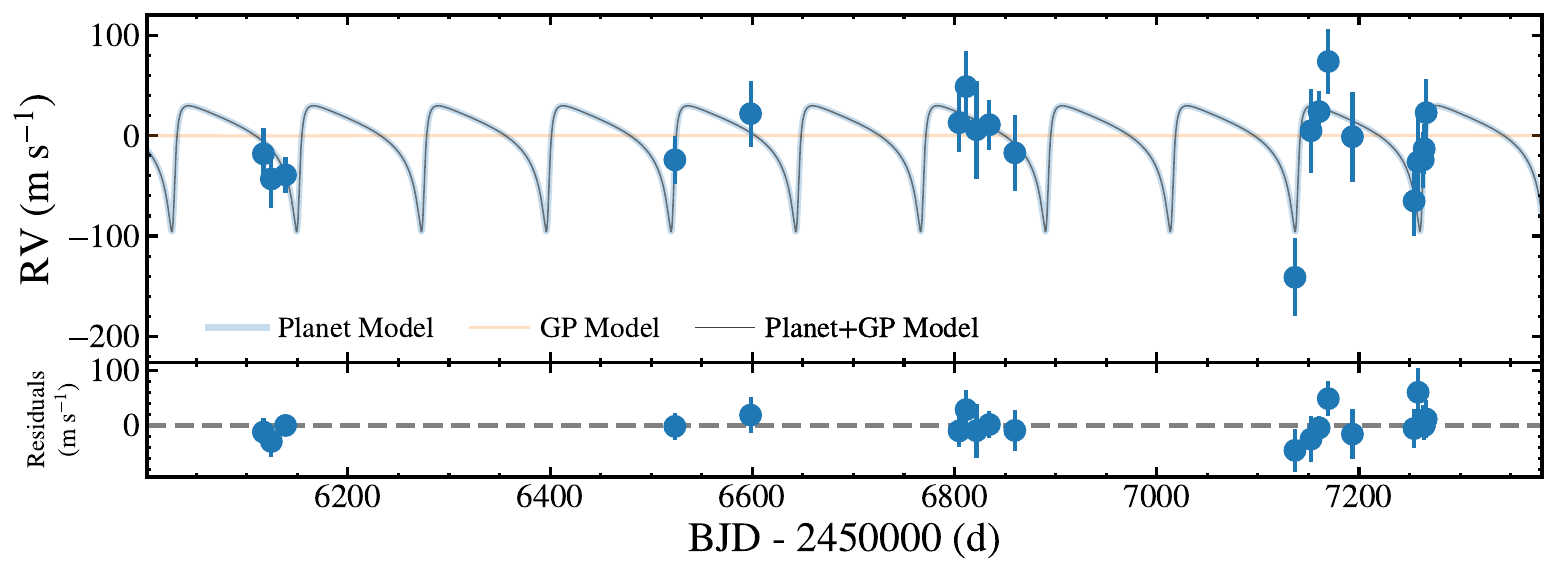}%
    
    \caption{The results of each model fit to the Kepler Quarters 14--17 photometry and CAFE RVs from \citet{Mancini2016} for Kepler-539. The panels are the same as in \autoref{fig:GJ_3021_results}.}
    \label{fig:Kepler-539_results}
\end{figure*}

\begin{figure*}
    \centering
    \hspace{0 mm}
    \includegraphics[width=0.8\linewidth]{Figures/K2-29_lc_gp.pdf}%

    \hspace{-3.25mm}
    \includegraphics[width=0.8175\linewidth]{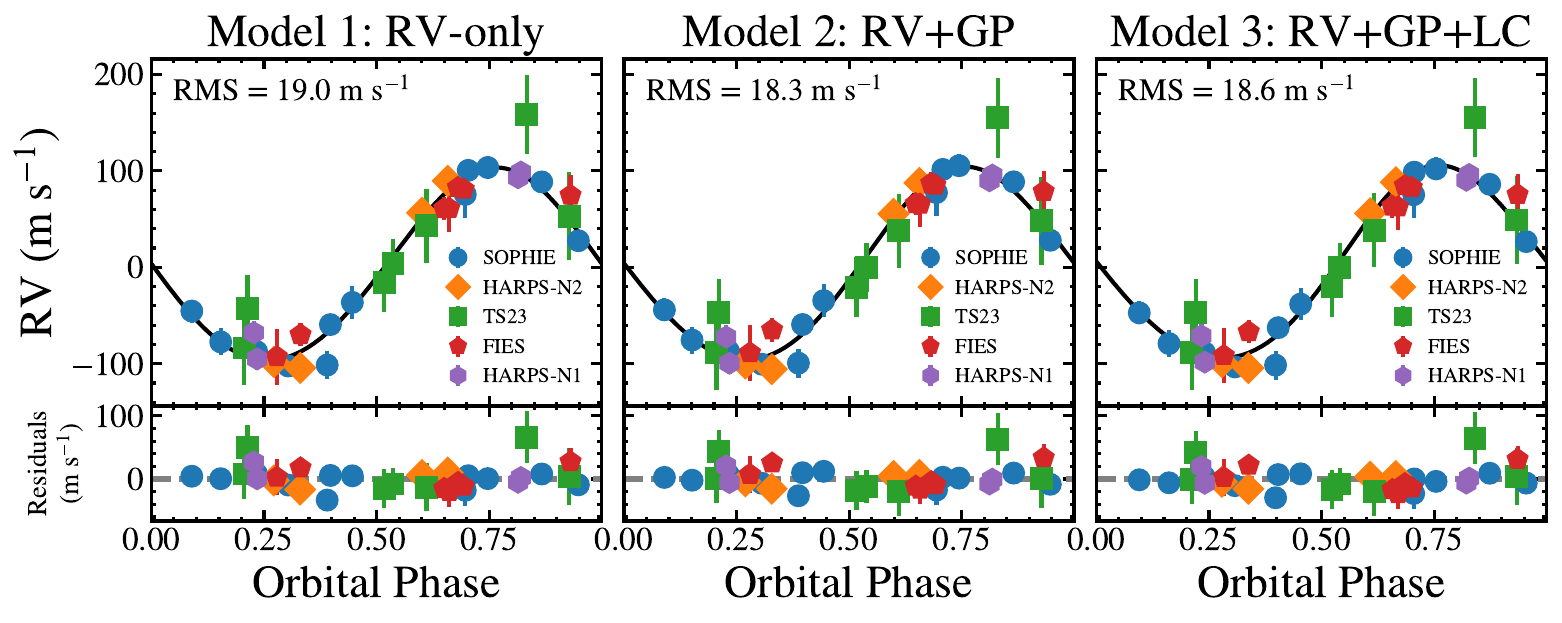}%

    \hspace{6.5mm}
    \includegraphics[width=0.785\linewidth]{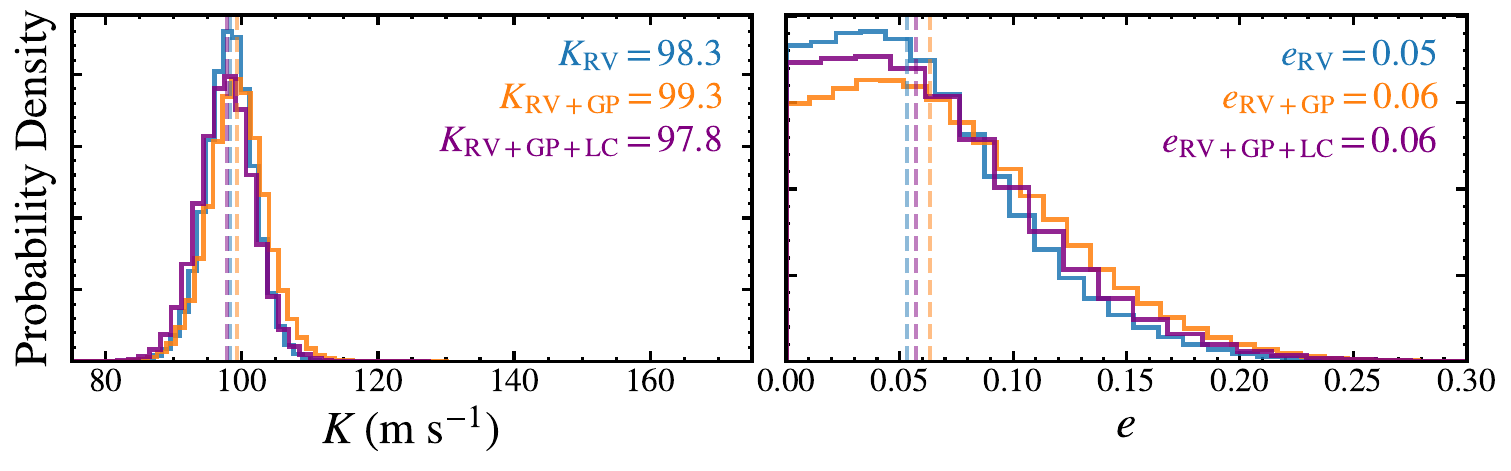}%

    \hspace{-2.5mm}
    \includegraphics[width=0.815\linewidth]{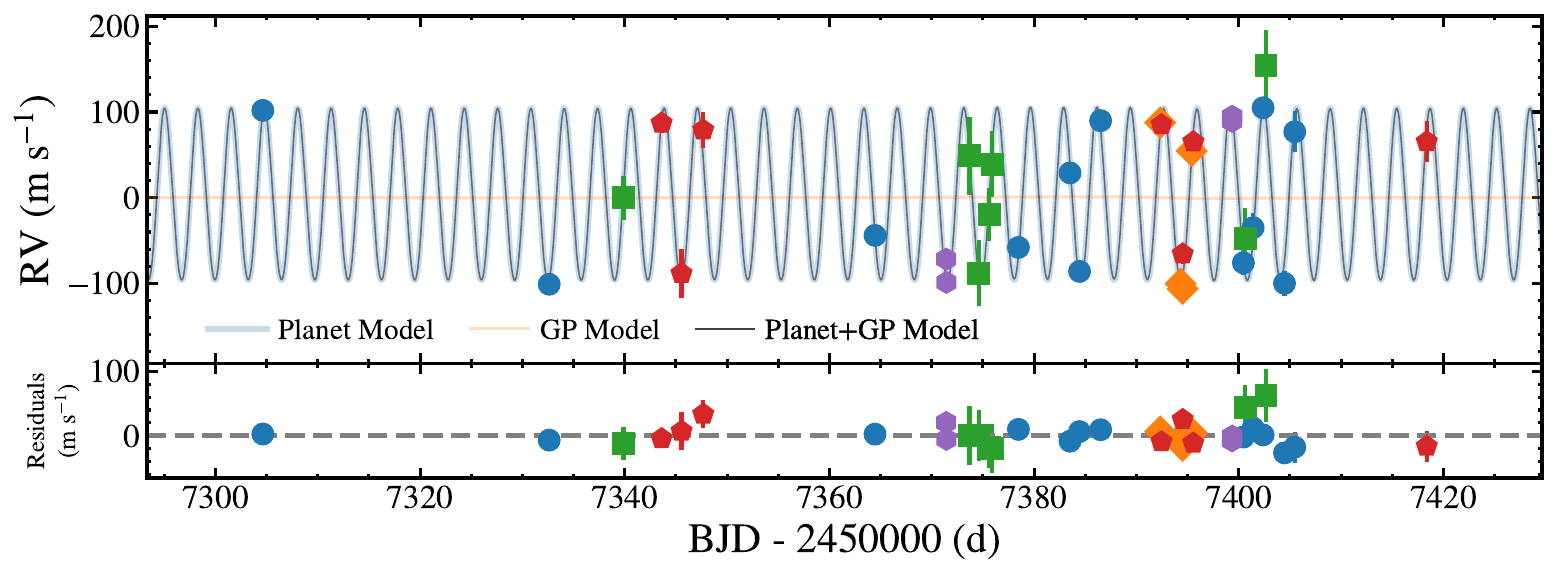}%
    
    \caption{The results of each model fit to the K2 Campaign 4 photometry and SOPHIE, HARPS-N, T223, and FIES RVs from \citet{Johnson2016} and \citet{Santerne2016} for K2-29. The panels are the same as in \autoref{fig:GJ_3021_results}.}
    \label{fig:K2-29_results}
\end{figure*}

\begin{figure*}
    \centering
    \hspace{0 mm}
    \includegraphics[width=0.8\linewidth]{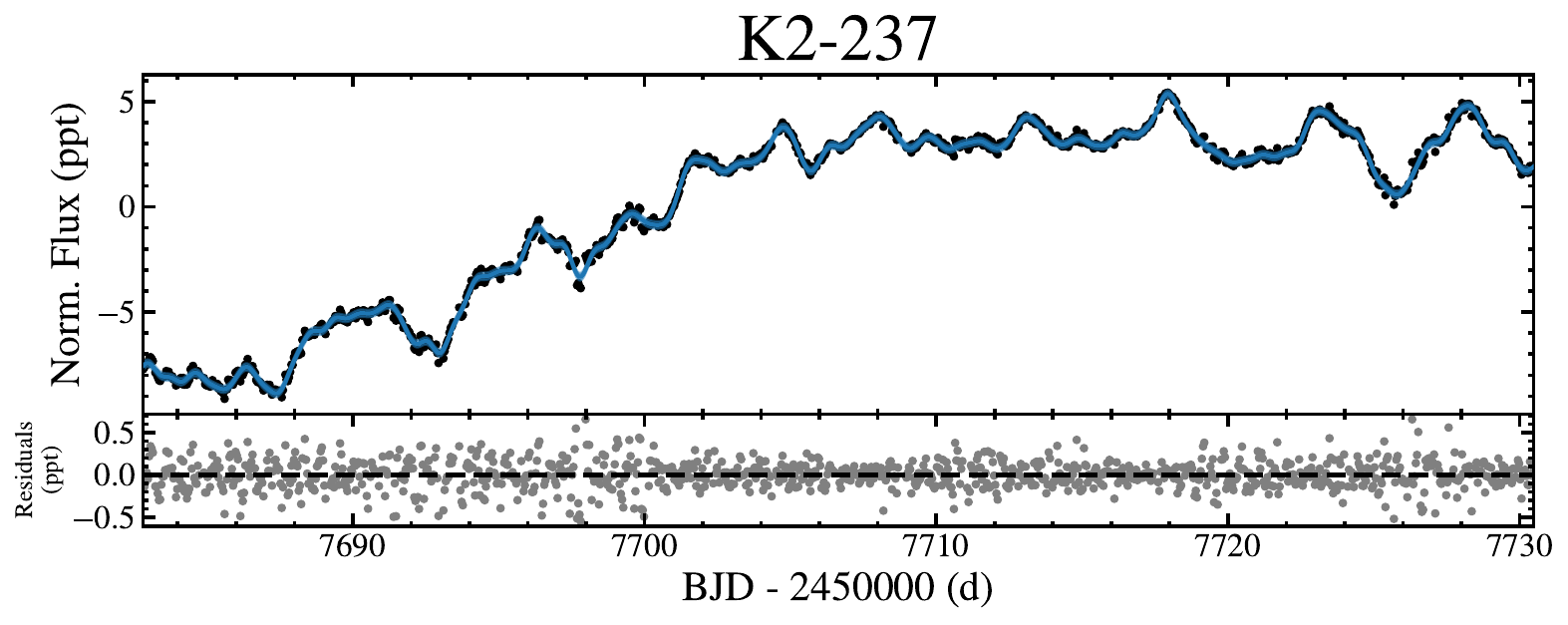}%

    \hspace{-3.25mm}
    \includegraphics[width=0.8175\linewidth]{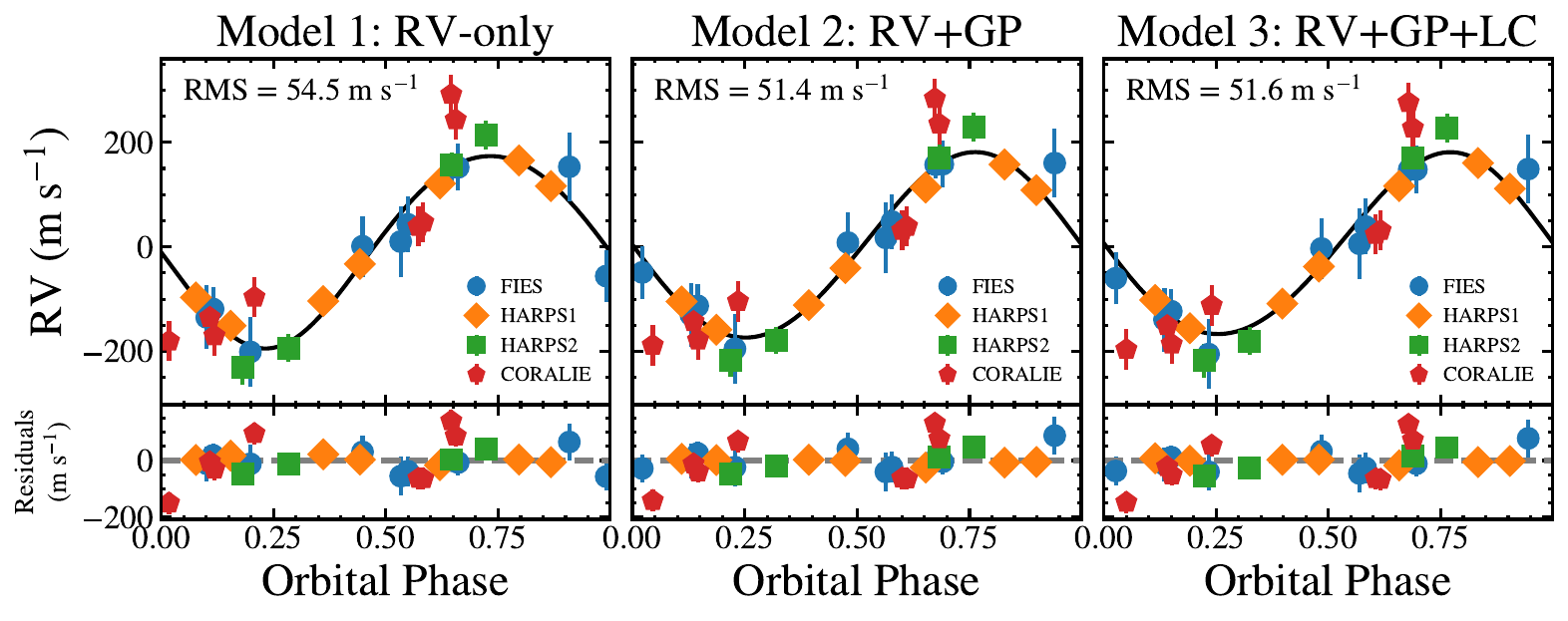}%

    \hspace{6.5mm}
    \includegraphics[width=0.785\linewidth]{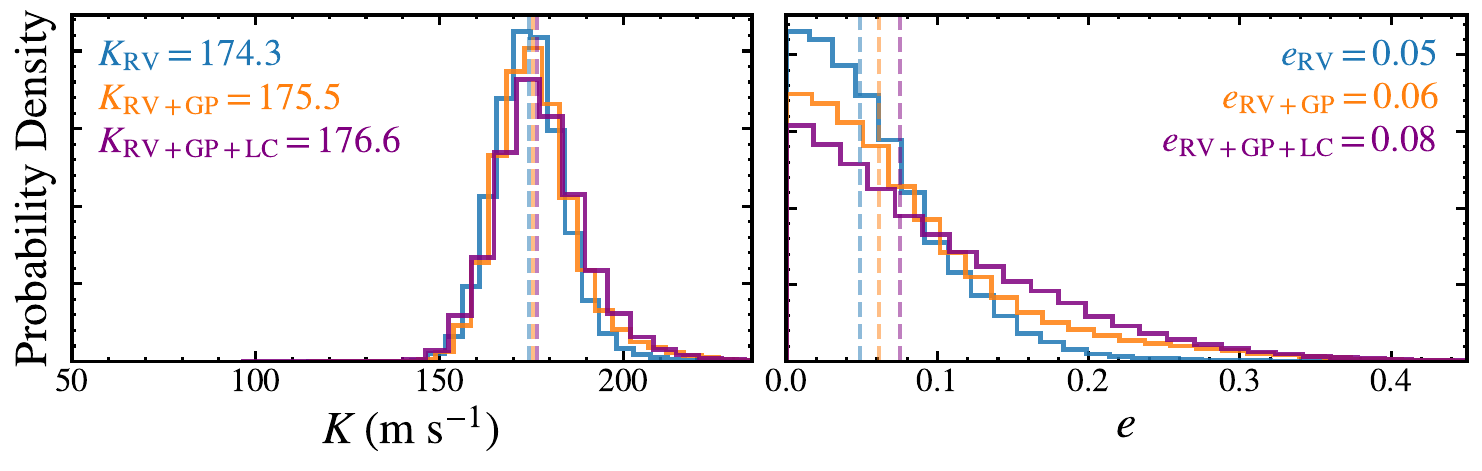}%

    \hspace{-2.5mm}
    \includegraphics[width=0.815\linewidth]{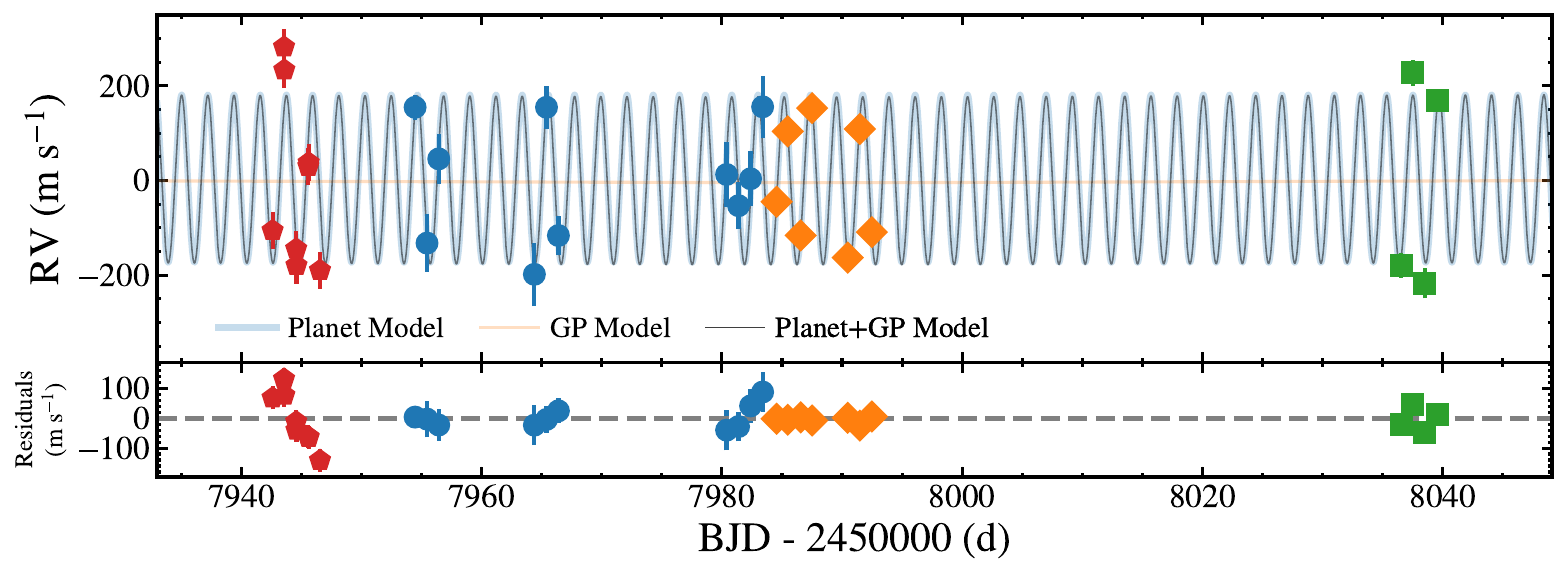}%
    
    \caption{The results of each model fit to the K2 Campaign 11 photometry and FIES, HARPS, and CORALIE RVs from \citet{Soto2018} and \citet{Smith2019} for K2-237. The panels are the same as in \autoref{fig:GJ_3021_results}.}
    \label{fig:K2-237_results}
\end{figure*}

\begin{figure*}
    \centering
    \hspace{0 mm}
    \includegraphics[width=0.8\linewidth]{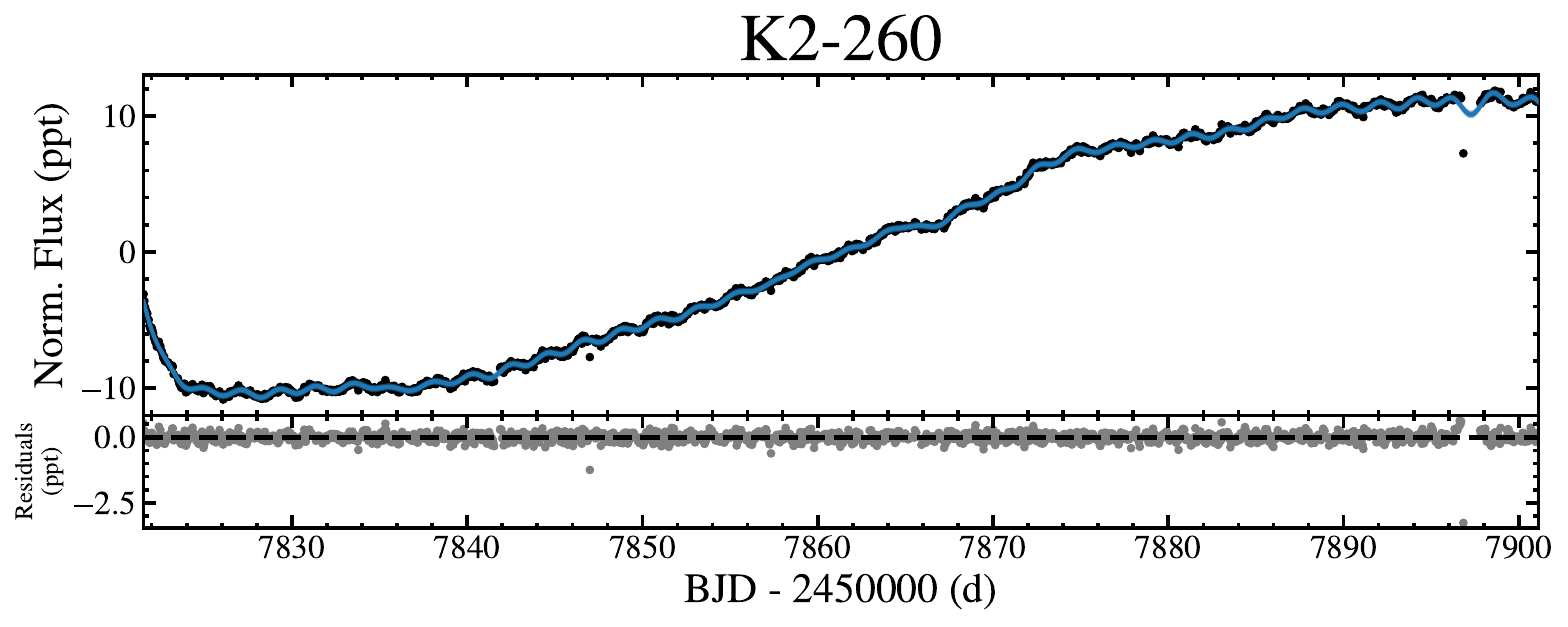}%

    \hspace{-3.25mm}
    \includegraphics[width=0.8175\linewidth]{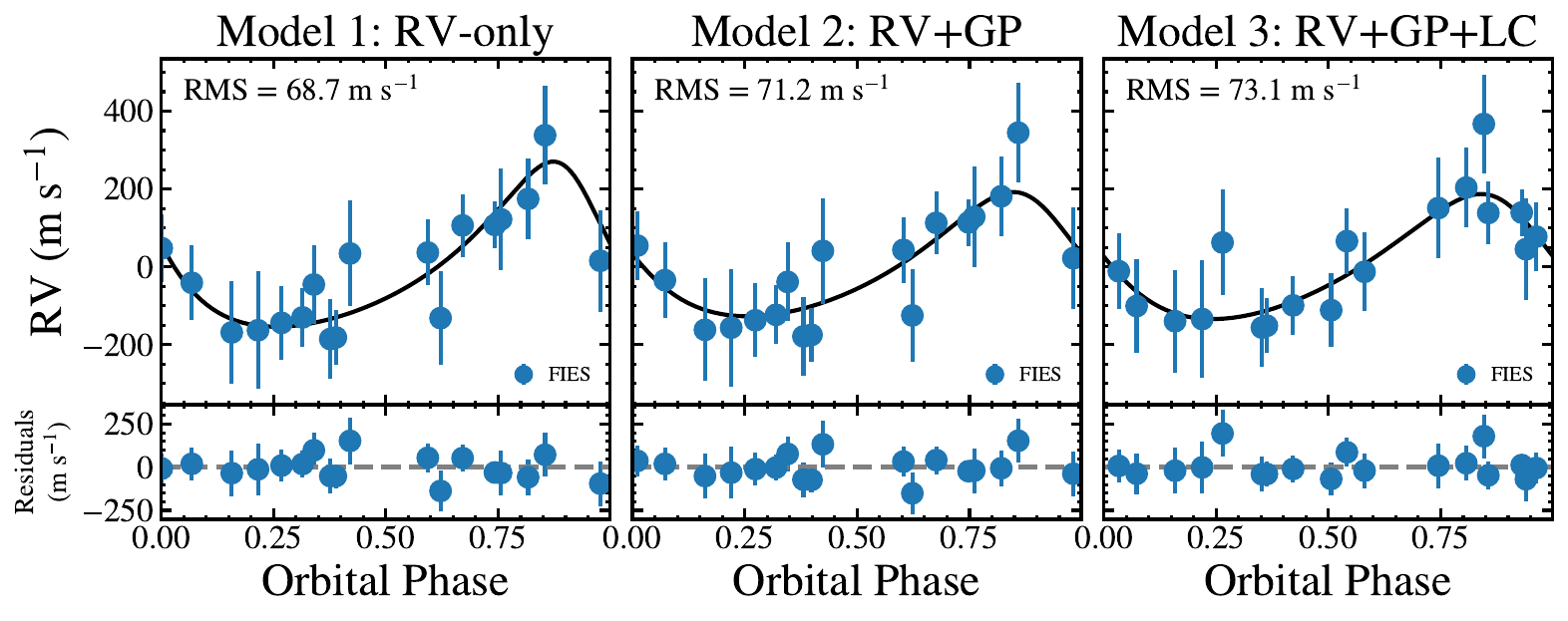}%

    \hspace{6.5mm}
    \includegraphics[width=0.785\linewidth]{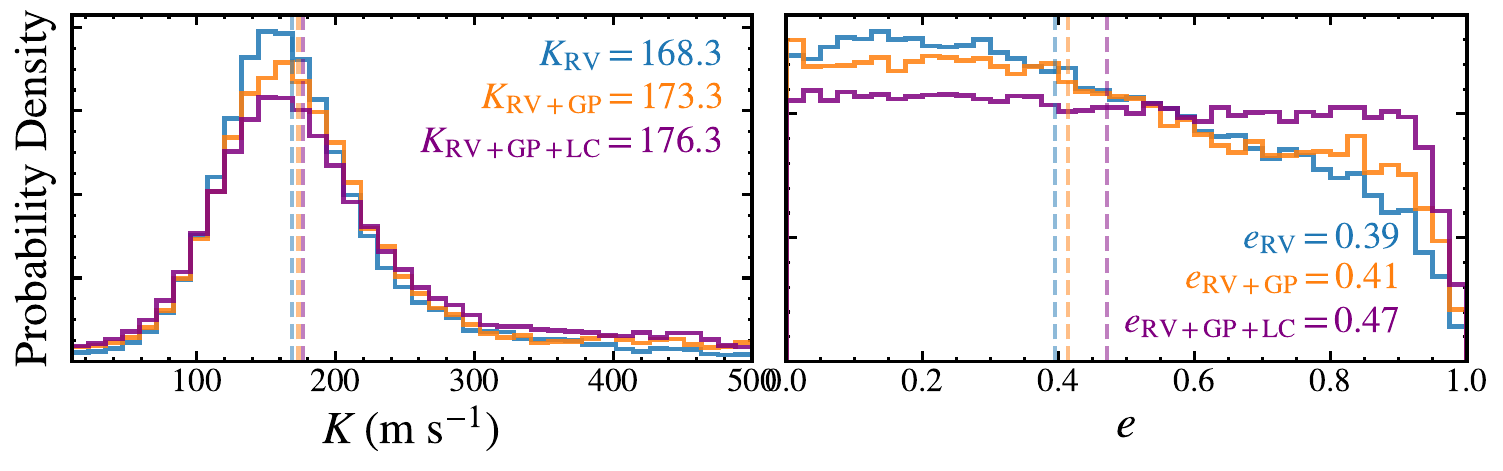}%

    \hspace{-2.5mm}
    \includegraphics[width=0.815\linewidth]{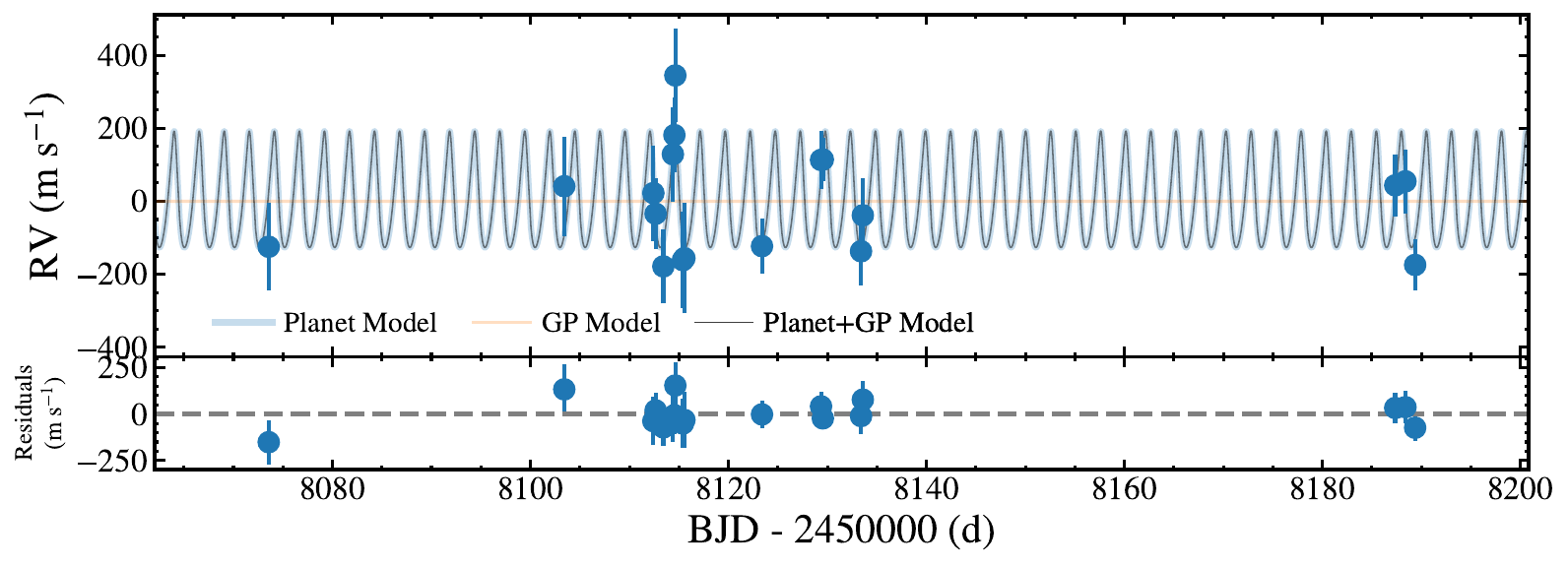}%
    
    \caption{The results of each model fit to the K2 Campaign 13 photometry and FIES RVs from \citet{Johnson2018} for K2-260. The panels are the same as in \autoref{fig:GJ_3021_results}.}
    \label{fig:K2-260_results}
\end{figure*}

\begin{figure*}
    \centering
    \hspace{0 mm}
    \includegraphics[width=0.8\linewidth]{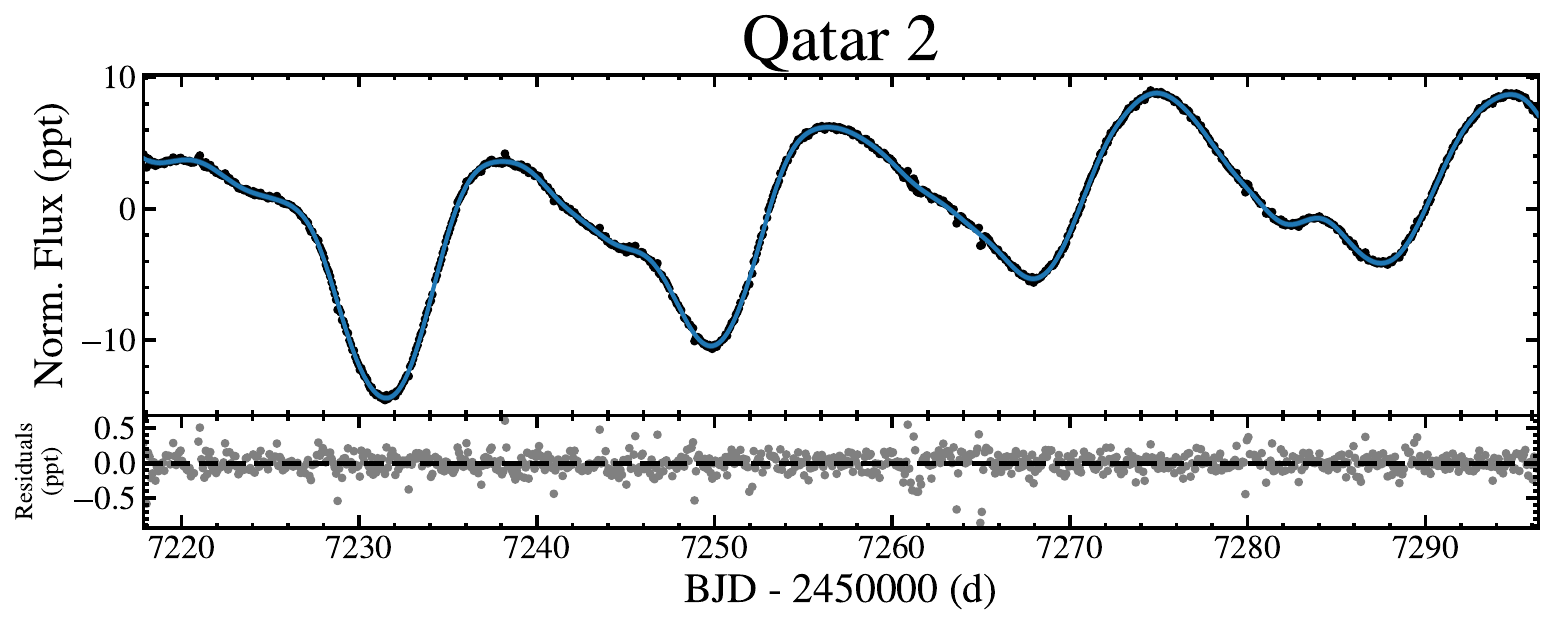}%

    \hspace{-3.25mm}
    \includegraphics[width=0.8175\linewidth]{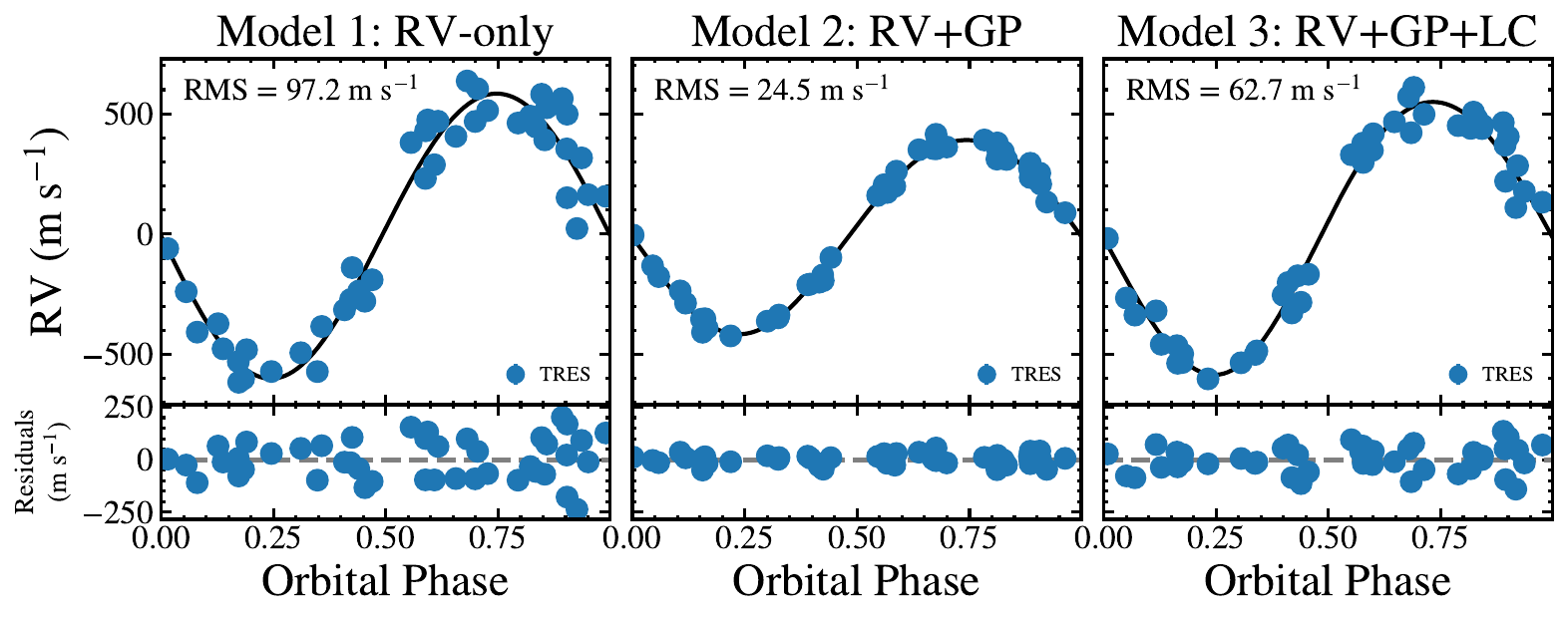}%

    \hspace{6.5mm}
    \includegraphics[width=0.785\linewidth]{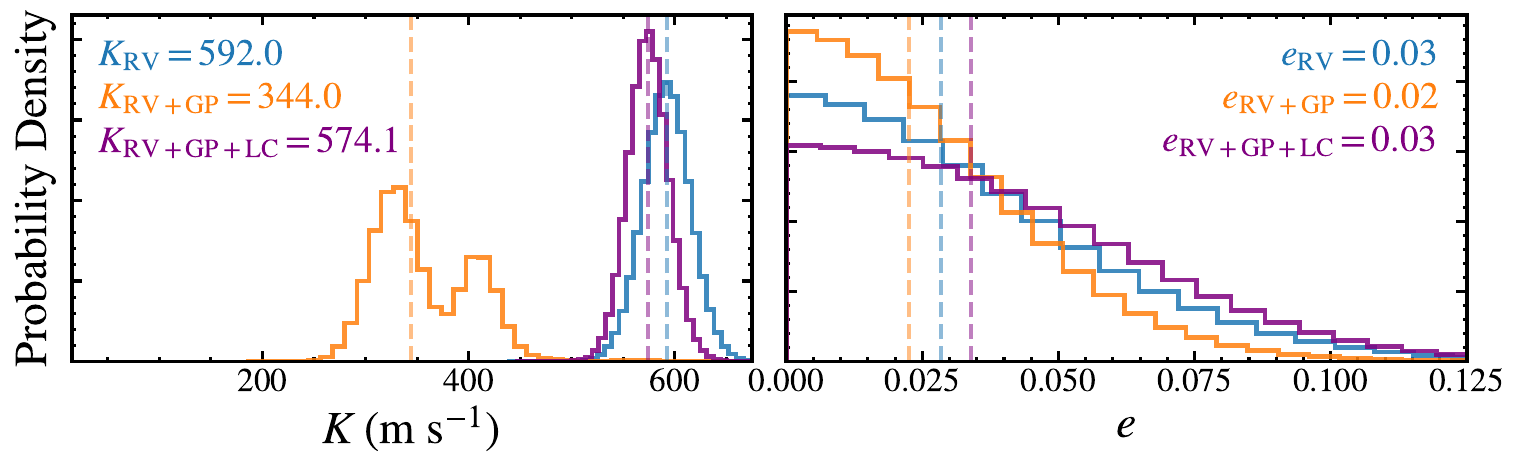}%

    \hspace{-2.5mm}
    \includegraphics[width=0.815\linewidth]{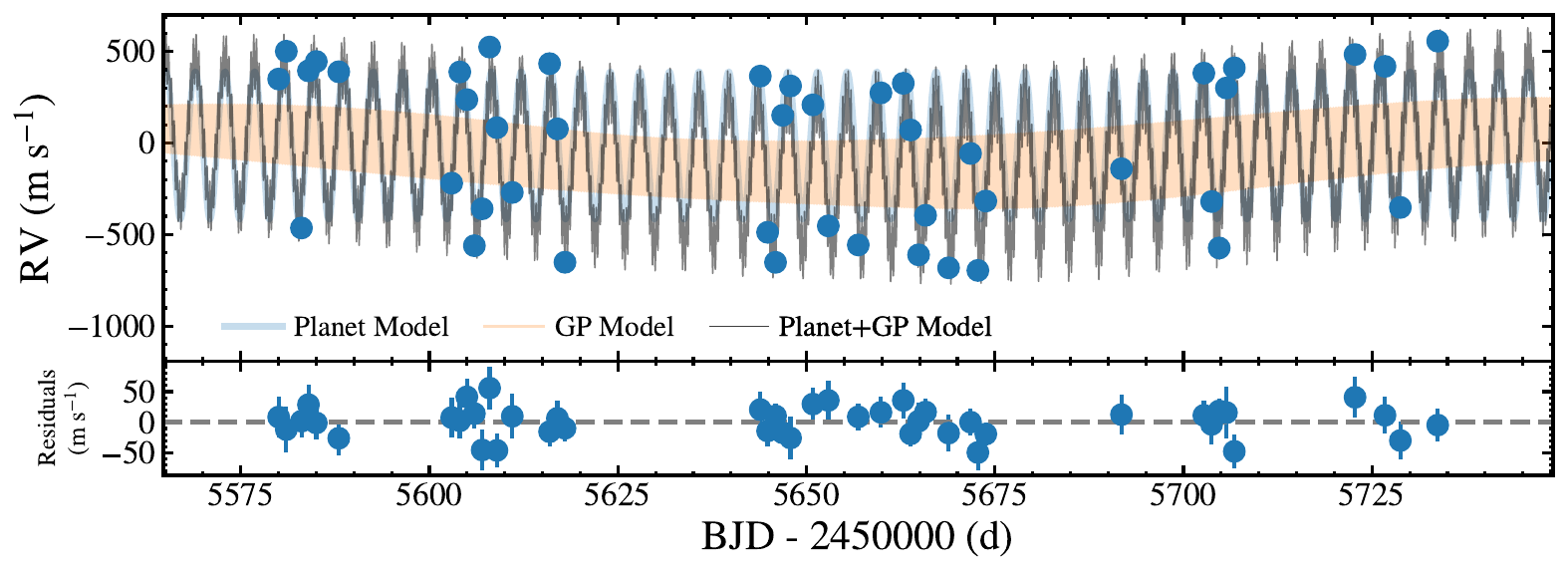}%
    
    \caption{The results of each model fit to the K2 Campaign 6 photometry and TRES RVs from \citet{Bryan2012} for Qatar 2. The panels are the same as in \autoref{fig:GJ_3021_results}.}
    \label{fig:Qatar_2_results}
\end{figure*}

\begin{figure*}
    \centering
    \hspace{0 mm}
    \includegraphics[width=0.8\linewidth]{Figures/WASP-180_A_lc_gp.pdf}%

    \hspace{-3.25mm}
    \includegraphics[width=0.8175\linewidth]{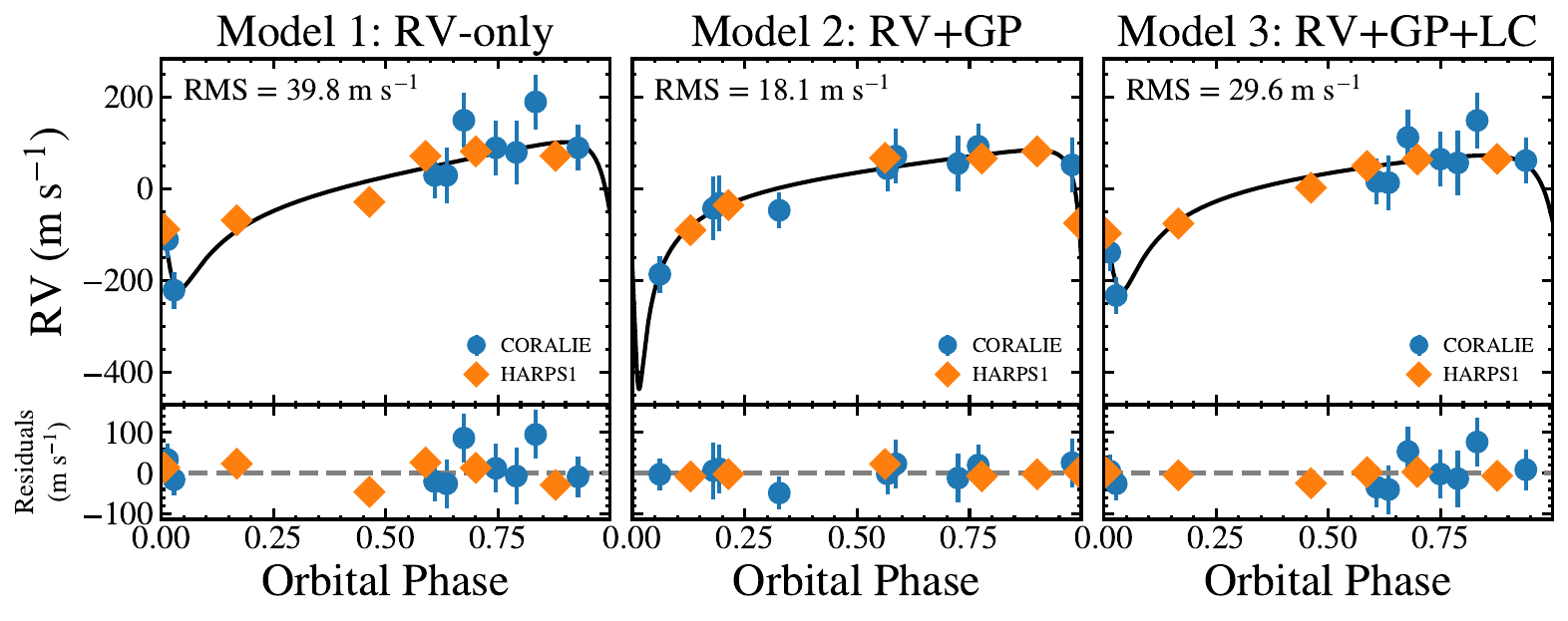}%

    \hspace{6.5mm}
    \includegraphics[width=0.785\linewidth]{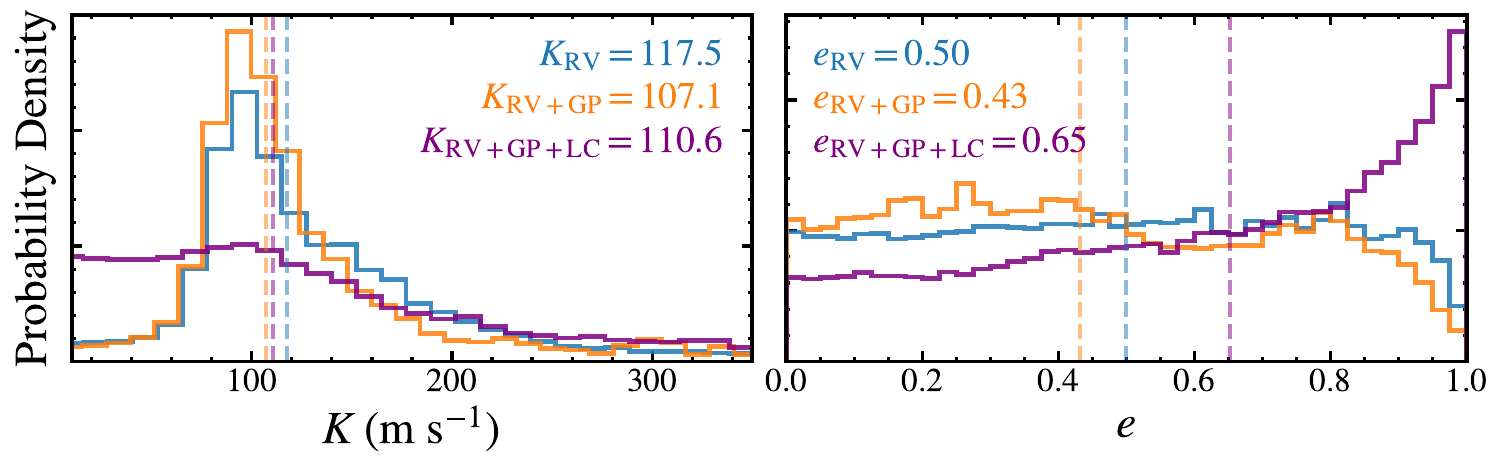}%

    \hspace{-2.5mm}
    \includegraphics[width=0.815\linewidth]{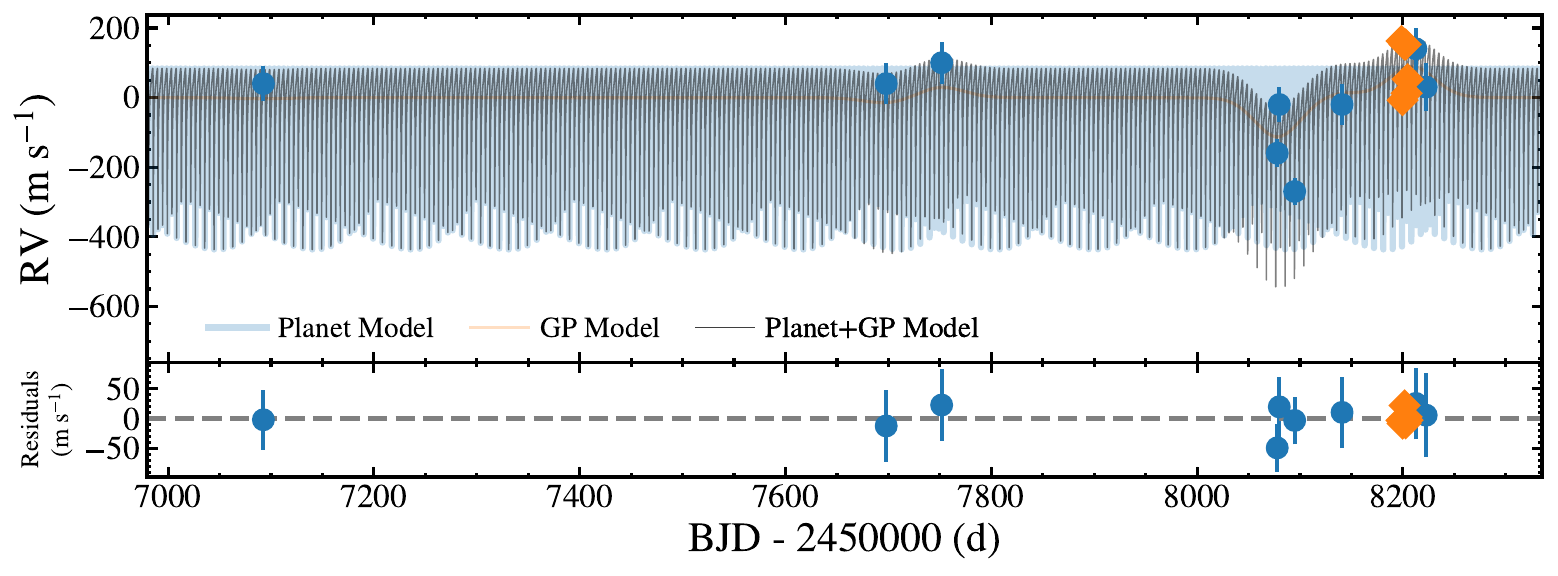}%
    
    \caption{The results of each model fit to the TESS Sector 34 photometry and CORALIE and HARPS RVs from \citet{Reinhold2020} for WASP-180 A. The panels are the same as in \autoref{fig:GJ_3021_results}.}
    \label{fig:WASP-180_A_results}
\end{figure*}

Figures~\ref{fig:GJ_3021_results}--\ref{fig:WASP-180_A_results} show the model fit results for each system in our sample. The top panel displays the TESS, Kepler, or K2 photometry, best-fit GP model, and residuals. The RVs, phased RV curve (with stellar activity component of the model removed), and residuals for each of the three models are shown in the second panel. The corresponding posterior distributions for the RV semi-amplitude and eccentricity are presented in the third panel. Finally, the bottom panel shows all RVs over time, the best-fit planet, GP, and combined models for the Model 2 framework, and the fit residuals to illustrate how stellar activity is approximated by the GP.

\clearpage

\bibliography{gp_rv_lc}{}
\bibliographystyle{aasjournal}

\end{document}